\documentclass[prb,twocolumn]{revtex4-2}

\usepackage{}
\usepackage{hyperref}
\usepackage{epsfig}
\usepackage{graphicx}
\usepackage{amsfonts,amssymb}
\usepackage{amsmath}
\usepackage{color}
\usepackage {times}
\definecolor{refcolor}{rgb}{0.0,0.0,1.0}

\hypersetup{
    colorlinks=true,
    linkcolor=refcolor,
    citecolor=refcolor,
    urlcolor=refcolor,
    breaklinks=true,
}
\begin{document}

\title{Nonequilibrium dynamics of suppression, revival, and 
loss of charge order\\
in a laser-pumped electron-phonon system}

\author{Sankha Subhra Bakshi$^1$,  
Debraj Bose$^1$, Arijit Dutta$^2$ and Pinaki Majumdar$^1$}

\affiliation{$^1$~Harish-Chandra Research Institute 
(A CI of Homi Bhabha National Institute), 
Chhatnag Road, Jhusi, Allahabad 211019\\
$^2$~Institut für Theoretische Physik, Goethe-Universität, 
60438 Frankfurt am Main, Germany
}
\date{\today}

\begin{abstract}
An electron-phonon system at commensurate filling often displays charge order 
(CO) in the ground state. Such a system subject to a laser pulse shows a wide 
variety of behaviour. A weak pulse sets up low amplitude oscillations in the 
order parameter, with slow decay to a slightly suppressed value. A strong pulse 
leads to destruction of the charge order with the order parameter showing rapid, 
oscillatory, decay to zero. The regime in between, separating the weak pulse 
CO sustained state from the strong pulse CO destroyed state, shows an initial 
rapid decay of the order parameter, followed by a low amplitude quiescent state, 
and the rise to a finite steady state value over a timescale $\tau_{cr}$. 
The steady state value goes to zero and $\tau_{cr}$
diverges as the pulse amplitude is increased towards a critical strength. 
We provide a complete characterisation of the dynamics 
in this nonequilibrium problem for varying electron-phonon coupling and pulse 
strength, highlight the multiple gap closing and opening transitions that occur, 
and suggest a simple dynamical model incorporating the pulse induced 
nonequilibrium electron population. 
\end{abstract}
\maketitle

\section{Introduction}

Recent experiments have started to explore the quantum dynamics
in correlated electron systems 
\cite{Bloch,Beaure,Carpene08,Beaud,Fiebig,Iwai,
Matsuzaki,cervantes2019,Okamoto,cho2015,Ichikawa,Liu,Budden,
Fausti,Han2015,Zhang20,Makler21,Schmitt11,
Wolf14,Hellmann10}.
The method involves imparting
energy to the electron system via an electromagnetic pulse and
studying the evolution by probing time dependence of the
photoemission spectra or optical conductivity \cite{Carpene09,Sobota21,Eich14,Gedik17,
Devereaux16,Mitrano20}. Although all experimental
systems are in some thermal environment, at short enough time
the pulse driven systems probe quantum states at
energies much higher than what temperature alone would allow.
To that extent one probes ``athermal'' dynamics.
The excess energy
imparted by the pulse can lead to various nonequilibrium
phenomena including melting of order \cite{Carpene08, 
Beaud, Iwai, Matsuzaki, cervantes2019, Fiebig,Okamoto,cho2015},
the emergence of new order \cite{Fausti, Ichikawa, Liu, Budden} 
, and ``transitions'' in transport and
optical behaviour \cite{Okamoto,Iwai}
- all of these arising from a non thermal
population of excitations.

For systems coupled to a thermal bath, at long time the added
energy is dissipated  and one attains the equilibrium state,
but the transient dynamics can be very interesting.
For a system that is nominally thermally isolated,
i.e, cannot dissipate the added energy, 
neither the transient nor the steady
state is known a priori. In particular, one key question
is whether the long time state admits an equilibrium
description with an effective temperature. Finally,
a system that is periodically driven can display steady
states which do not have an equilibrium counterpart 
\cite{Rech,Oka,Bukov}.

Our focus is on the dynamics in a ``thermally isolated''
charge ordered electron-phonon (EP) system 
subject to a laser pulse, and the spatio-temporal 
dynamics that arises as a result.
One expects that as the added energy increases the
charge order will be progressively suppressed, and
ultimately destroyed. That is indeed the answer at
long times. What is far less obvious is the dynamics
associated with this process. This paper provides a
real space, real time, description of this process
and suggests a phenomenology to explain the
microscopic answers.
Before summarising our method and key results we touch
upon what is known about the equilibrium charge ordered
state in EP systems, 
and what recent 
pump-probe experiments have revealed about dynamics.

Strong electron-phonon interaction leads to the formation of
an electron-phonon bound state - the lattice polaron - which 
serves as the basic degree of freedom at strong coupling
\cite{frohlich,latt-pol,Setvin21}.
At commensurate electron filling the polarons can order
into a `charge ordered' (CO) state, involving periodic modulation
of the bond lengths and electron density
\cite{adams,Batrouni21}.
The equilibrium physics of such CO systems is well understood 
\cite{McKenzie96,Pradhan15,chen2016,costa2020}.

Recent experimental advances allow strong 
perturbation of the CO state by application of a laser pulse
(the `pump') and probe the resulting electron-phonon dynamics via 
time and angle resolved photoemission spectroscopy (trARPES) 
\cite{Sobota21,Eich14,Schmitt11,Gedik17} or resonant inelastic 
x-ray scattering (RIXS) \cite{Devereaux16,Mitrano20}.
The pump adds excess energy to the electrons which,
due to the presence of EP interaction, gets 
transferred in part to the lattice. As a result, 
suppression and eventually melting of charge order 
can be observed.  Such experiments have been done on 
1T-TaS$_2$ \cite{cho2015}, tritellurides\cite{Wolf14} 
and the cuprates \cite{Mitrano20} in their charge ordered phases.
Depending on the strength 
of the photo-excitation the following features are observed. 
(i)~Weak photo-excitation leads to a damped periodic modulation 
of the electronic gap  as well as the  
electronic temperature
\cite{Mitrano20, Wolf14}. 
(ii)~With increasing photo-excitation one sees dynamics where 
the charge order vanishes at short times and then recovers slowly
to a suppressed value. This is accompanied by  
an insulator-metal-insulator
transitions as a function of time \cite{Makler21,Zhang20}. 
(iii)~At strong photo-excitation the CO melts rapidly resulting 
in an insulator-metal (I-M) transition \cite{Hellmann10, Han2015}.

The dynamics in response to a weak pump pulse can 
be understood via linear response theory 
\cite{Weinberger05, Seibold17}, along with the 
system's response to a probe pulse 
\cite{Freericks09, Eckstein08, Galina88}. However, 
the strongly perturbed system displays dynamics that 
require the solution of the full time-dependent problem. 
For this, one can employ a phenomenological 
time-dependent Ginzburg-Landau theory \cite{Dolgirev20, Zong20}, 
or fully microscopic approaches.  

The microscopic difficulty is in handling widely 
different electron and phonon timescales at strong coupling. 
For EP systems methods like exact diagonalization (ED) 
\cite{Junichi19} and dynamical mean field theory (DMFT) \cite{Freericks10,Freericks16} 
provide accurate results on
electronic timescales but cannot access the real space correlations. 
ED calculations are severely limited by system size. 

Given the energy scale difference between electrons and phonons
in the adiabatic regime, and the related need to retain a large
tower of  phonon states, the approach could be to treat the 
phonons classically.  
This approach has been used
within a Monte Carlo scheme recently to explain the ringing 
in the EP system on photoexcitation \cite{Freericks22}. 
The method 
still requires iterative diagonalization to evolve the system

and for a large system size diagonalization even for 
`non interacting' electrons become too costly to capture the 
rich dynamics of the order parameter.
Our method, below, allows access to `long times'  at modest 
computational cost and captures all the key features of order 
parameter dynamics in a charge ordered system.

We use a ``mean field dynamics'' (MFD) scheme
to study the dynamics of the CO state in the half-filled (spinless)
Holstein model in two dimensions. We set up coupled equations of 
motion for the expectation values
of the phonon displacement operator ${\hat x}_i(t)$ and the electron 
bilinear ${\hat \rho}_{ij}(t) = c^{\dagger}_i(t) c_j(t)$.
The equations close when one factorises the electron-phonon
interaction term.
This is a $\mathcal{O} (N^2)$ process for
each time step and the number of timesteps required range
from $10^6$ to $10^8$.
Despite its apparent simplicity the method captures the
effect of strong EP coupling and spatial correlations
accurately. It goes beyond the `static phonon' 
(or adiabatic) approximation, widely used before
\cite{Weber}. An approach similar to ours has been used recently 
to study the double exchange model \cite{Chern21}.

We work at intermediate EP coupling (below the
single polaron threshold) and probe the pulse strength
$(E_0)$ dependence of the dynamics. 
Our primary indicators are the phonon amplitude at the
ordering wavevector, $x_{\bf Q}(t)$, and the charge 
density wave field $n_{\bf Q}(t)$, where 
${\bf Q} = (\pi/a_0, \pi/a_0)$
on our square lattice with spacing $a_0$. 
We set $a_0=1$.
Our main results are the following:

{\bf (1).}~{\it Long time state:}
With increasing pulse amplitude $E_0$ the long time state 
attained by the system changes from charge ordered to charge 
disordered at a critical  amplitude $E_0^c$. 
At the EP coupling where we study the problem 
$E_0^c$ also demarcates a gap closing transition in the long 
time electronic state.

{\bf (2).}~{\it Transient response:}
There are three regimes in terms of pulse strength: 
(i)~For  $E_0 \ll E_0^c$ the indicators  $\vert x_{\bf Q}\vert$ 
and $\vert n_{\bf Q} \vert$ show oscillatory decay towards a 
suppressed long time value. (ii)~For $E_0 \rightarrow E_0^c$,  
these indicators first show a drop towards zero over a timescale 
of few phonon oscillations, remains small for a time $\tau_{cr}$, 
and then rise towards a finite steady state value. (iii)~For  
$E_0 > E_0^c$, the indicators decay monotonically to zero.  

{\bf (3).}~{\it `Critical' slowing down:}
For $E_0 \rightarrow E_0^c$ from below, 
$\vert x_{\bf Q}(t \rightarrow \infty)\vert^2 $ and 
$\vert n_{\bf Q}(t \rightarrow \infty)\vert^2 $ vanish as 
$(E_0^c - E_0)^{\alpha}$ 
where $\alpha \sim  0.7$.
We find that $\tau_{cr}$ diverge as $(E_0^c - E_0)^{-\nu}$, where
$\nu \sim 0.45$.

{\bf (4).}~{\it Time dependent spectrum:}
Constructing a density of states (DOS) from the
instantaneous eigenvalues we find that the DOS remains  
gapped at all times when  $E_0 \ll E_0^c$, and ungapped 
at all times for $E_0 > E_0^c$. For $E_0 \rightarrow E_0^c$ 
the DOS first shows gap closing and then reopening as a function 
of time. 

{\bf (5).}~{\it Phenomenology:} 
The coupled dynamics of electrons and 
phonons can be approximated by the dynamics of the phonons
in the presence of a force arising from a {\it nonequilibrium}
electron population. Parametrising this population distribution 
based on the full dynamics and using a Langevin equation 
allows us to capture both the suppression-revival
dynamics as well as the ``critical behaviour''. 

This paper is organised as follows.
In Section II we define the model, explain the dynamical
scheme, and set out the parameter space. In Section III we
describe the classification of order parameter response
into three regimes. Section IV provides a detailed
characterisation of the order parameter response in terms
of a few parameters. In Section V we examine
the dynamics of 
the overall system, including the  time dependence 
of the 
energy distribution over momentum modes, the electronic
density of states, and the population of `excited'
electrons. Based on these we construct an effective
classical model for the phonon dynamics in Section VI,
allowing access to large spatial scales and long times.
Section VII discusses issues related to our analytic
and numerical approximations and Section VIII concludes
the paper. 

\section{Model and Method}

\subsection{Model and evolution equation}

The Holstein model is given by:
\begin{eqnarray}
H~~ &=& H_e+H_{ph}+H_{ep} \cr
\cr
&=&  \sum_{ij} t_{ij}\hat{c}_i^{\dagger}\hat{c}_j 
+ \sum_i (\frac{\hat{p}_i^2}{2M} +\frac{K\hat{x}_i^2}{2} ) 
 - g\sum_i {\hat n}_i \hat{x}_i
\nonumber
\end{eqnarray}
For the $t_{ij}$ we 
consider nearest neighbour hopping
on a square lattice. 
We start by writing the Heisenberg equation 
for ${\hat x}_i$, leading to the family below.
Setting $\hbar =1$,
\begin{eqnarray}
{ {d {\hat x}_i} \over {dt}} ~& = & - i [{\hat x}_i, H] = 
{ {\hat p}_i \over {m} } \cr
\cr
{ {d {\hat p}_i} \over {dt}} ~&=& -i [{\hat p}_i, H] 
= -K {\hat x}_i + g  {\hat n}_i 
\nonumber
\end{eqnarray}
Now there are different options. (i)~One can take an average on 
the left and right hand side of the equations and replace 
$\langle {\hat n}_i(t) \rangle$ by its expectation in the 
instantaneous phonon background $x_i(t)$.
This is the `adiabatic evolution' (AE) 
scheme, and involves iterative diagonalisation
of the electron problem in `classical' 
phonon backgrounds $\{x_i\}$ to compute the force on the
phonons.  
Alternately,
(ii)~one can go beyond the adiabatic scheme 
and write an equation of motion
for the $ {\hat n}_i(t)$ itself, and so on, and close the 
hierarchy at some order. Following this route:
\begin{eqnarray}
{ {d {\hat n}_i} \over {dt}} ~&=& -i [ {\hat n}_i, H] = 
-i \sum_{mn} t_{mn}  [c^{\dagger}_i c_i, c^{\dagger}_m c_n] 
\cr
&=& -i \sum_{\alpha} 
t_{i \alpha} (c^{\dagger}_{\alpha} c_i - c^{\dagger}_i c_{\alpha})
= \sum_{\alpha}  {\hat J}_{i \alpha} \cr
\cr
{{d {\hat J}_{i \alpha }} \over {dt}}   
&=& -i \sum_{mn} t_{mn} [ {\hat J}_{i\alpha}, c^{\dagger}_m c_n] 
-i g \sum_j {\hat x}_j [{\hat J}_{i\alpha}, c^{\dagger}_j c_j]
\nonumber
\end{eqnarray}
Using 
$
[c^{\dagger}_{\alpha} c_{\beta},  c^{\dagger}_m c_n] = 
\delta_{n \alpha}  c^{\dagger}_m c_{\beta} - \delta_{m \beta}
c^{\dagger}_{\alpha} c_n
$
and simplifying the commutators, we obtain:
$$
{{d {\hat J}_{i \alpha }} \over {dt}} 
= - t_{i \alpha} \sum_{mn} t_{mn} {\hat f}^{i \alpha}_{mn}
- g t_{i \alpha} \sum_j {\hat x}_j {\hat h}^{i \alpha}_{j}
$$
where the $f$ and $h$ are fermion bilinears.
The first term comes purely from the hopping, the second
term however involves a cross coupling between electron
and phonon variables.
The equations are not closed and now we need to know the
dynamics of mixed objects like 
${\hat x}_j {\hat h}^{i \alpha}_{j}$, etc.
This leads to 
the Bogoliubov-Born-Green-Kirkwood-Yvon (BBGKY) 
hierarchy.  We truncate the hierarchy by taking 
the expectation value of the LHS and RHS and
approximating:
$
\langle {\hat x}_j {\hat h}^{i \alpha}_{j} \rangle
\approx  \langle {\hat x}_j \rangle \langle 
{\hat h}^{i \alpha}_{j} \rangle
$ 
The factorisation leads to a closed family of equations.
This  `non adiabatic evolution' (NAE) scheme
involves two processes:
(i)~The phonons are ``driven'' by $- K x_i + g n_i$, with 
$n_i = \rho_{ii}$, where $\rho$ is the one body
``density operator'', and (ii)~the density operator 
$\rho_{ij}$ is non locally correlated 
and driven by the $x_j$.

\begin{figure}[t]
\centerline{
\includegraphics[width=6.6cm,height=3.7cm]{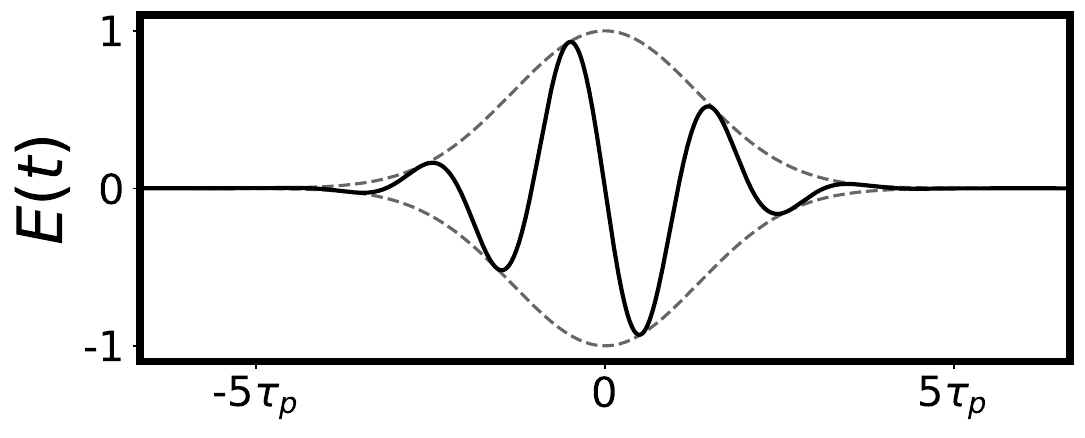} }
\caption{
Laser pulse: We set the pulse width to $\tau_p=\tau_0/10$, the pulse
frequency $\Omega_p=5 \Omega_0$, and vary the magnitude of the electric
field $E_0$ from $0.01  -1$.
We plot the time dependence of the electric field, normalized by $E_0$.
}
\end{figure}

Instead of writing a two step equation for $n_i$ and current we
could have directly written an equation for $\rho_{ij}$, factorised
it as above, and used the local component $\rho_{ii}$ in the
phonon equation.  The coupled equations would be:
\begin{eqnarray}
m { {d^2 {\hat x}_i} \over {dt^2}} ~& = & 
 - K {\hat x}_i + g  {\hat \rho}_{ii} \cr
\cr
{ {d {\hat \rho}_{ij} } \over {dt}} 
~& = & -i [{\hat \rho}_{ij},\tilde{H}] = 
-i([{\hat \rho}_{ij},H_e] + [{\hat \rho}_{ij},H_{ep}]) 
\cr
&=& -i [{\hat \rho}_{ij},H_e] + i g\sum_m 
[{\hat \rho}_{ij}, {\hat n}_m]{\hat x}_m
\nonumber
\end{eqnarray}
We take the expectation value of both sides and
factorise.

\begin{figure}[b]
\centerline{
\includegraphics[width=4.8cm,height=4.6cm]{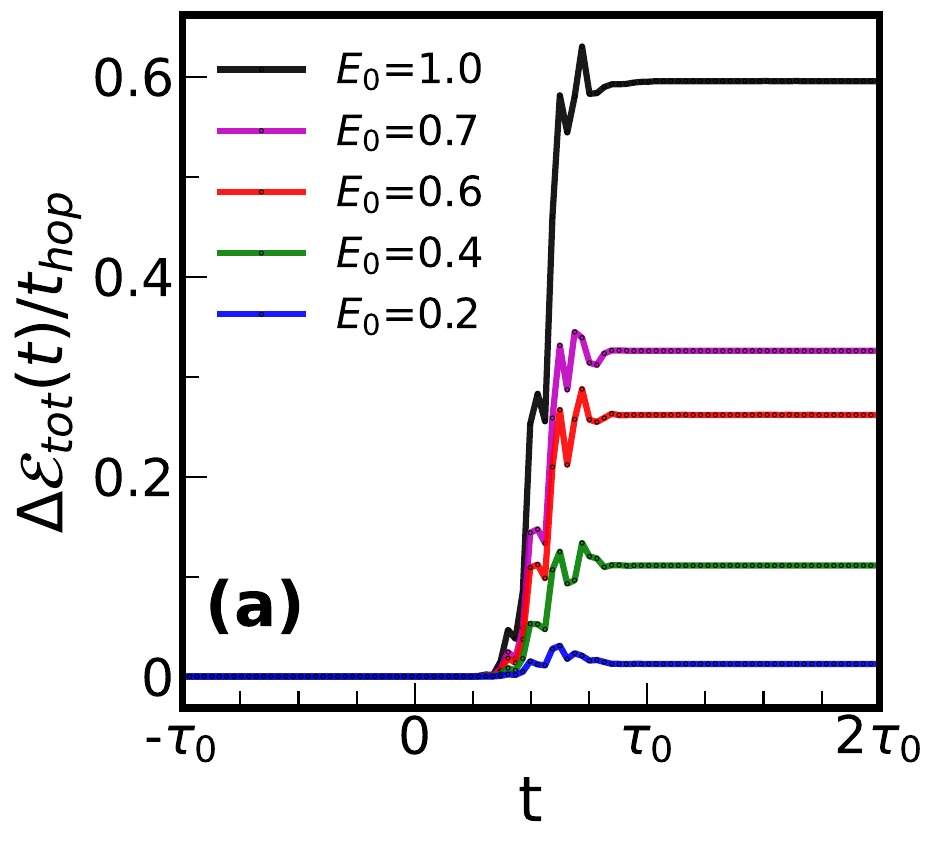}
\includegraphics[width=3.9cm,height=4.6cm]{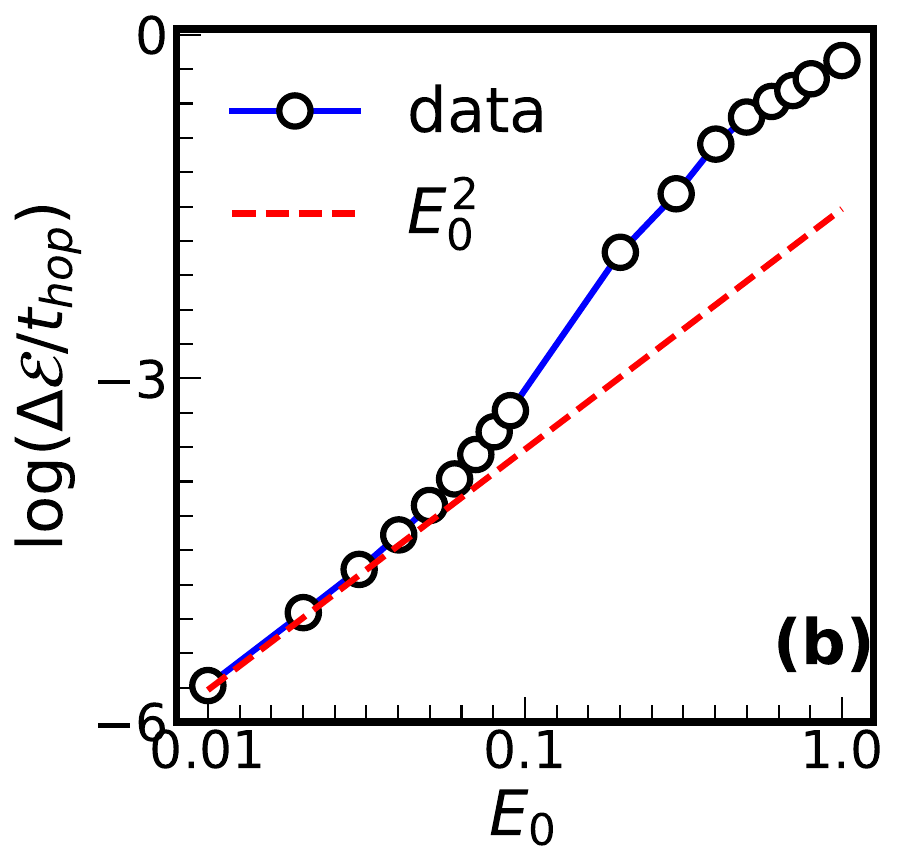}
}
\caption{(a)~The time dependence of the change 
in the total energy $\Delta\mathcal{E}(t)$.
The energy of the unperturbed system is $\mathcal{E}(0)$ and pumping injects
an amount of energy $\Delta \mathcal{E}$.  Since the post pulse dynamics is
conservative, total energy $\mathcal{E}_{tot}(t) 
= \mathcal{E}(0) + \Delta \mathcal{E}(t)$ should
be constant for $t \gg \tau_p$. The
panel confirms this and also quantifies $\Delta \mathcal{E}(E_0)$. (b).~The
dependence of $\Delta \mathcal{E}$ on $E_0$, showing the `linear response'
regime, $\Delta \mathcal{E} \propto E_0^2$, and beyond.
}
\end{figure}

\subsection{Generating the initial configuration}

We consider the equilibrium system to be in its ground
state. At half filling, $n=0.5$, the Holstein model with
nearest neighbour hopping has a charge ordered ground
state at all $g$. The CO is driven by a Fermi surface
instability at small $g$ and the virtual hopping of polarons
at large $g$. We obtain the reference CO state by minimising 
the total energy with respect to a periodic $x_i$ field. 
At the minimum, $\{ x_i^0 \}$, say,  we use the 
the eigenvectors  of $H\{ x_i^0 \}$ to compute the averages 
$\rho_{ij}^0 = \langle c^{\dagger}_i c_j \rangle_0$.

\subsection{Modeling the laser pulse}

We include the laser field via
Peierls substituion. This leads to the time dependent
hopping term:
\begin{eqnarray}
H_e(t) &=& 
\sum_{ij} t_{ij} (e^{i \int_{\vec{r_j}}^{\vec{r_i}}
\vec{A}(t) \cdot \vec{dr}  } c_i^{\dagger} c_j + h.c)
\cr
{\vec E}(t) & = & -\frac{d\vec{A}(t)}{dt} 
= 
\vec{E_0}exp[- ( {{t-t_{0}} \over {\tau_p} })^2]  
cos(\Omega_pt) 
\nonumber
\end{eqnarray}
$E_0$ is the strength of the field,
$\tau_p$ and the $\Omega_p$ are the width of the
pulse and frequency of the incident wave respectively. 
The electric field is taken in ($\vec{x}+\vec{y}$) direction. 
We set $t_0 = 0$. 
The time dependence of the electric field
associated with the laser pulse is shown in Fig.1.

\subsection{Parameter space}

In the rest of the paper we will use $t$ as the time 
variable and denote the hopping $t_{ij}$ as $-t_{hop}$ for
nearest neighbours, and set  $t_{hop}=1$.
We set $K=1$ and the oscillator mass to $M=25$ so that 
the local phonon frequency is $\Omega_0 = \sqrt{K/M} 
= 0.2$. As a result the ``adiabaticity'' parameter
$\gamma = \Omega_0/t_{hop} = 0.2$, small but not
negligible.
We measure time in units of $\tau_0 = {2 \pi}/\Omega_0$.
We focus on the half-filled case $n=0.5$.
In the adiabatic regime the energy of a polaron 
is $E_p \sim  -g^2/{2K}$. A polaron forms when this equals
the band bottom energy $-4t_{hop}$ (in two dimensions). This
leads to the definition of a dimensionless EP coupling
strength: $\lambda = {g^2}/{8Kt_{hop}}$. 
Single polaron formation, in the adiabatic limit, 
corresponds to $\lambda \sim 1$. In the paper we work
mainly with $\lambda = 0.38$, intermediate coupling but well
below the single polaron threshold.
We set $\Omega_p/\Omega_0 = 5$ and $\tau_p=\tau_0/10$.
Having fixed $n$, $t_{hop}$, $\Omega_0$, $\lambda$, 
and the pulse
parameters $\tau_p$ and $\Omega_p$, we study the response 
of the half filled CO state to a laser pulse for varying 
pulse amplitude $E_0$.

\subsection{Numerical techniques}

Most of our data are on system size N = $16 \times 16$.
We use the Runge-Kutta 4 (RK4) method to solve 
$N^2+2N$ coupled first order differential equations 
for $x_i$, $p_i$ and $\rho_{ij}$. 
Our step parameter is set to $\delta t = \tau_0/1500$
our total run length  $\tau_{max}$ is at least  
$6000 \tau_0$. In the critical pulse 
regime, where the system undergoes a transition from 
the CO state to a charge disordered state, simulation
timescales needed to access the steady state grow
rapidly. There we have used 
$\tau_{max} \sim 10^5 \tau_0$
(on system size $ 12 \times 12$).

\begin{figure*}[t]
\centerline{
\includegraphics[width=14cm,height=6.0cm]{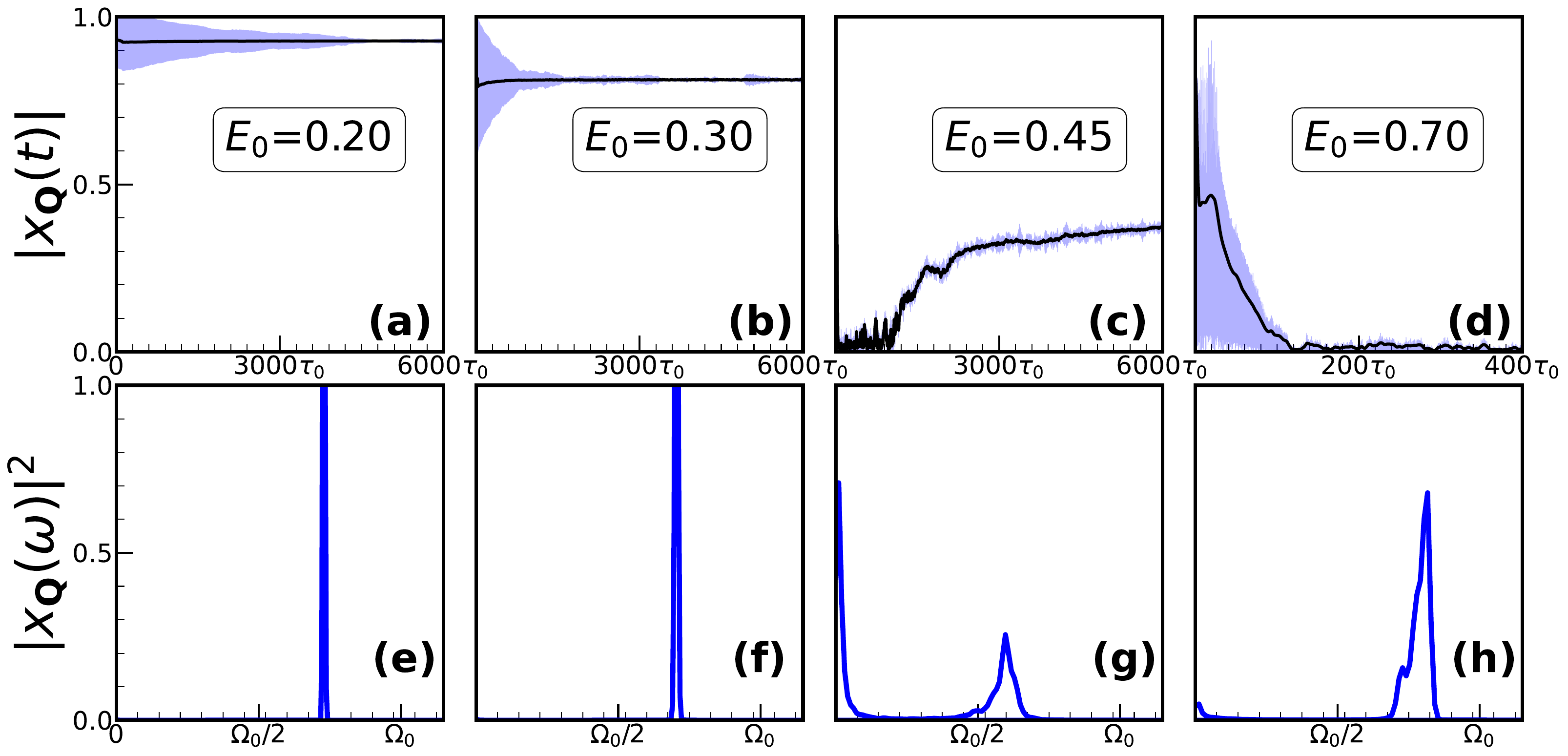}
}
\caption{
Overview of dynamics of the order parameter mode $|x_{\bf Q}|$ for different
$E_0$.  Upper row: panels (a)-(d) show real time dynamics:  $|x_{\bf Q}(t)|$.
(a)-(b)~Weak pulse regime, roughly $E_0 = 0-0.35$, where $x_{\bf Q}(t)$
oscillates with a single frequency and decays to a suppressed CO state.
(c)~Critical pulse regime, roughly $E_0 = 0.35-0.50$, where $x_{\bf Q}(t)$
is rapidly suppressed, stays small for a time window, and then rises to a
low amplitude long time state.  (d)~Strong pulse regime: $E_0 >0.5$, where
the CO gets destroyed within a few $\tau_0$ while the phonon ringing continues
for a longer time. 
Lower row: $|x_{\bf Q}(\omega)|^2$, where $x_{\bf Q}(\omega)$
is the Fourier transform of  $x_{\bf Q}(t)$ over the whole time window.
(e)-(f)~The weak pulse $|x_{\bf Q}(\omega)|^2$ has primary feature at the
electronically renormalised phonon frequency $\Omega_{\bf Q}$. There is a
small broadening of the lines due to mode coupling induced decay. (g)~In the
critical pulse regime the spectrum has a broad feature at finite frequency
and another feature around $\omega =0$.  (h)~Spectrum in the strong pulse
window, where the real time behaviour is oscillatory decay.  System size
$16 \times 16$.
}
\end{figure*}

\subsection{Indicators}

Our basic output is the time series for $x_i(t)$. Based
on this we can compute various correlation functions of
the phonon variables. 
We can compute the
instantaneous electronic density of states (DOS) 
from the electronic eigenvalues $\epsilon_n$ in a
background $x_i(t)$.  We also have access to the
equal time correlation $\rho_{ij}(t) = \langle
c^{\dagger}_i(t) c_j(t) \rangle$ and can compute the 
`occupation' of levels $\epsilon_n(t)$ from this.
The spatial Fourier transforms related to phonon
distortions and density are, respectively:
$$
x_{\bf q}(t) = {1 \over N} \sum_{ij} e^{i {\bf q}. {\bf r}_i } x_i(t), 
~~~~
n_{\bf q}(t) = {1 \over N} \sum_{ij} e^{i {\bf q}. {\bf r}_i } n_i(t) 
$$
where $n_i(t) = \rho_{ii}(t)$.
There are the corresponding Fourier transforms to
frequency:
$
x_{\bf q}(\omega)
= \int_{0}^{t_{max}} dt e^{-i \omega t} x_{\bf q}(t) $.
We also calculate the following instantaneous distributions
$$
N(\omega, t) = 
{1 \over N} \sum_n \delta(\omega - \epsilon_n(t)),~~ 
P(x,t) = {1 \over N} \sum_i \delta(x - x_i(t))
$$
The occupation function $f(\omega, t)$ for the instantaneous eigenstates is defined
by:
$$
f(\omega, t) N(\omega,t)  = \sum_n \rho_{nn}(t) \delta(\omega-\epsilon_n(t))
$$
Where the $\rho_{nn}$ is the expectation value of the
density operator associated with the $n$-th eigenstate with energy 
$\epsilon_{n}$: $\rho_{nn}(t) = 
\sum_{ij} U^*_{in}(t)U_{jn}(t)\rho_{ij}(t)$, where $U(t)$ are 
the instantaneous eigenvectors.


\section{Global behaviour of order parameter}

Before entering into the detailed characterisation of the
dynamics we attempt a broad classification of regimes that
occur for varying $E_0$. 
In the absence of an external drive our 
method of evolution  strictly conserves the energy. 
If the pre-pulse energy of the system is ${\cal E}(0)$
and the pump adds an energy $\Delta {\cal E}$ then the
energy ${\cal E}(t)$ for $t \gg \tau_p$ is 
${\cal E}_{final} = {\cal E}(0) + \Delta {\cal E}(E_0)$
independent of time.
In Fig.2(a) we depict the time evolution of 
$\mathcal{E}(t)$ with varying strength $E_0$.
Fig.2(b) shows $\Delta \mathcal{E}$ against $E_0$, 
revealing that at low $E_0$, it is proportional to 
$\sim E_0^2$ as expected from linear response.

Fig.3 illustrates the different dynamical regimes that
arise in response to pulses of increasing strength.
The data is for $\lambda =0.38$, the results have a 
similar trend at other $\lambda$.
The charge ordered state corresponds to $O(1)$ value of the
Fourier transform
$x_{\bf Q} = (1/N) \sum_i x_i e^{i {\bf Q}.{\bf r}_i}$,
the order parameter of the CO state.
The upper row in Fig.3  
shows the time dependence of $\vert x_{\bf Q}(t) \vert$.
The pre-pulse value is normalised to $1$ for convenience. 
Panels (a)-(b) are in the `weak pulse' regime where we
mainly see oscillatory decay 
to a long time value that is suppressed with respect to
the pre-pulse (equilibrium) value. We will call
this response ``weak oscillatory suppression'' (WOS).
Panel (c)
show results in the `critical pulse' regime where 
$\vert x_{\bf Q}(t) \vert $ 
initially drops almost to zero, stays there
for some time, and then `revives' heading towards a 
finite long time value. We call this ``strong 
suppression and revival'' (SSR) dynamics.
Panel (d) shows $\vert x_{\bf Q}(t) \vert$ in the strong 
pulse
regime where we see a monotonic decay of the oscillation
envelop to zero and there is no revival. This is simply 
``monotonic suppression'' (MS) dynamics.

The lower row in Fig.3 shows the time Fourier transform
of the corresponding upper panels, taken over the interval
$0 - 6000 \tau_0$. 
The character of $\vert x_{\bf Q}(\omega) \vert^2$
can be understood from the behaviour of $x_{\bf Q}(t)$.

When the energy added by the pulse is very small 
as in (e)-(f) 
it leads to weakly damped oscillation of the various
normal modes of the CO state. The 
oscillation frequency would be $\Omega^0_{\bf q}$, 
the electronically renormalised phonon dispersion
about the CO state (discussed later). 
The amplitude of oscillation of the
modes will 
depend on the specific nature of the perturbation.
The order parameter mode will have a response of the
form $x_{\bf Q} \sim A + B cos(\Omega^0_{\bf Q}t)$
The lineshape will have a sharp feature 
at $\Omega^0_{\bf Q}$ as observed in (e).  
With growing excitation  mode coupling leads 
to a damping, as seen in (b), and 
$ x_{\bf Q} 
\sim A + B e^{-t/\tau_w} cos(\Omega_{\bf Q}t)$. 
We denote the decay time in the weak pulse regime as
$\tau_w$, and $\Omega_{\bf Q}$ is the oscillation frequency
now renormalised by $E_0$.
The damping rate $\Gamma_w = \tau_w^{-1}$ increases 
with $E_0$. However,  $\tau_w \gg \tau_0$ so 
the damping $\Gamma_w \ll \Omega_{\bf Q}
\sim \Omega_0$. 

For pulse strength in the critical window the general
feature of $x_{\bf Q}(t)$ is rapid suppression at small
times, a seeming quiescent period, and then a revival. 
The behaviour is accompanied
by weak oscillation about the mean curve. Ignoring the
sharp drop at small times, the longer time behaviour can be
described roughly by 
$ x_{\bf Q}(t)
 \sim A e^{-(\tau_{cr}/t)^{\beta}}  + B cos 
(\Omega_{\bf Q}t)$. 
The `waiting time' in the low amplitude state is
decided by $\sim \tau_{cr}$, while the approach
to the long time state has a power law character
$ 1 - (\tau_{cr}/t)^{\beta}$.
The Fourier transform, panel (g),  
involves a low energy feature of 
width $\sim (\tau_{max} - \tau_{cr})^{-1}$
and a broadened line around $\Omega_{\bf Q}$.%

In (h), the strong pulse regime, there is 
a stable oscillation amplitude at short times 
followed by a monotonic decay to zero.
A simple analytic form for $x_{\bf Q}(t)$ 
is $A (1 - e^{-(\tau_s/t)^{\gamma}} ) cos(\Omega_{\bf Q}t)$.
The low frequency feature quickly vanishes as
$E_0$ increases in this regime.

We collect together the expressions for
$x_{\bf Q}(t)$:
\begin{eqnarray}
 x_{\bf Q}(t) &\sim&  A + B e^{-t/\tau_w} cos(\Omega_{\bf Q}t)
~~~~~~~~~~~~~~~(weak)
\cr
\cr
& \sim &  A e^{-(\tau_{cr}/t)^{\beta}}  + B cos (\Omega_{\bf Q}t) 
~~~~~~~~~~(critical)
\cr
\cr
&\sim& A (1 - e^{-(\tau_s/t)^{\gamma} }) cos(\Omega_{\bf Q}t)
~~~~~~~~~~(strong)
\nonumber
\end{eqnarray}

Fig.4(a) shows the global phase diagram in the 
$\lambda - E_0$
plane, highlighting both the long time state and
the dynamical regimes.
It is based on analysis of the kind shown in
Fig.3, now carried out for several $\lambda$. The
blue region corresponds to surviving CO,
while the red region corresponds to a state with 
CO destroyed in response to the pulse. In the ordered
region, dark blue (WOS) corresponds to the weak oscillatory
suppression defined earlier, while light blue (SSR) 
corresponds to strong suppression and revival. 
Red region (MO) is
charge disordered, where the dynamics is of
monotonic suppression, and is separated from the ordered regime 
by the nonequilibrium phase transition. A true 
phase transition, with $\vert x_{\bf Q}(t \rightarrow \infty)\vert 
\rightarrow 0$, as $E_0$ tends to some $E_0^c$, can
be seen only as $L \rightarrow \infty$ and 
$t_{max} \rightarrow \infty$. 
The boundary in Fig.4(a) is drawn via
extrapolation of $L$ dependent data at
a few sizes,   $12 \times 12$ to $20 \times 20$.

\begin{figure}[b]
\centerline{
\includegraphics[width=4.4cm,height=4.4cm]{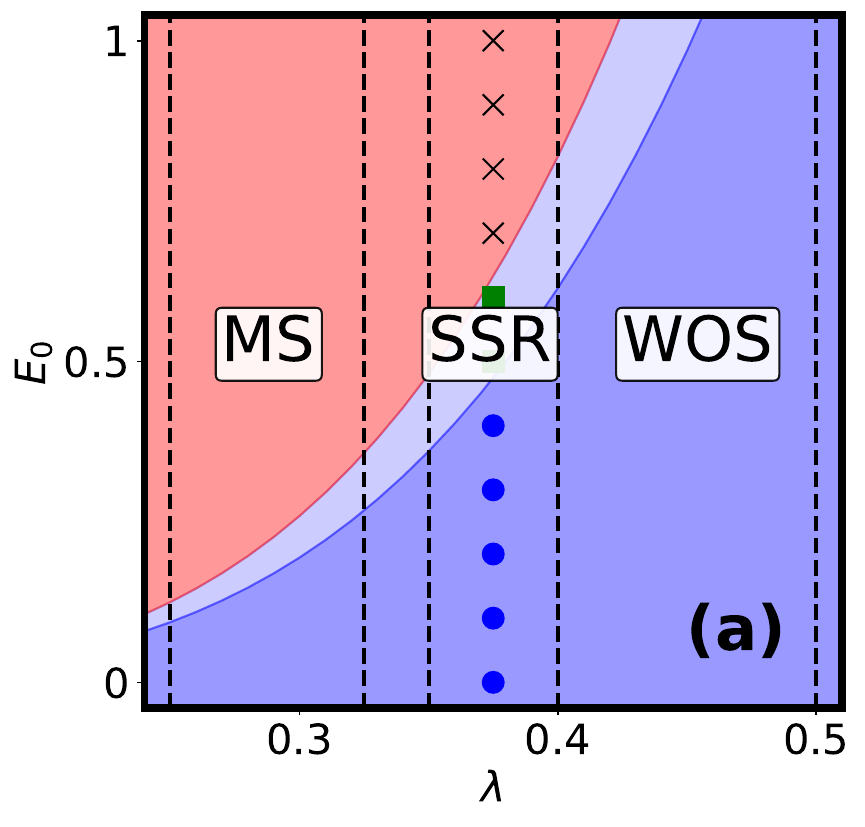}
\includegraphics[width=4.4cm,height=4.4cm]{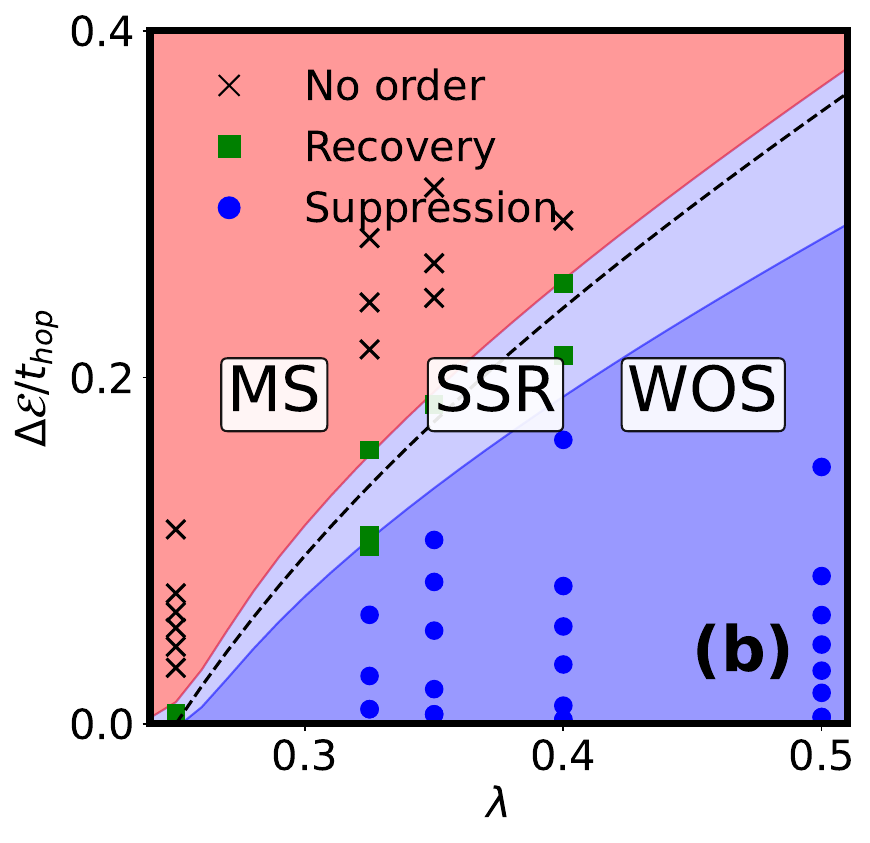}
}
\caption{
(a) $\lambda - E_0$ global phase diagram in terms of the
long time state and the dynamics that arises in response to a pulse.
In the blue regions the long time state has charge order.
In dark blue, $|x_{\bf Q}|$ shows an oscillatory decay towards
its long time CO state.  In region light blue, $|x_{\bf Q}|$
shows an initial decay to zero and then revives towards a low
amplitude CO state at long times.
In red region, $|x_{\bf Q}|$ shows monotonic decay
to zero with no revival.
The boundary between regions blue and red is a phase transition.
(b)~Phase diagram in terms of energy absorbed, $\Delta {\cal E}$.
The dotted line, based on 
high electronic temperature calculation
(see text), roughly captures the transition line.
}
\end{figure}

Fig.4(b) shows the phase diagram in terms of $\lambda$ and the energy
absorbed.  We make a rough estimate of the transition line
(dotted) using a electronic temperature model introduced 
later (Section VI). Briefly, the excess energy 
$\Delta\mathcal{E}$ associated with population transfer to
the upper-level destroys the double well structure 
in $x_{\bf Q}$ at $\Delta\mathcal{E}\sim\Delta\mathcal{E}_c$.
We estimate this value in the following way:
We assume a finite electronic temperature $T_{el}$ and using a
uniform distortion in ${\bf Q}$-mode we calculate the
effective potential by evaluating the eigen values $\epsilon_n$,
$$V_{eff}(x_{\bf Q}, T_{el})={1\over N} \sum_{n} 
\epsilon_n F(\epsilon, T_{el})~+\frac{1}{2}K x_{\bf Q}^2$$
$F(\epsilon, T_{el})$ being the Fermi function.
This looses it's double well structure at $T_{el}\sim T_{el}^c$ 
and at $T_{el}^c$
we calculate the change in energy as 
$\Delta \mathcal{E}_c = \text{min}(V_{eff}(x_{\bf Q},T_{el})-\text{min}(V_{eff}(x_{\bf Q},0))$.
This shows if the system absorbs energy 
$\Delta\mathcal{E}<\Delta\mathcal{E}_c$ the long time state shows a
finite $|x_{\bf{Q}}|$.


\section{Detailed dynamics of order parameter}

\subsection{Weak pulse regime}

Fig.5 shows in detail the phonon and charge density 
dynamics at the ordering wavevector ${\bf Q}$ for
a weak pulse, $E_0 = 0.30$. For both the phonon and
the density field 
there is an initial drop, on the scale of a few $\tau_0$.
This is followed by an oscillatory decay to a long time state.
Panels (a) and (c) show 
$\vert x_{\bf Q} \vert$ and $\vert n_{\bf Q} \vert$, 
respectively, over the whole run, while
the insets show the time dependence at short times 
$\sim 15 \tau_0$. As we have argued, 
$
x_{\bf Q}(t)
\sim A + B e^{-t/\tau_w} cos(\Omega_{\bf Q}t)
$
This simple function does not capture the oscillations (and
the `beating pattern') at $t \gg \tau_w$ but seems adequate for
an overall description. Also, to capture the $t=0$ state
we should have $A + B =1$.
Panels (b) and (d) show the respective Fourier transforms,
the main panel showing the transform of the full time series
and the inset showing the transform of the short time response.

\begin{figure}[b]
\centerline{
\includegraphics[width=4.8cm,height=3.3cm]{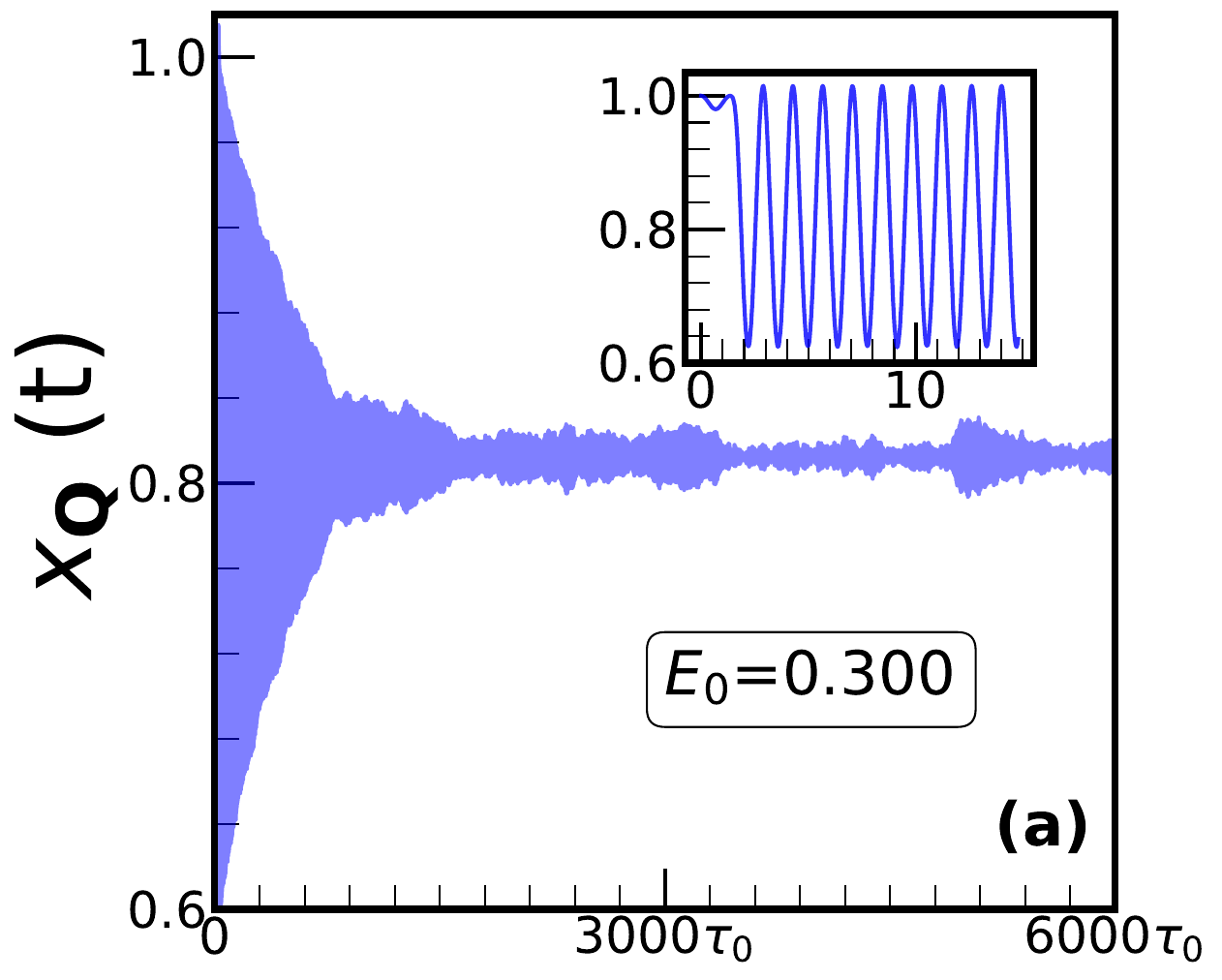}
\includegraphics[width=4.0cm,height=3.3cm]{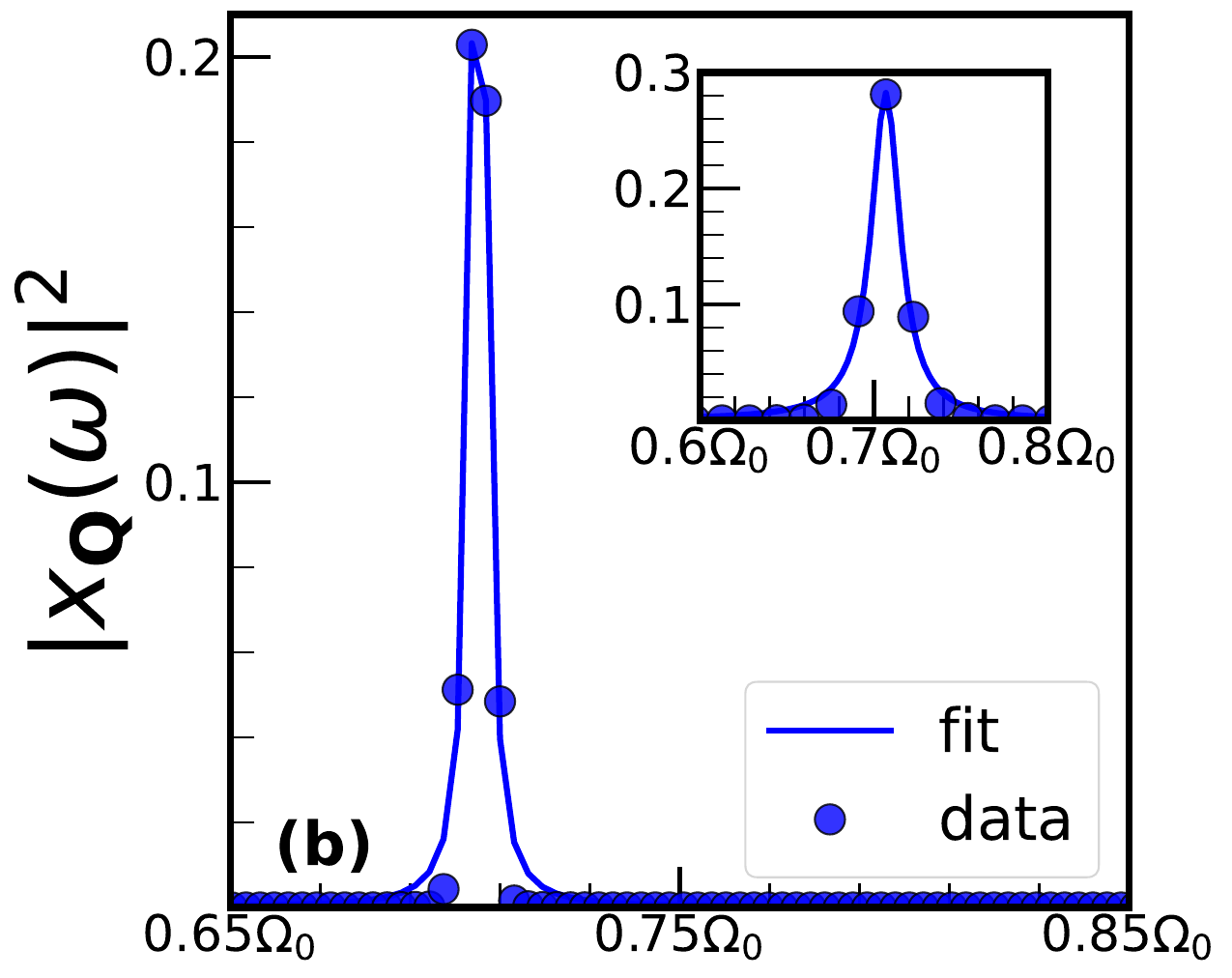} }
\centerline{
\includegraphics[width=4.8cm,height=3.3cm]{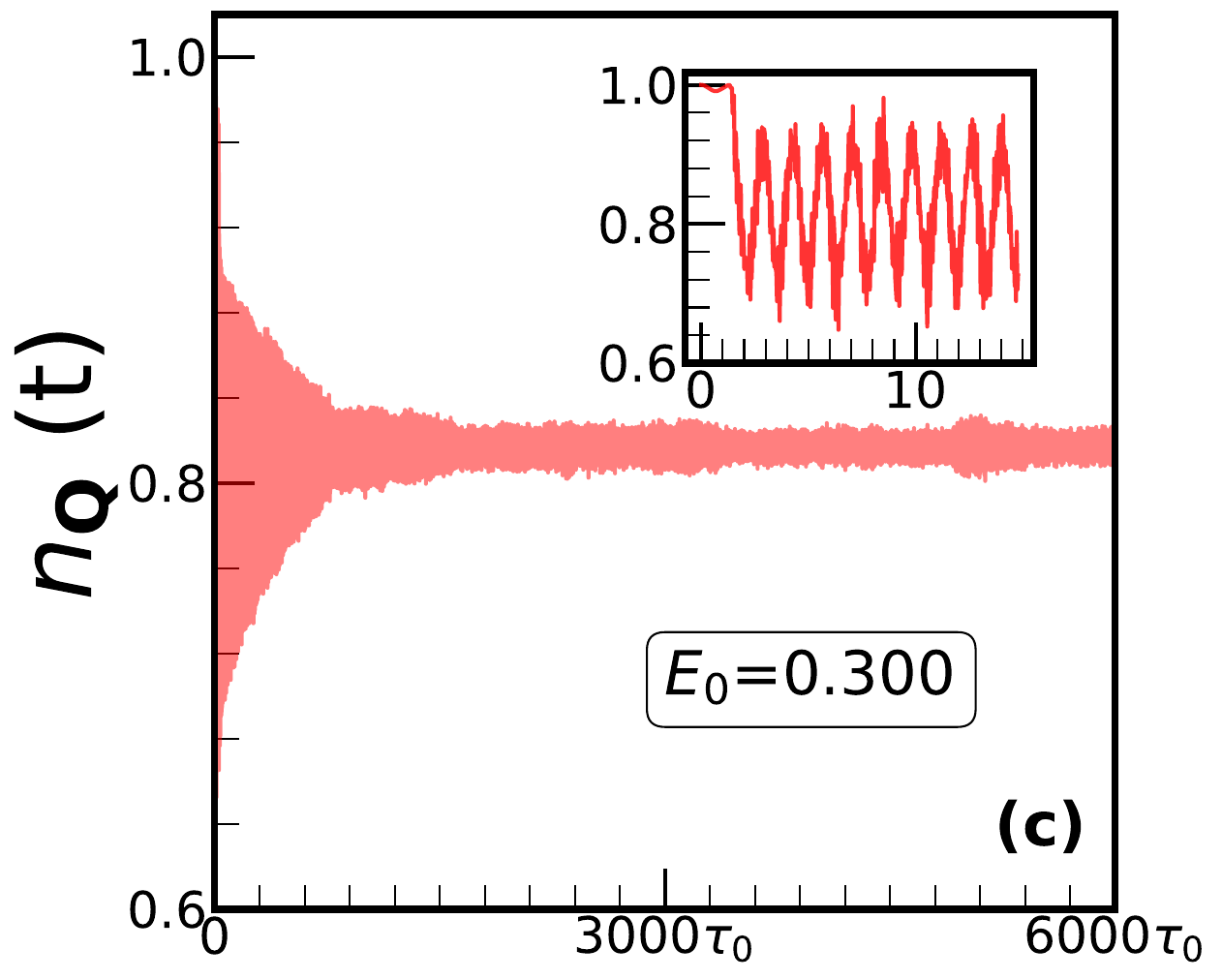}
\includegraphics[width=4.0cm,height=3.3cm]{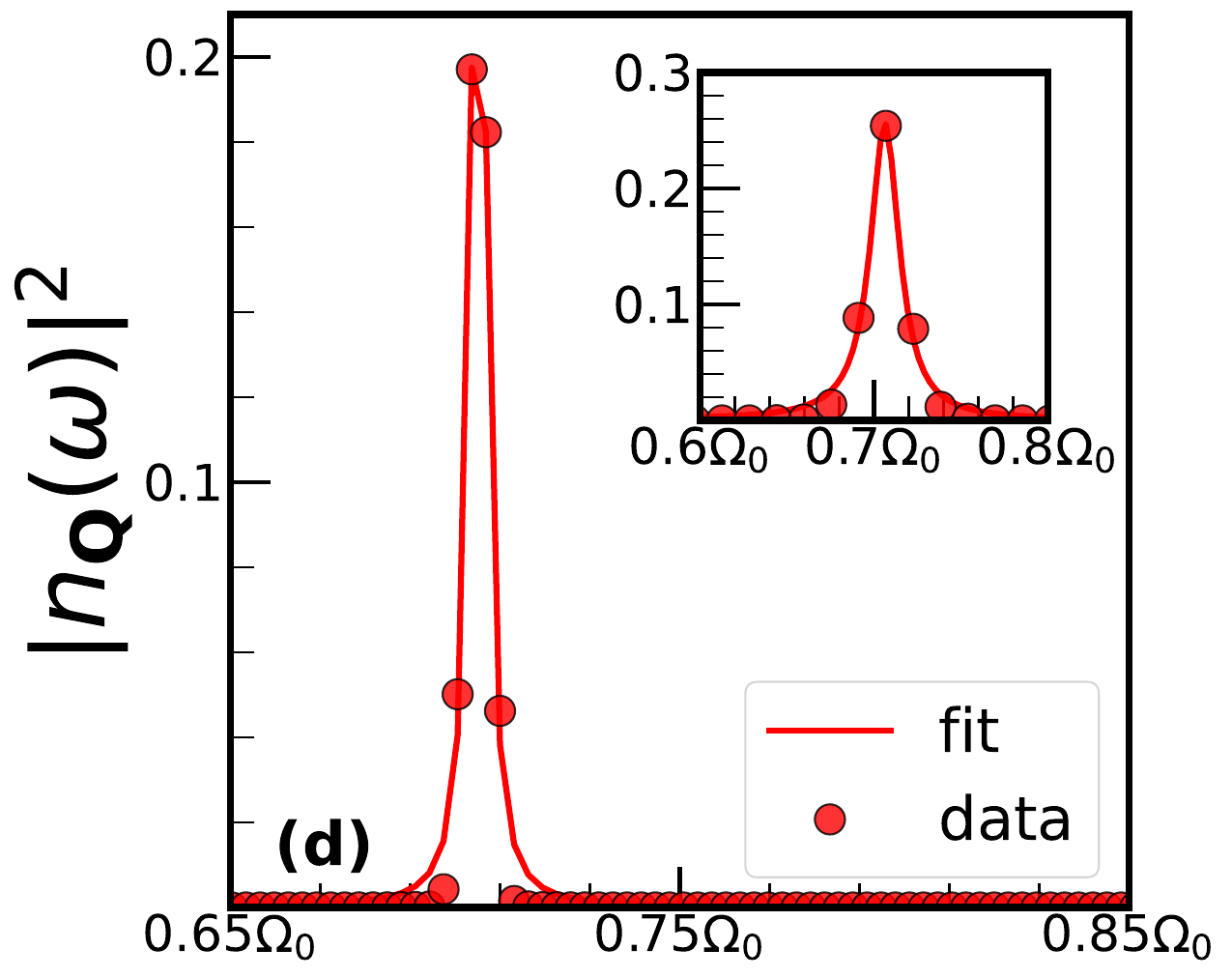} }
\caption{Weak pulse regime: (a)~shows $|x_{\bf Q}(t)|$
with the inset highlighting the very short time behaviour.
(b) Shows  $|x_{\bf Q}(\omega)|^2$, obtained by Fourier
transform over the entire time window, while the inset
shows the transform of data over the interval $t
\sim [0-15] \tau_0$.
(c) and (d) show corresponding data for $|n_{\bf Q}(t)|$
and  $|n_{\bf Q}(\omega)|^2$.  Note the quick drop in
$|n_{\bf{Q}}(t)|$ at very short time.  
During this time, 
the electron dynamics is primarily governed by 
the pump, and the density modulation
can reduce despite the presence of a 
lattice distortion.}
\end{figure}
\begin{figure}[t]
\centerline{
\includegraphics[width=4.4cm,height=4.2cm]{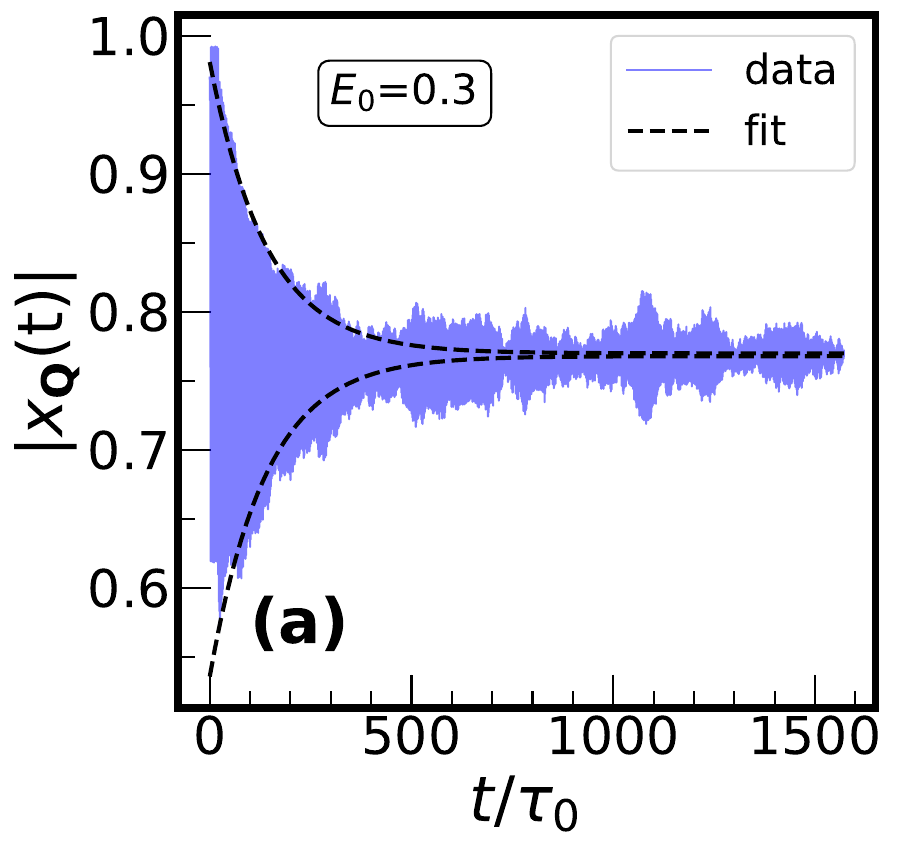}
\includegraphics[width=4.2cm,height=4.2cm]{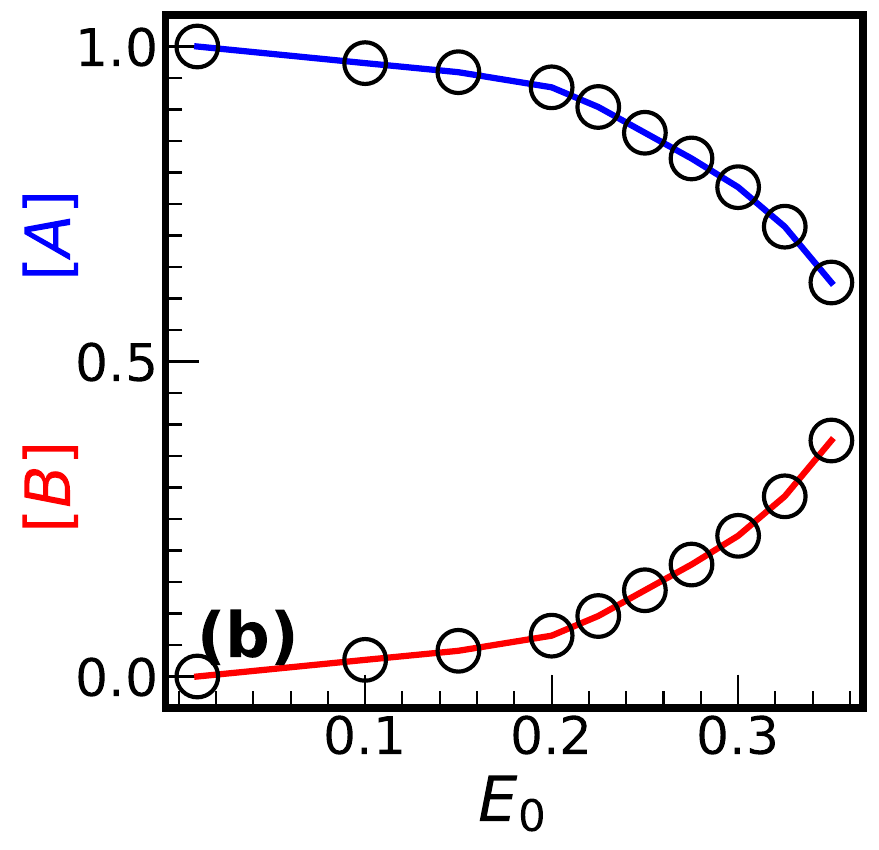} }
\centerline{
~~
\includegraphics[width=4.4cm,height=4.2cm]{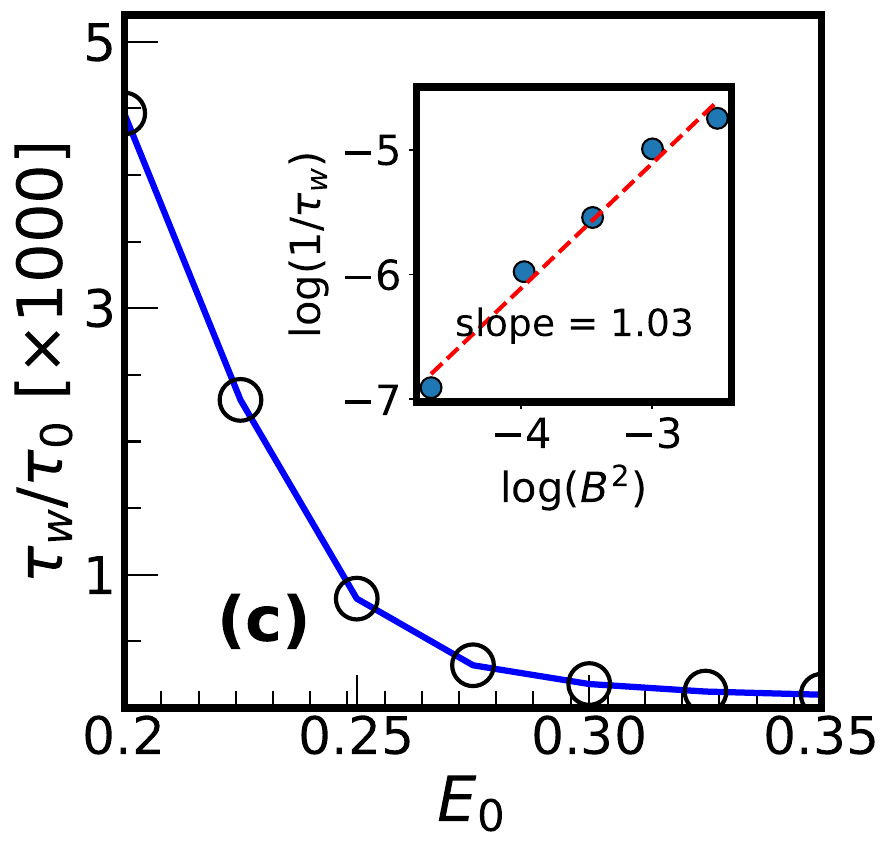}
\includegraphics[width=4.2cm,height=4.2cm]{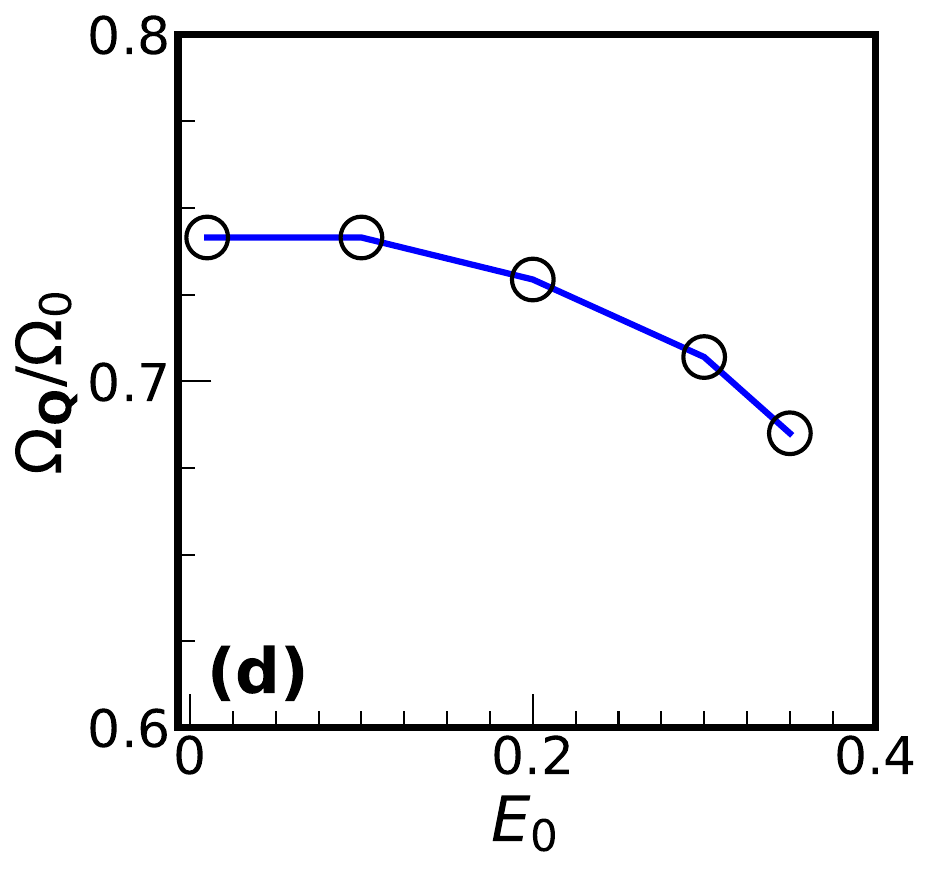} }
\caption{ Characterising dynamics in the weak pulse regime.
We fit the $|x_{\bf Q}(t)|$ to the form
given in the text. (a)~Shows the fit to the envelop. (b)~Shows
the amplitude $A(E_0)$, with $A$ being the steady state
value of $|x_{\bf Q}(t)|$ at the ordering wave vector ${\bf Q}$. 
Note the sharp downward trend already apparent at $E_0 \sim 0.35$.
(c) Shows the timescale $\tau_w$ over which the main oscillations
decay. It diverges as $E_0 \rightarrow 0$ when the dynamics
reduces to decoupled normal modes. In the inset we plot 
log($B^2$) with log($\tau_w^{-1}$) and it shows a slope 
$\sim$1.03 showing the relationship $\tau_w^{-1} \sim 1/B^2$. 
(d) Shows the oscillation
frequency $\Omega_{\bf Q}$ which shows a downward trend, akin
to what one observes on raising temperature. $\Omega_{\bf Q}$
has electronic renormalisation effects built into it even at
$E_0=0$.
}
\end{figure}
In Fig.6 panel (a) shows the fit to the $\vert x_{\bf Q} \vert$ 
envelop using the fitting function above.
According to the fitting function, the $t \rightarrow \infty$
value of $\vert x_{\bf Q} \vert$ is $A$ and there will be a
peak in the spectrum around $\Omega_{\bf Q}$
with width $\tau_w^{-1}$. 
Panel (b) shows the $E_0$ dependence of $A$ and $B$.
$A$ falls from $1$ at $E_0=0$ to $\sim 0.6$ at $E_0 \sim 0.35$
while $B$ rises from zero to about $0.4$.

Panel (c) shows $\tau_w(E_0)$,
which has a remarkable decrease with increasing $E_0$. As $E_0
\rightarrow 0$ the weakly perturbed lattice has only
undamped normal mode oscillations, and  $\tau_w \rightarrow \infty$.
Since $\tau_w^{-1}$ emerges from anharmonicity, 
in the weak pulse limit we expect it to vary as a
power $B^{\alpha}$.
A plot of $log($B$)$ versus  log($\tau^{-1}_{w}$) suggests
that $\tau^{-1}_w \sim B^2$.

The oscillation frequency, $\Omega_{\bf Q}$,
shows only weak $E_0$ dependence. 
The frequency for low amplitude oscillations on the CO state
is given by $\Omega_{\bf Q} = \sqrt{\Omega_0^2 + (g^2/m)\Pi^0_{\bf Q}}
$, where $\Pi^0_{\bf Q}$ is the density response function of the
CO state at momentum ${\bf Q}$ \cite{sauri20}. 
Electron-phonon coupling gives a dispersion to the
phonons, and also lowers the frequency from the bare value
$\Omega_0$. The effect of the added energy due to $E_0$ 
is to additionally
reduce $\Omega_{\bf Q}$, akin to what one observes in the
effect of temperature \cite{sauri19}.

\begin{figure}[b]
\centerline{
\includegraphics[width=4.8cm,height=3.2cm]{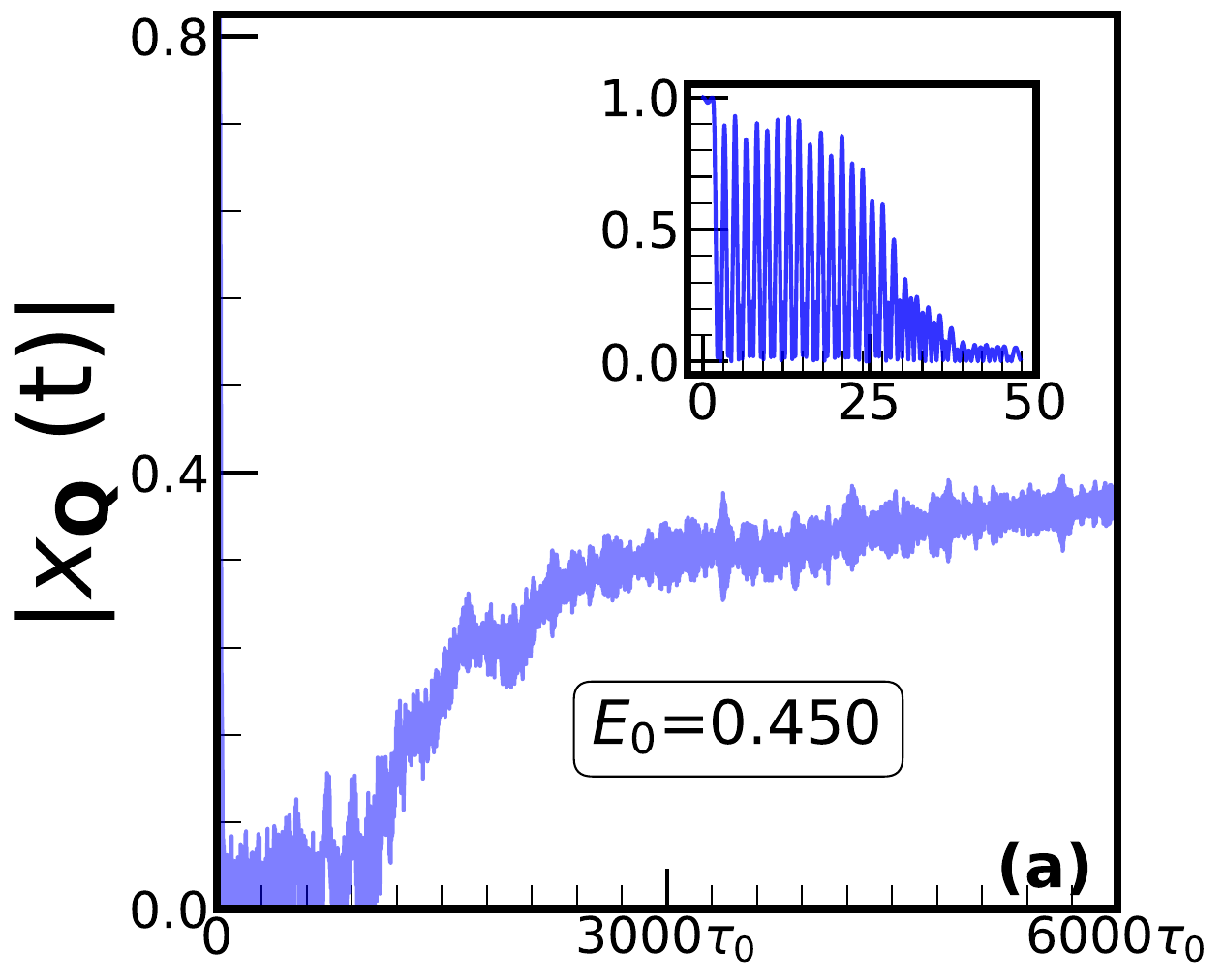}
\includegraphics[width=4.0cm,height=3.2cm]{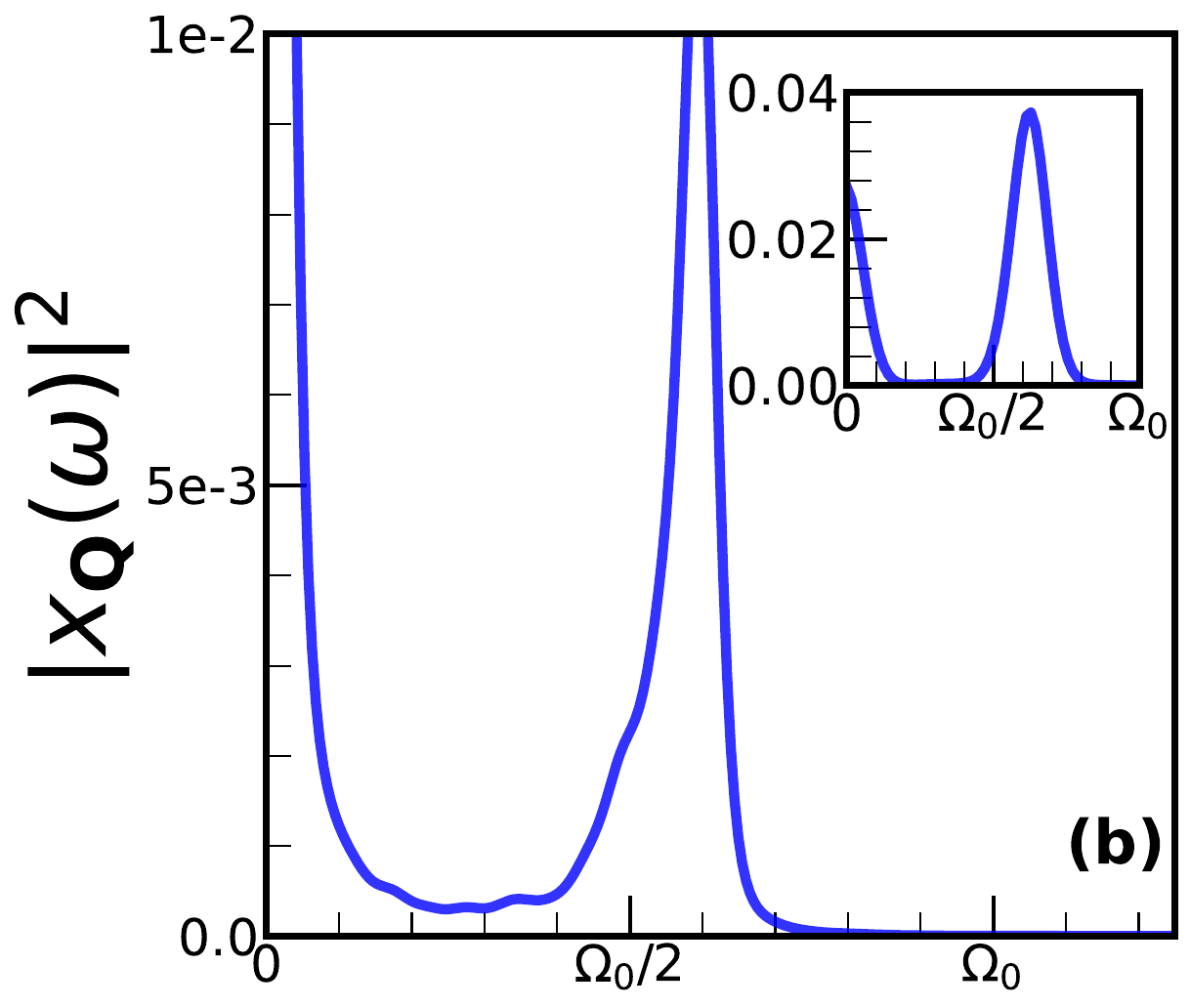} }
\centerline{
\includegraphics[width=4.8cm,height=3.2cm]{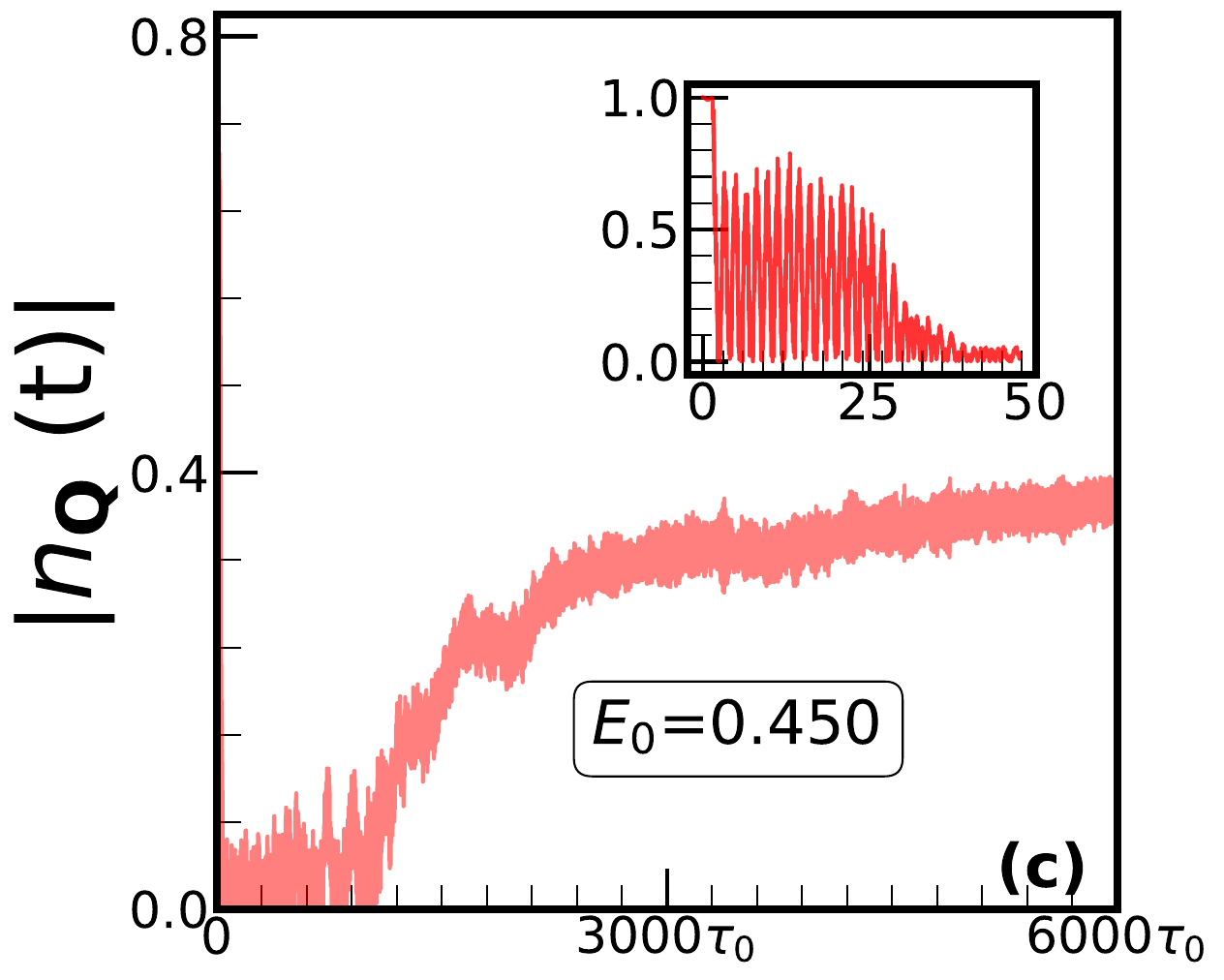}
\includegraphics[width=4.0cm,height=3.2cm]{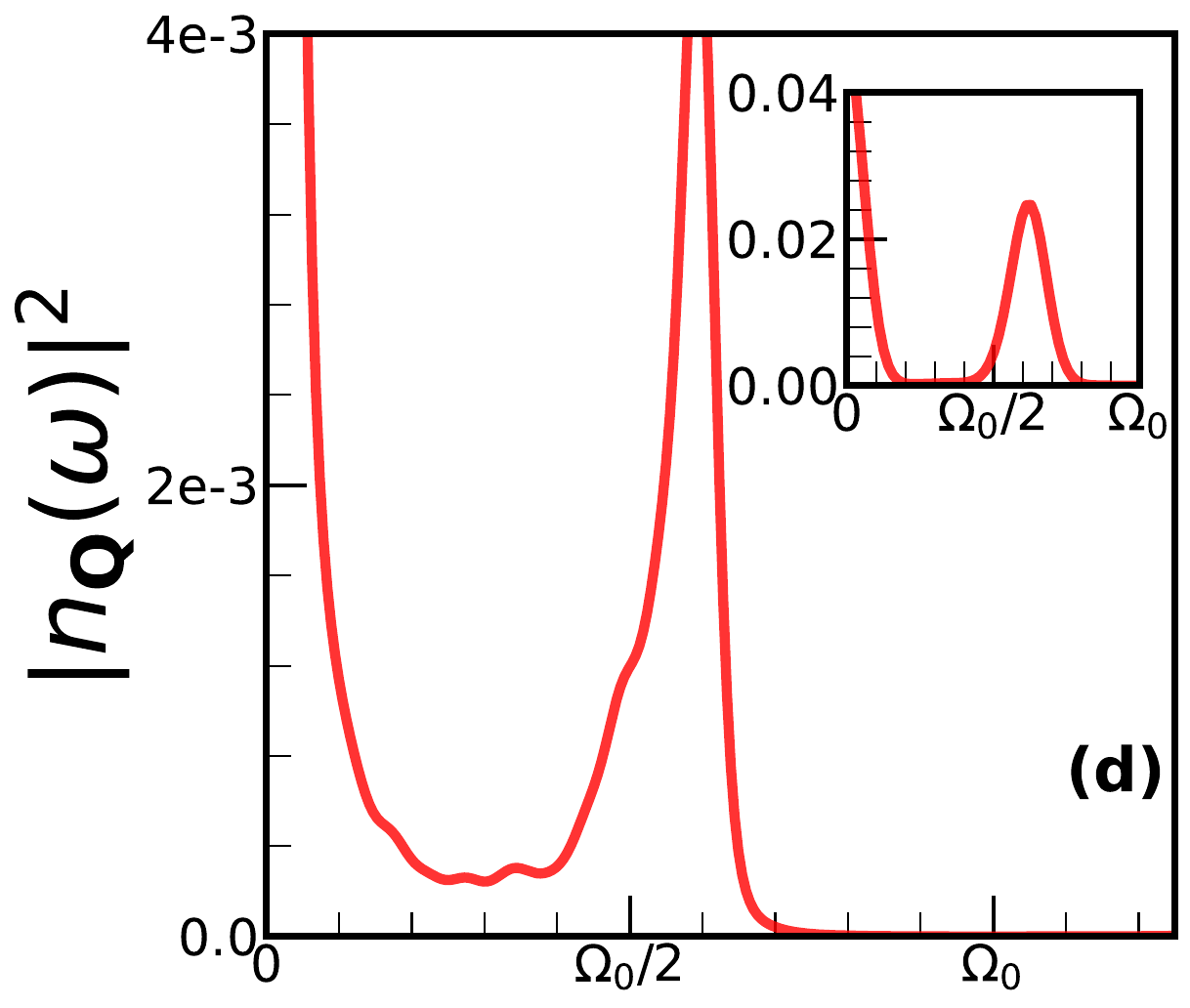} }
\caption{Critical pulse strength regime: (a)~and (c) shows real time
behaviour of $|x_{\bf{Q}}(t)|$ for the phonons and the electron density,
respectively.  Corresponding insets show the behaviour at short times,
$t \lesssim 50 \tau_0$. At short time the ordering amplitude in both phonons
and electrons decay to zero. Then there is a very low amplitude
quiescent state, followed by a `revival' - rising to the long time value
as $|x_{\bf{Q}}(t)| \sim e^{-(\tau/t)^{\beta}}$.  The respective lineshapes,
(b) and (d), show a peak near zero as well as broad finite frequency
feature at $\Omega_{\bf Q}$.  }
\end{figure}
\begin{figure}[t]
\centerline{
~
\includegraphics[width=4.4cm,height=4.0cm]{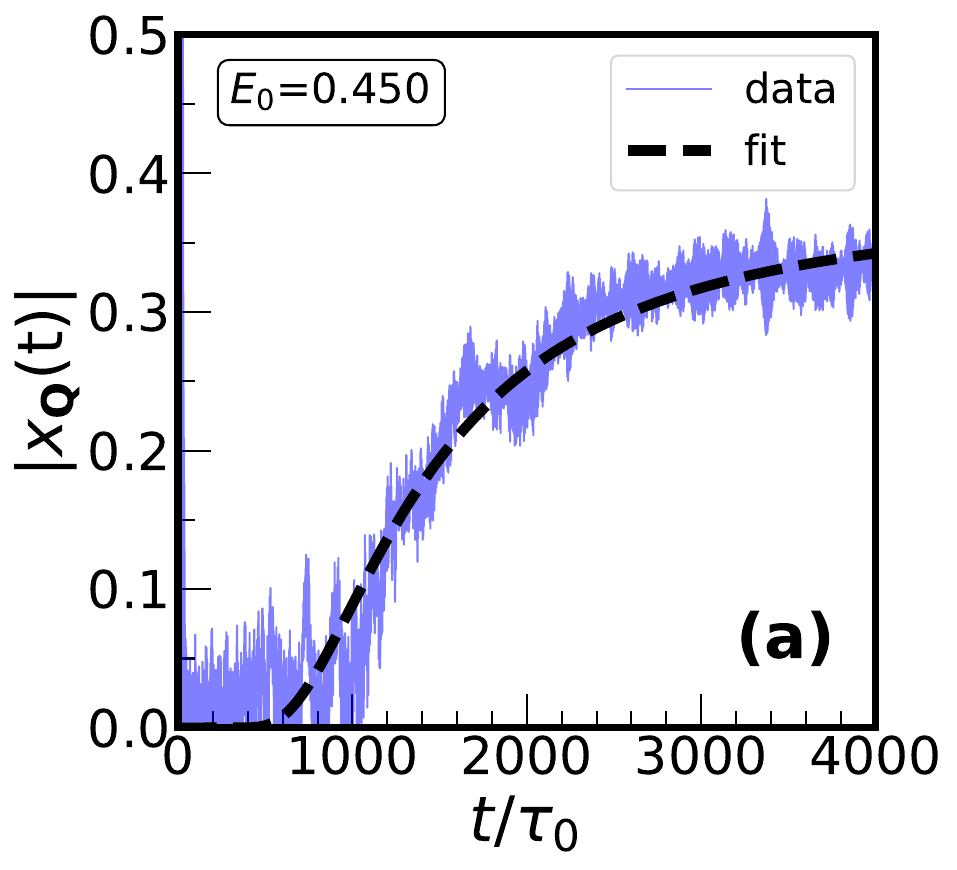}
\hspace{-.4cm}
\includegraphics[width=4.2cm,height=4.0cm]{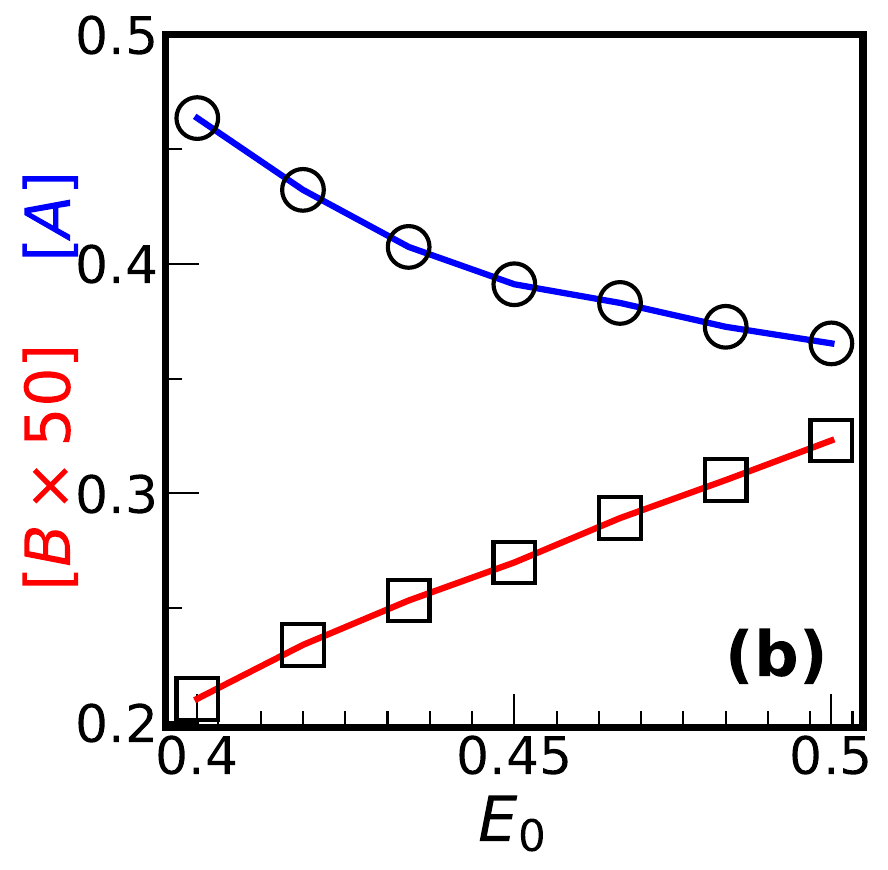} }
\centerline{
\includegraphics[width=4.2cm,height=4.0cm]{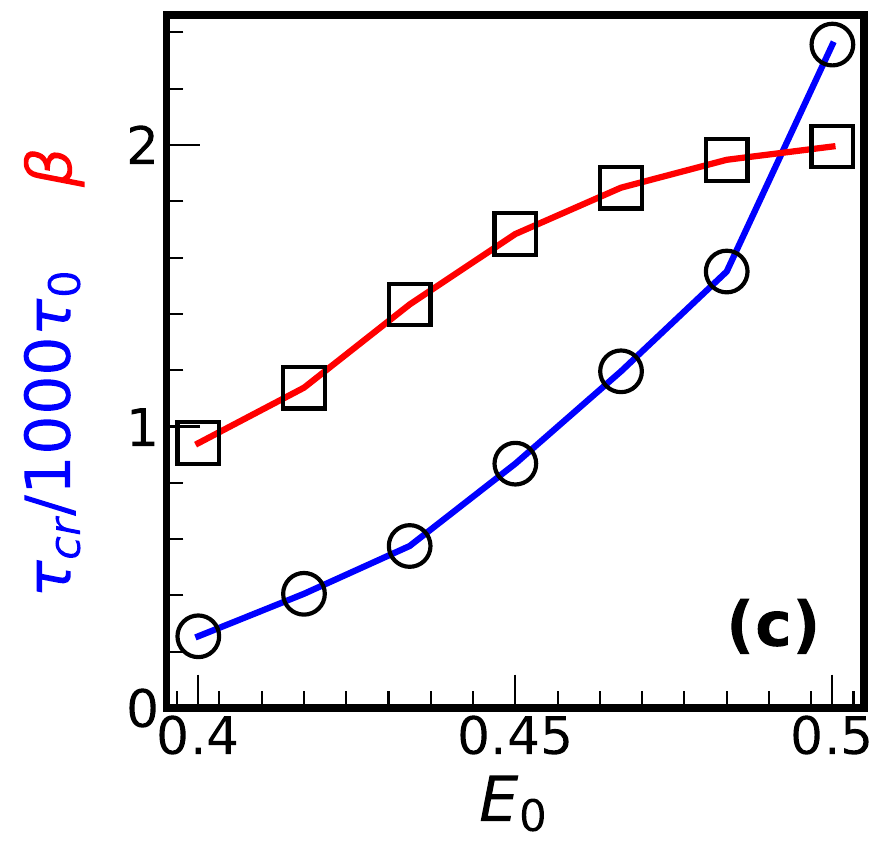}
\includegraphics[width=4.2cm,height=4.0cm]{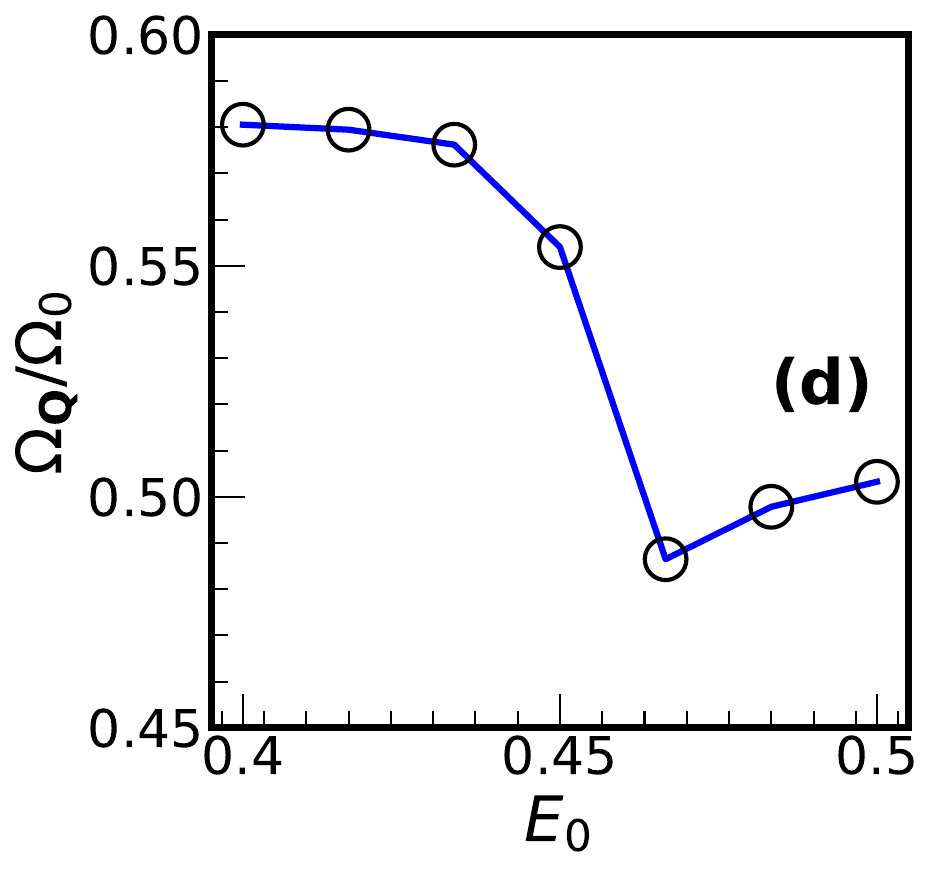} }
\caption{Characterising dynamics in the critical pulse regime. We fit
$x_{\bf Q}(t)$ to the form $A  e^{-(\tau_{cr}/t)^{\beta}}
+ B cos(\Omega_{\bf Q}t)$.  (a)~Comparison of the growth part
$A  e^{-(\tau_{cr}/t)^{\beta}}$ of the fit function with the actual data at
$E_0 = 0.45$.  (b)~The amplitude $A$ (blue) that decides the steady state value
of $|x_{\bf Q}|$ and $B$ (red) that decides the size of oscillations. Note
that $B$ is one order of magnitude smaller than $A$ for the
system size we have studied here. (c)~The increase in $\tau_{cr}$ (blue)
by more than an order of magnitude while $E_0$ varies only from $0.41$ to $0.47$,
and the exponent $\beta$ (red)
which stays between $0-1$. 
(d)~The frequency $\Omega_{\bf Q}$ shows non-monotonic
behaviour, dropping and then rising with increasing $E_0$.}
\end{figure}

\subsection{Critical pulse strength regime}

For $E_0 \gtrsim 0.35$ the nature of $\vert 
x_{\bf Q}(t)\vert$ changes.
Beyond a few cycles of oscillatory response 
both the phonon and charge density fields are 
suppressed to roughly $\sim 1-2 \%$ of their pre-pulse value, 
and persist in this state - with small oscillations
- for a timescale that we call $\tau_{cr}$. 
Beyond this  both  $\vert x_{\bf Q}(t) \vert $ and 
$\vert n_{\bf Q}(t) \vert$
`revive', reaching a finite but low amplitude long time
state. Fig.7.(a) and (c) show the time dependence of the 
two structure factors at $E_0 = 0.45$, the main panels show 
the overall time dependence while the inset shows the initial 
drop that occurs over $\sim 25 \tau_0$.
A description of the form 
$
x_{\bf Q}(t) \sim 
A e^{-(\tau_{cr}/t)^{\beta}}  + B cos (\Omega_{\bf Q}t)
$
captures multiple features of the complex dynamics.
Beyond the initial drop the function above captures the
following features of the data: 
(i)~The low amplitude quiescent
state, at $t \lesssim \tau_{cr}$, is described by 
exponentially small contribution from the first term,
and small oscillations $B \vert cos(\Omega_{\bf Q}t) \vert$, 
with  $B \ll 1$.
(ii)~As $t$ goes beyond
$\tau_{cr}$ the $A$ term makes a significant contribution,
seen in the rise of the mean curve.
The exponential saturates for $t \gg \tau_{cr}$ but
long term oscillations persist.
(iii)~Finally, the mean
value of $\vert x_{\bf Q} \vert$ rises as 
$1 - (\tau_{cr}/t)^{\beta}$
at long times, a power law rather than exponential rise
to the steady state.

Panels (b) and (d) show $\vert x_{\bf Q}(\omega) \vert$ and 
 $\vert n_{\bf Q}(\omega) \vert$, respectively. There are the 
usual features around $\Omega_{\bf Q}$ 
There is also an interesting low energy feature, whose
width is $\sim (\tau_{max} - \tau_{cr})^{-1}$.

Fig.8 panel (a) compares the actual data to a fit of the form
assumed. We have shown only the envelop since showing the
oscillations is not feasible over $6000 \tau_0$. 
As we have stated the single scale $\tau_{cr}$ describes both
the low amplitude quiescent state as well as the power law
rise to the steady state. Panel (b) shows the amplitude $A(E_0)$.
In a later figure we will show the size dependence of this 
result, allowing us to extract the critical behaviour. 
(c) shows the rapid rise of the `delay time' $\tau_{cr}$ that 
is needed for revival of the CO state. (d) Shows the exponent
$\beta$ that controls the power law approach to the steady
state. (e) and (f) show the amplitude $B$, controlling the
oscillations, and the primary frequency $\Omega_{\bf Q}$.

\begin{figure}[b]
\centerline{
\includegraphics[width=4.4cm,height=4.0cm]{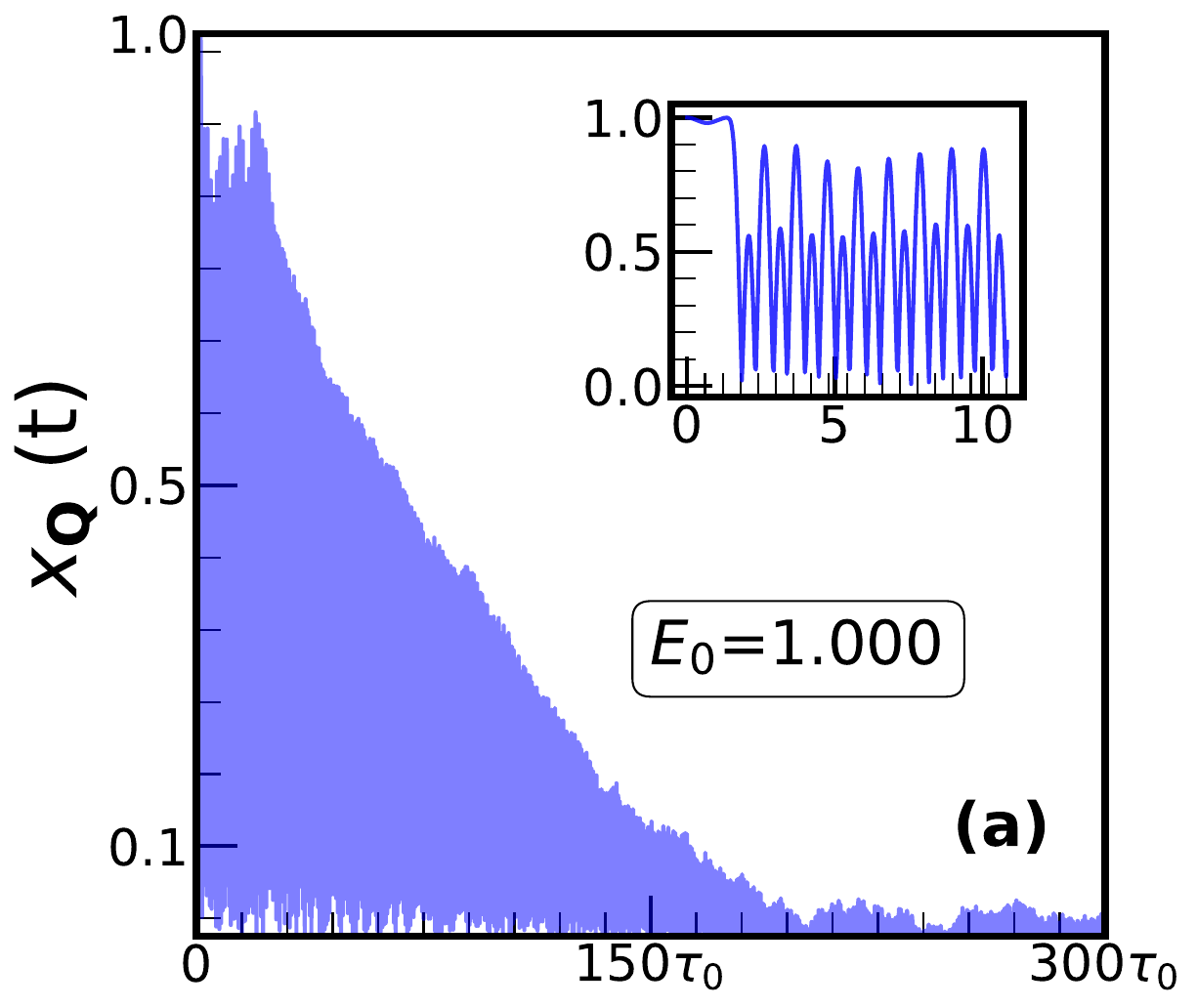}
\includegraphics[width=4.0cm,height=4.0cm]{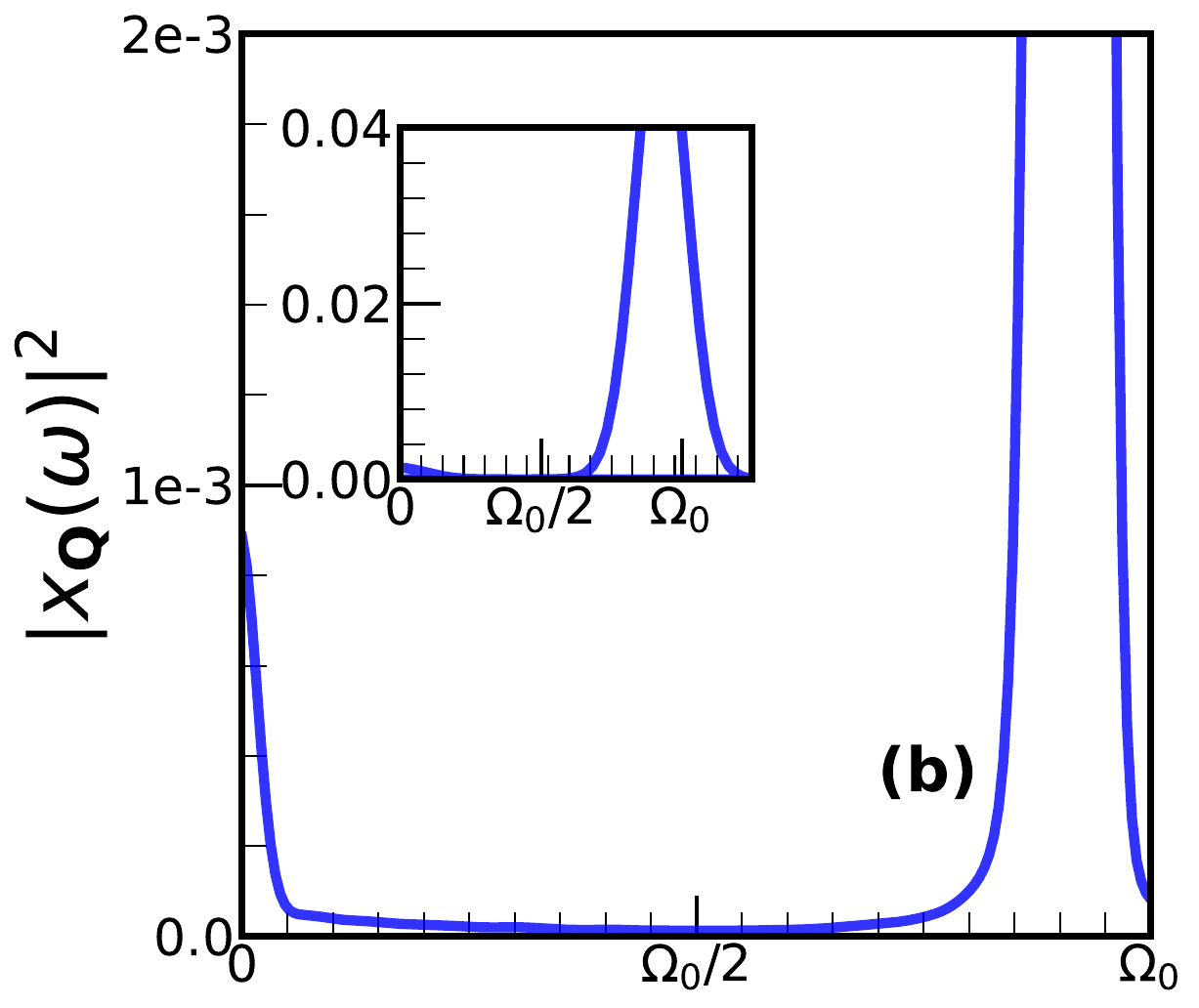}
~~}
\centerline{
\includegraphics[width=4.3cm,height=4.0cm]{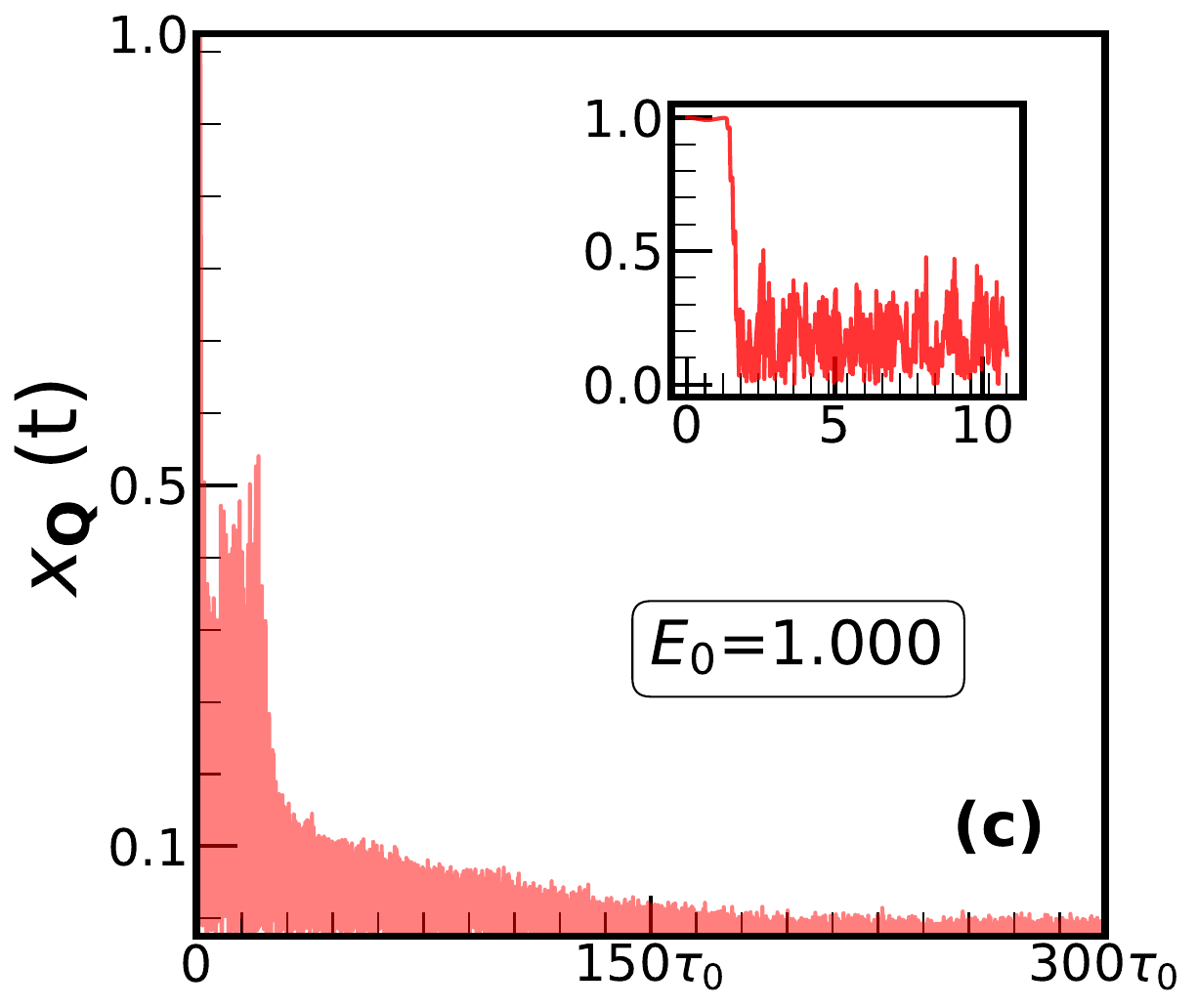}
\includegraphics[width=4.2cm,height=4.0cm]{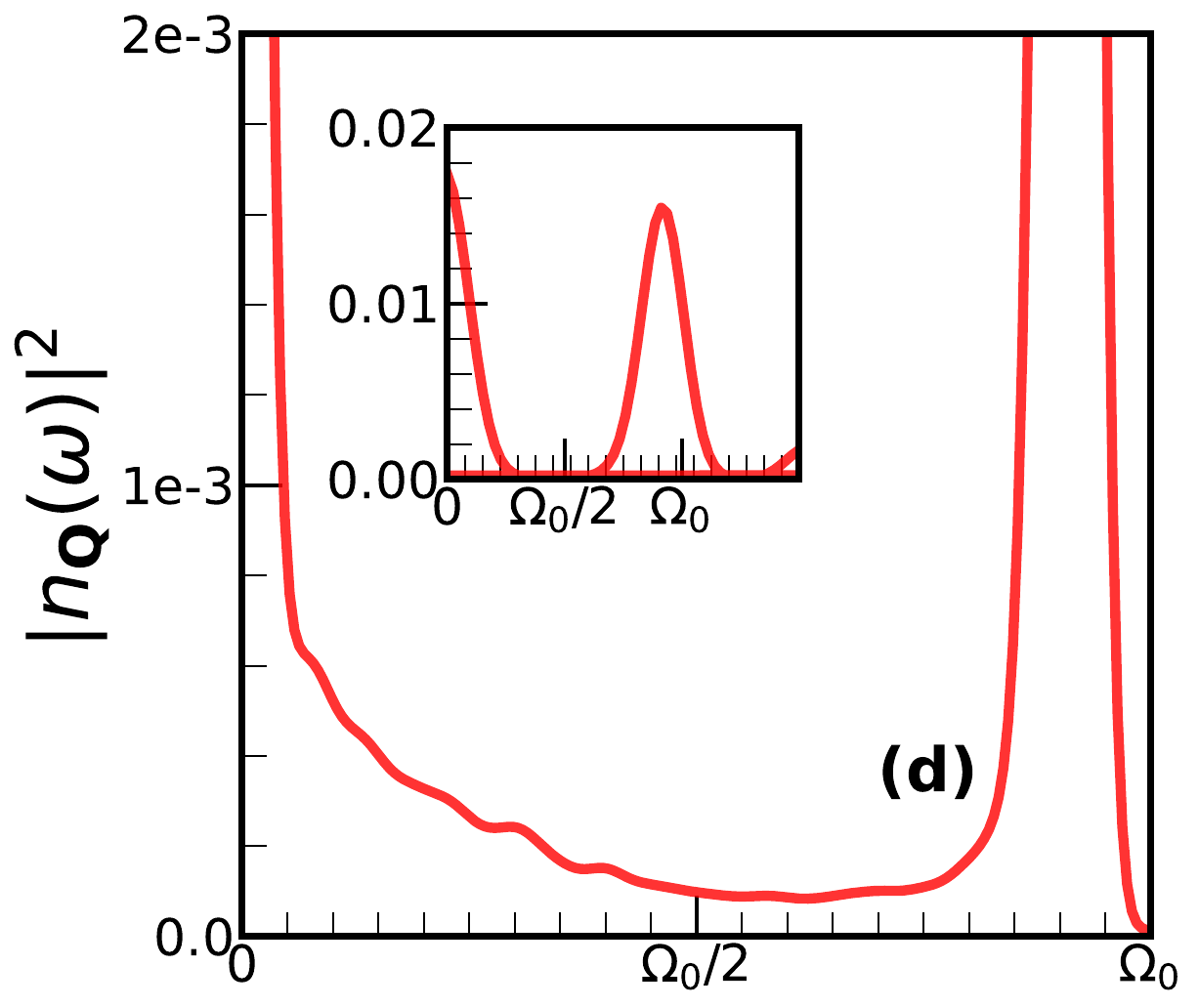}
}
\caption{
Strong pulse regime: (a)~The relatively slow decay of
$|x_{\bf Q}(t)|$ contrasted with (c)~the rapid
suppression of $|n_{\bf Q}(t)|$.
(b) and (d) show the Fourier transforms of $x_{\bf Q}(t)$
and $n_{\bf Q}(t)$, respectively.
}
\end{figure}
\begin{figure}[t]
\centerline{
\includegraphics[width=4.3cm,height=4.0cm]{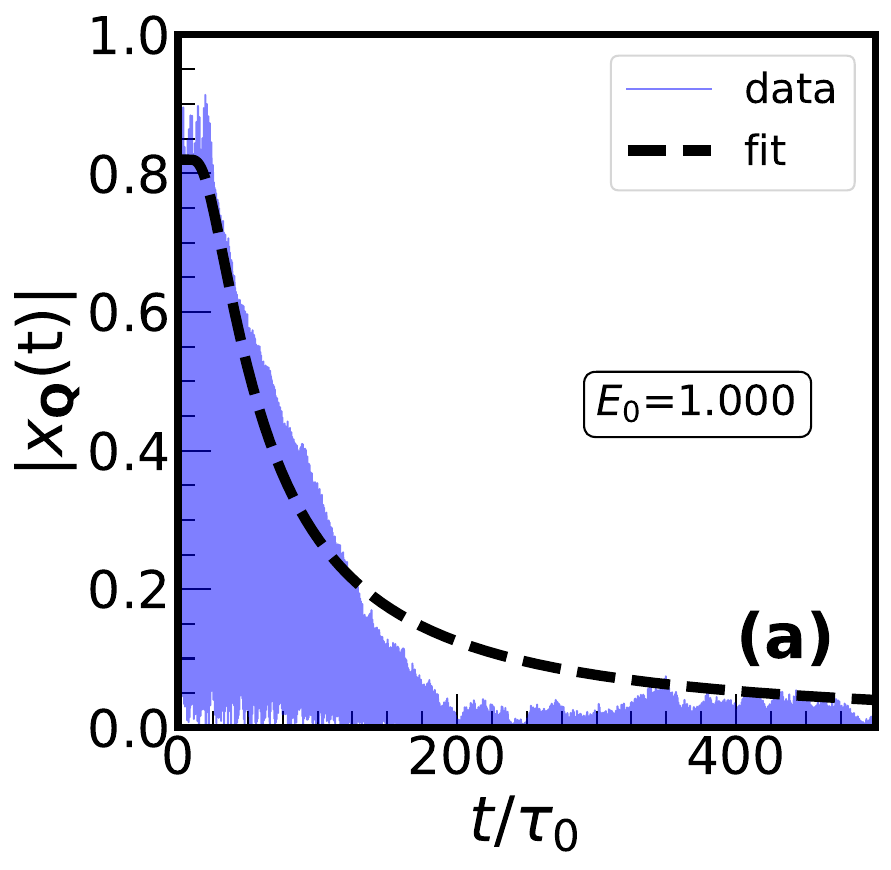}
\includegraphics[width=4.0cm,height=4.0cm]{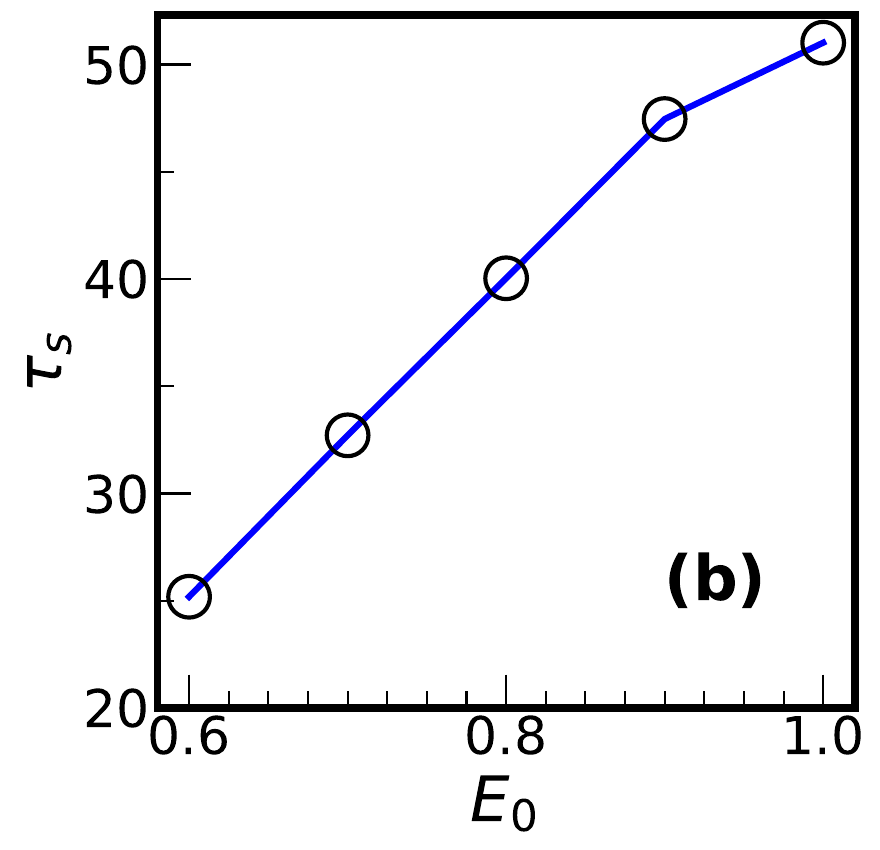} }
\centerline{
\includegraphics[width=4.2cm,height=4.0cm]{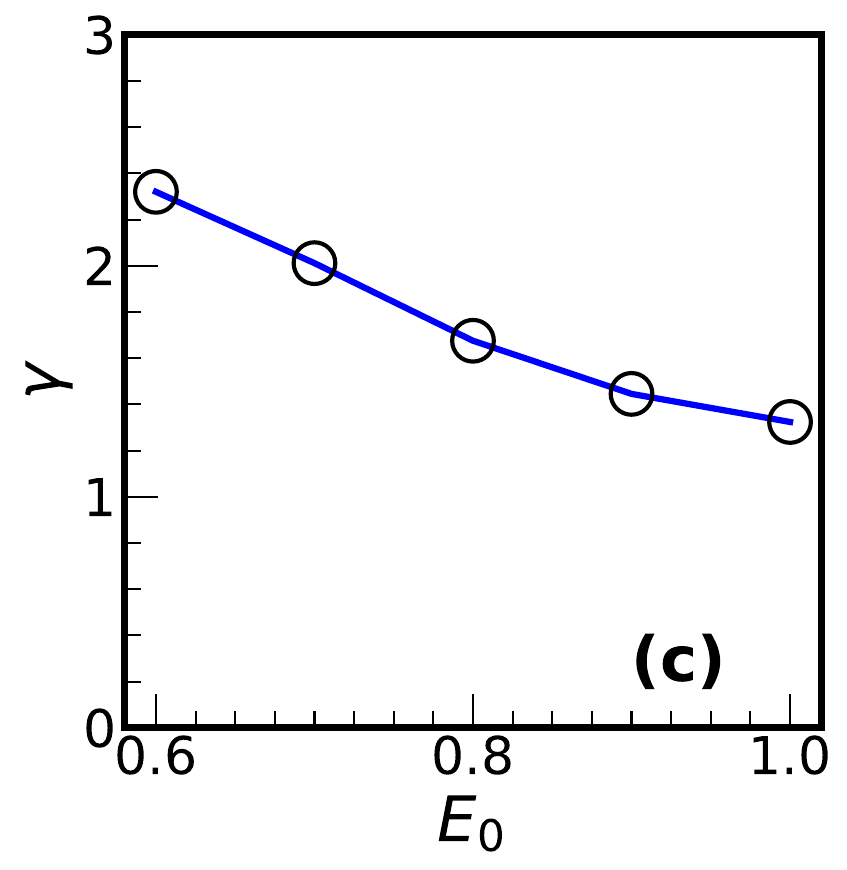}
\includegraphics[width=4.0cm,height=4.0cm]{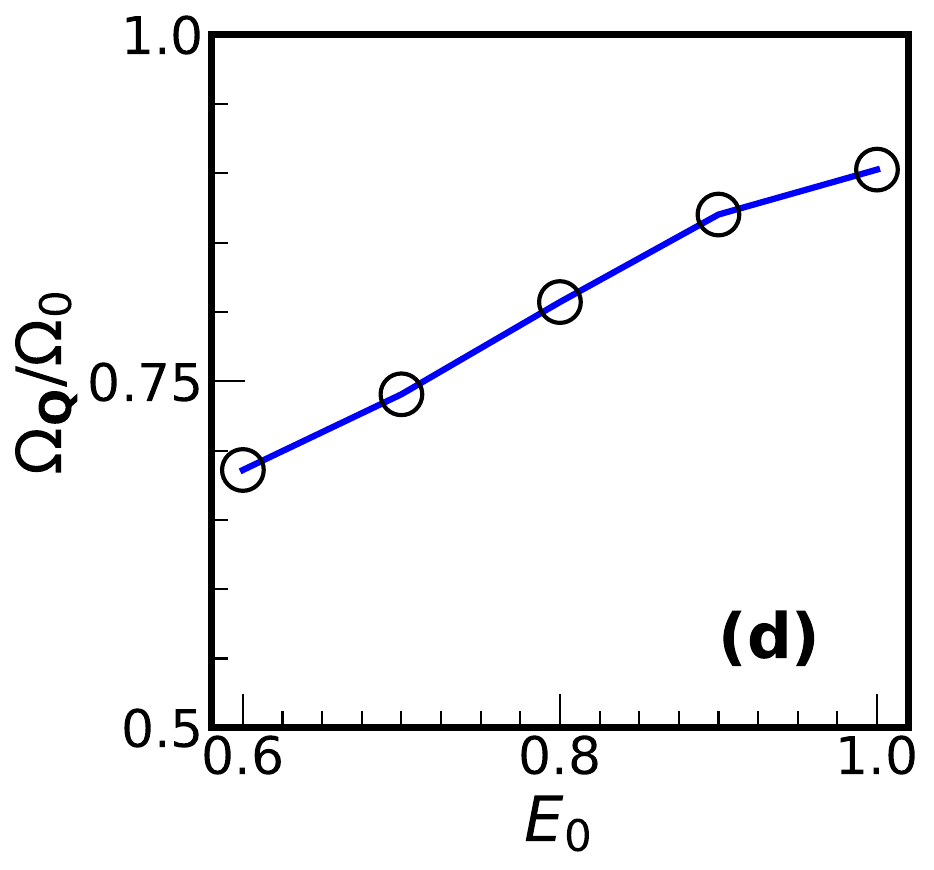} }
\caption{
Strong pulse regime: fit to $|x_{\bf Q}(t)|$, and
fitting parameters. (a)~Shows the fit to the envelop,
capturing both the initial flat part and the longer term
decay using the same timescale $\tau_s$.
(b) and (c)~Shows monotonic increase of $\tau_s$
and decrease of the exponent $\gamma$ with $E_0$.
(d)~Shows the increase in the oscillation frequency
towards $\Omega_0$ as $E_0$ increases.
We are not showing the amplitude $A$, which
stays almost constant at $A \sim 0.88$.
}
\end{figure}

\subsection{Strong pulse regime}

Fig.9 shows the detailed dynamics at $E_0 = 1.0$, which is 
in the strong pulse regime. Panel (a) shows $\vert x_{\bf Q}(t)
\vert$,
it has a modest drop on the scale of $\tau_0$, then a
period of roughly constant amplitude 
oscillation for about $25 \tau_0$,
and finally a decay to zero that seems to be fit by a power law.
For our fitting function: 
$ x_{\bf Q}(t) \sim 
A~(1-e^{-(\tau_s/t)^{\gamma}}) ~cos(\Omega_{\bf Q}t)$,
the $t \ll \tau_s$ regime is of fixed amplitude oscillation,
$\vert A~cos(\Omega_{\bf Q}t) \vert$
while for $t \gg \tau_s$ we get a behaviour 
$\vert A (\tau_s/t)^{\gamma}~cos(\Omega_{\bf Q}t) \vert$.

The response in the charge sector, panel (c), is very different.
$|n_{\bf Q}|$ falls sharply from $1$ to $\sim 0.15$ within
$t \sim \tau_0$, then has roughly fixed amplitude oscillations
for $t \lesssim \tau_s$, and then abruptly collapses.
We do not see any prominent power law tail (at least at
this value of $E_0$) unlike in $|x_{\bf Q}|$.
Panels (b) and (d) show the corresponding Fourier transforms.
Both of these have the usual peaks at $\Omega_{\bf Q}$.

Fig.10 shows the fit to the strong pulse data and the
variation of the fit parameters with $E_0$.
Panel (a) shows the fit to the actual data using the
assumed functional form. Both the fixed amplitude part
and the ``power law'' decay in $x_{\bf Q}$ are
captured reasonably by the fit function. Panel (b) shows
the timescale $\tau_{s}$ which {\it increases} initially
with increasing $E_0$ (past $E_0^c$) and then tends
to saturate. Panel (c) shows the exponent $\gamma$,
which reduces
from $\sim 2.5$ to $\sim 1.5$ as $E_0$ increases from
$0.6$ to $1.0$. In (d) $\Omega_{\bf Q}$
increases with $E_0$ tending to the bare $\Omega_0$
(independent local oscillators) at large $E_0$. 
The amplitude $A$ is almost flat in this regime at {\it $A=0.88$}. 

\begin{figure*}[t]
\centerline{
\includegraphics[width=12.4cm,height=3.4cm]{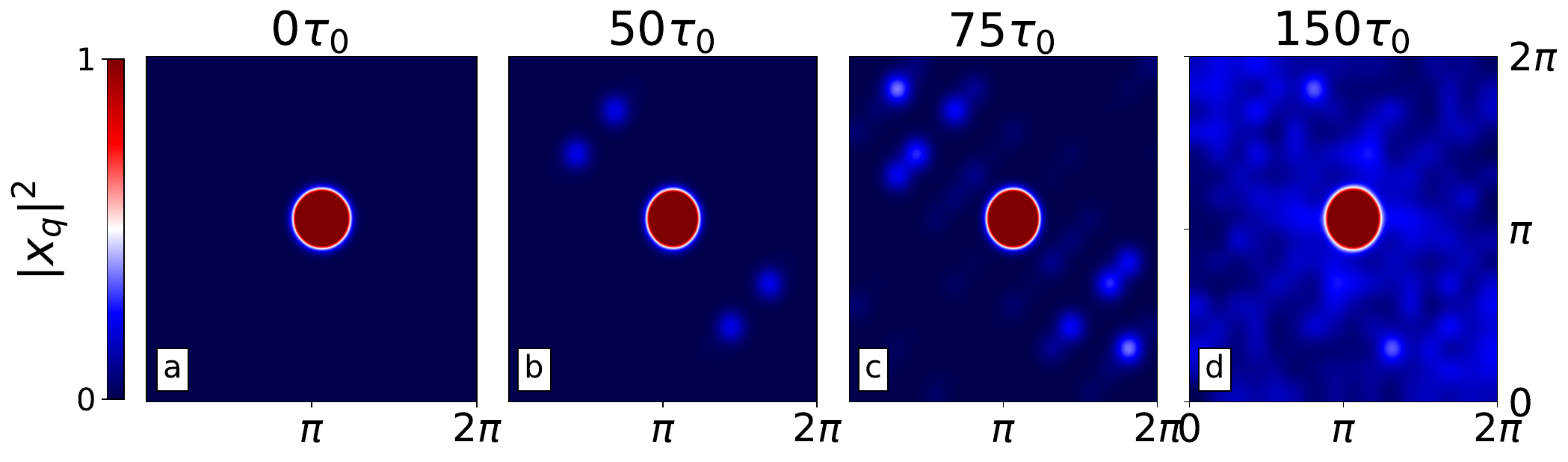} 
\includegraphics[width=4.30cm,height=3.2cm]{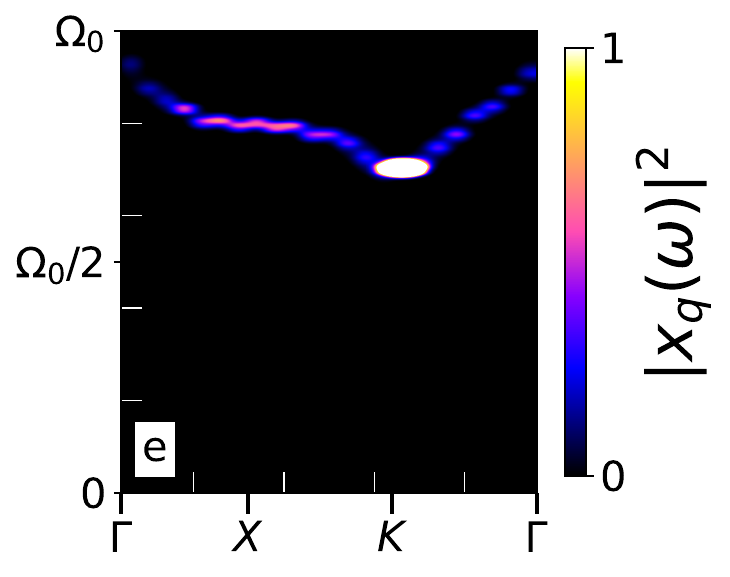} 
}
\centerline{
\includegraphics[width=12.4cm,height=3.4cm]{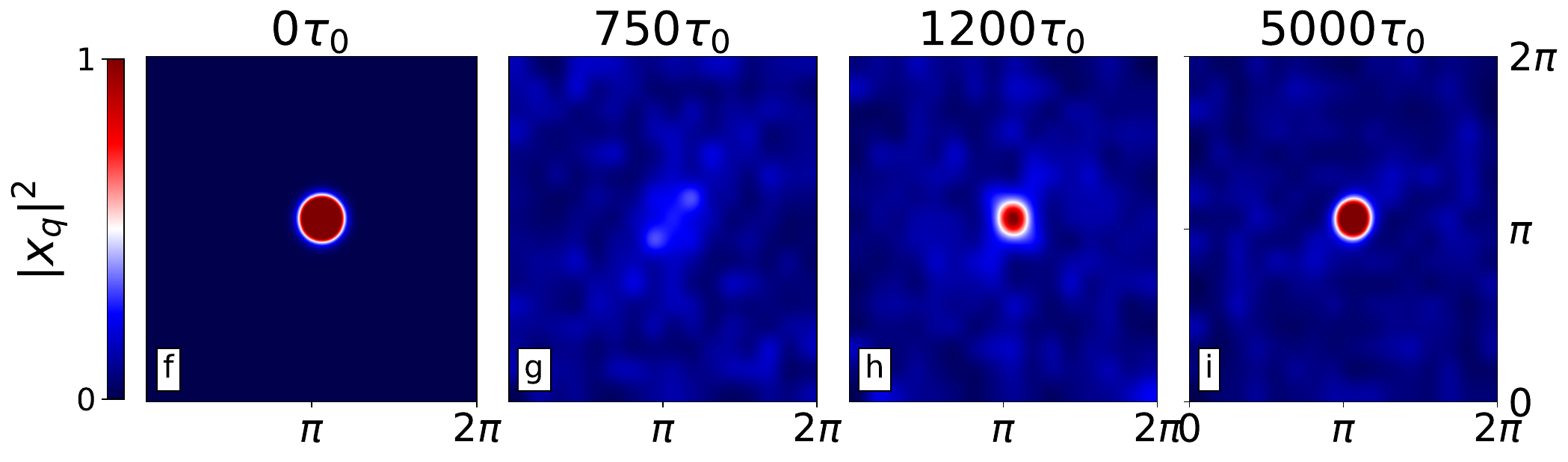} 
\includegraphics[width=4.30cm,height=3.2cm]{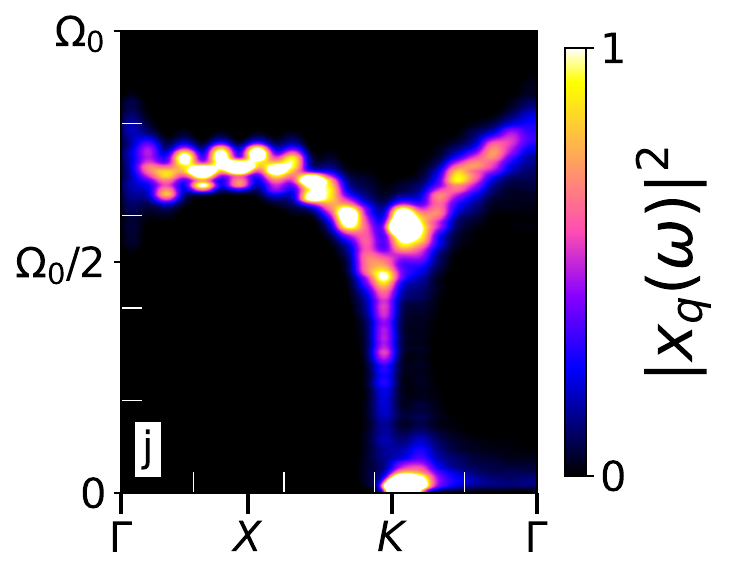} 
}
\centerline{
\includegraphics[width=12.4cm,height=3.4cm]{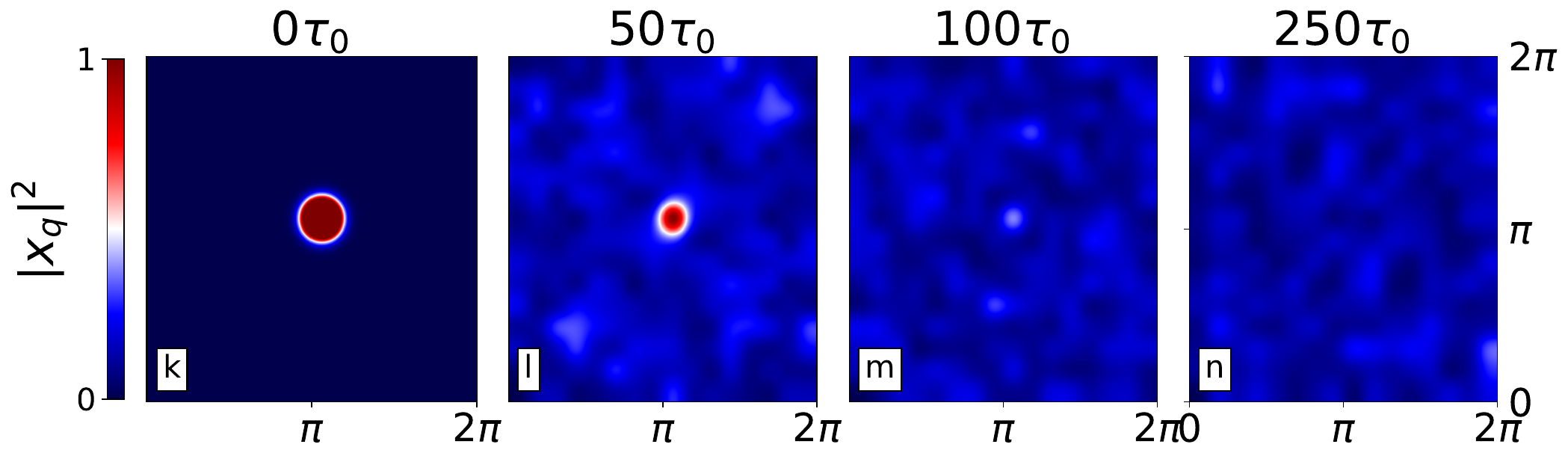} 
\includegraphics[width=4.30cm,height=3.2cm]{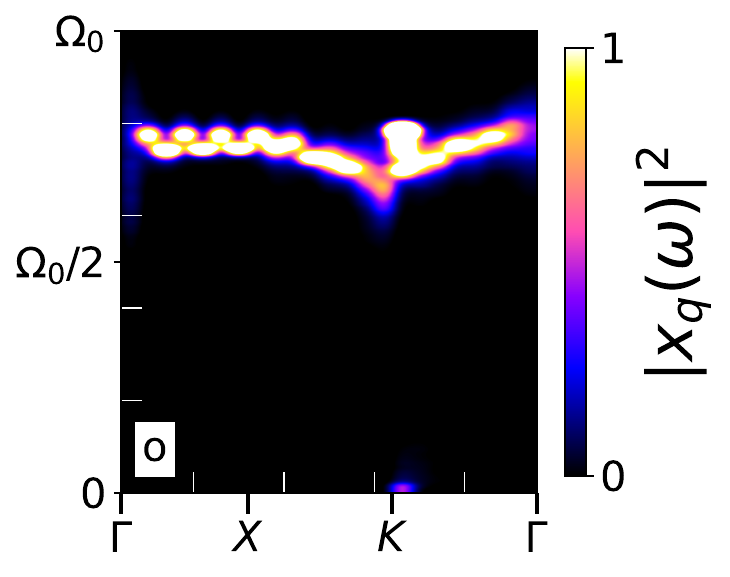} 
}
\caption{The time dependence of $|x_{\bf q}|^2$ over the entire 
Brillouin zone (BZ), and the associated frequency dependence, in 
the three pulse regimes. Top row: weak pulse ($E_0=0.2$). (a-d) 
With increasing $t$ first the intensity of ${\bf q}$ modes along
the diagonal increase and finally all ${\bf q}$ modes become active.
However the ${\bf q} = {\bf Q}$ mode retains its dominant position.
(e) shows the $|x_{\bf q}(\omega)|^2$, obtained via Fourier transform 
of  $x_{\bf q}(t)$ over the whole time range, on a selected path 
$\Gamma(0,0)-X(\pi,0)-K(\pi,\pi)-\Gamma(0,0)$ in $\textbf{q}$-space.
The response maps out the equilibrium phonon dispersion but with 
weight mostly centered at $K$-point.
Middle row: critical regime ($E_0=0.43$). (f-i)~There is a quick
transfer of energy to ${\bf q}$ modes all over the Brillouin zone
(BZ) and at $t=750 \tau_0$ there is no peak at ${\bf q} = {\bf Q}$.
By $1200 \tau_0$ there is  a recovery and by $5000 \tau_0$ the 
intensity approaches the long time asymptote.  (j)~Frequency 
dependence reveals phonon band softening and broadening and a 
significant dip near $K$. Bottom row: strong pulse ($E_0=0.7$).
The timescales here are shorter than in the upper rows. The
mode at ${\bf Q}$ initially transfers energy along the diagonal
of the BZ, which then cascades to the rest of the BZ. By
$t \sim 100 \tau_0$ the energy has been distributed almost evenly
over the BZ and there is no peak at ${\bf q} = {\bf Q}$.
(o)~The phonon-band starts 
flattening again and the mean frequency increases.}
\end{figure*}

\section{Dynamics of the overall system}

Till now the focus has been on the order parameter
mode at ${\bf q} = (\pi, \pi) \equiv {\bf Q}$. 
We now pay attention to the
other ${\bf q}$ modes to which energy is transferred 
from ${\bf Q}$ via
mode coupling, and also the electron population - 
which dictates the overall energy sharing between 
electrons and phonons.

\subsection{Dynamics of momentum modes}

We first look at evolution of
$\vert x_{\bf q}(t) \vert^2$ over the Brillouin zone (BZ)
for three representative values of pulse amplitude.
We also examine  $\vert x_{\bf q}(\omega) \vert^2$,
obtained by Fourier transforming $x_{\bf q}(t)$ 
over the whole time 
window, for scans across the BZ.

Our  $x_i(t)$ is `driven' by
the density variable $n_i(t)$. When the
$x_i$ and the corresponding density deviate only
slightly from the ideal periodic 
value we obtain normal mode
vibrations at frequencies given by:
$$
\Omega_{\bf q} = \sqrt{ \Omega_0^2 + (g^2/m) \Pi_0({\bf q})
} $$
where $\Pi_0$ encodes the density response of the CO state to
a change in $x$.

\begin{figure*}[t]
\centerline{
\includegraphics[width=14.0cm,height=11.9cm]{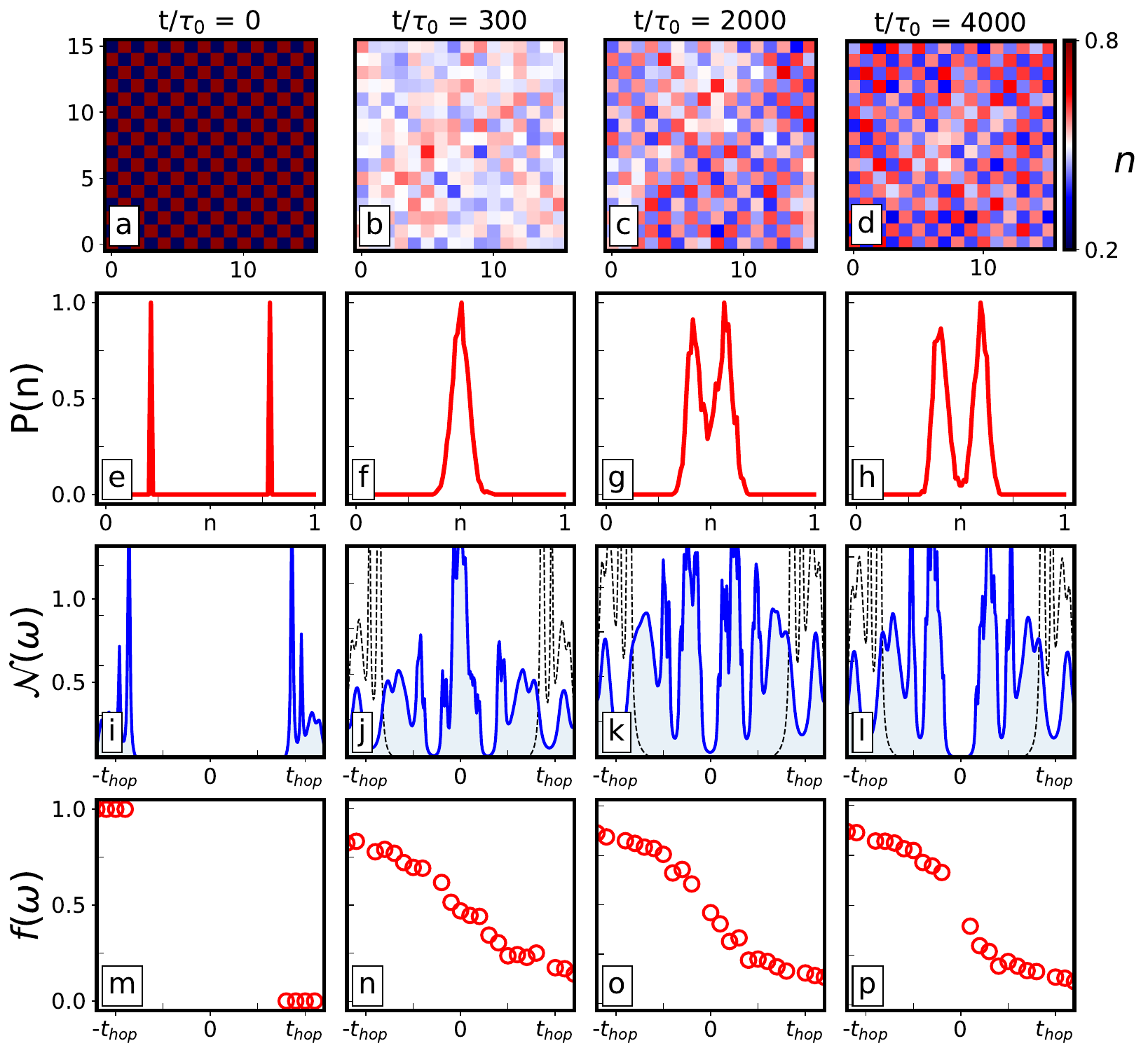}
}
\caption{Time dependence of spatial distribution and electronic 
properties in the critical pulse regime, $E_0 = 0.45$.
First row: spatial variation of electron density $n_i$. 
At $t=0$ we have perfect checkerboard order $n_i$ taking values 
$\sim 0.2$ and $0.8$ at alternate sites.  
Second row: the distribution 
of density $P(n,t)$ corresponding to the spatial maps above. 
The $P(n)$ is sharply bimodal. 
At $t=300 \tau_0$ the $n_i$ map show that charge modulations have 
essentially vanished, and $P(n)$ shows an unimodal distribution 
peaked around $n=0.5$. 
At $t = 2000 \tau_0$ we again see the re-emergence of charge 
modulations, a checkerboard structures, and a bimodal feature 
in $P(n)$. 
The last column show the `revived' charge density, now varying 
roughly between $0.4$ and $0.6$. This is roughly $1/3$ the
charge modulation in the $t=0$ state. 
The $P(n)$ has a clearer bimodal feature now.
Third row: density of states $\cal{N}(\omega)$ based on the 
instantaneous eigenvalues in the $x_i(t)$ background. 
At $t=0$ there is a gap $\sim 1.6 t_{hop}$. 
The gap closes around $\sim 300\tau_0$ and starts to reopen
around $\sim 2000\tau_0$. It reaches a saturation value 
$\sim 0.5t_{hop}$ around $\sim 4000\tau_0$. 
Fourth row: the occupation function $f(\omega, t)$.
The pre-pump distribution is the zero-temperature
Fermi function. The pump depletes low energy states and
creates population at significantly high energies. 
After the gap reopens a large upper-band population 
still retains. 
}
\end{figure*}

This undamped oscillatory behavior is observed only at
very weak pumping, as we have seen in Fig.3(a). It does
not capture mode coupling, that leads to damping, or to the
loss and revival phenomena that we observe in the critical
regime.  These effects involve the energy
distribution over all ${\bf q}$ as a function of time
and also the nature of electronic excitations.

Fig.11 shows maps of $\vert x_{\bf q}\vert$ at different times
for three values of $E_0$ and the associated  
$\vert x_{\bf q}(\omega) \vert^2$. 
In the left column, at $t=0$ (pre-pulse) and for 
all values of
$E_0$,  the only `bright' feature is at ${\bf q} = {\bf Q}$. 
Other modes are inactive.
The principal observations from the time dependence 
are the following:

(i)~{\it Weak pulse regime (top row):}
At short time, $\sim 50 \tau_0$, 
the first additional features show up along the diagonal of the
Brillouin zone (BZ). The peak at ${\bf Q}$ still remains 
the most prominent. At $75 \tau_0$ there are more excitations
along the diagonal and by $150 \tau_0$ the whole BZ is
`lit up' the brightest part being along the diagonal. 
This is related to the nature of the pump,
which is along $\vec{x}+\vec{y}$ and causes an
initial current in the  $\vec{x}+\vec{y}$ direction. 
The mode at ${\bf Q}$ remains the most prominent
at all times, diminishing slightly in intensity.
The Fourier transform $\vert x_{\bf q}(\omega) \vert^2$
essentially matches the phonon 
dispersion $\Omega_{\bf q}$ of the charge ordered phase, 
but with the brightest intensity at ${\bf q} = {\bf Q}$.
We are probing linear response.

(ii)~{\it Critical pulse regime (middle row):}  The spreading
of energy over the BZ happens quickly and suppresses the
peak at ${\bf Q}$. Till $\sim 1000 \tau_0$
there is no prominent feature in the BZ. Beyond this a peak
at ${\bf Q}$ reemerges and gains some weight. The intensity
at the longest time, $5000 \tau_0$, 
is far below what it was at $t=0$. 
The associated Fourier transform looks very different from
the linear response $\Omega_{\bf q}$, and is actually
reminiscent of what one observes in the equilibrium
problem near its thermal transition.

(iii)~{\it Strong pulse regime (bottom row):}  
The $x_{\bf Q}$ amplitude diminishes monotonically
with time and by $\sim 100 \tau_0$ there is no trace of the
peak at ${\bf Q}$. The corresponding $\vert x_{\bf q}(\omega) 
\vert^2$
shows an almost flat - momentum independent - oscillation, with
uniform distribution of amplitude across ${\bf q}$ 
similar to the equilibrium high temperature case.
\begin{figure}[t]
\centerline{
\includegraphics[width=8.4cm,height=3.9cm]{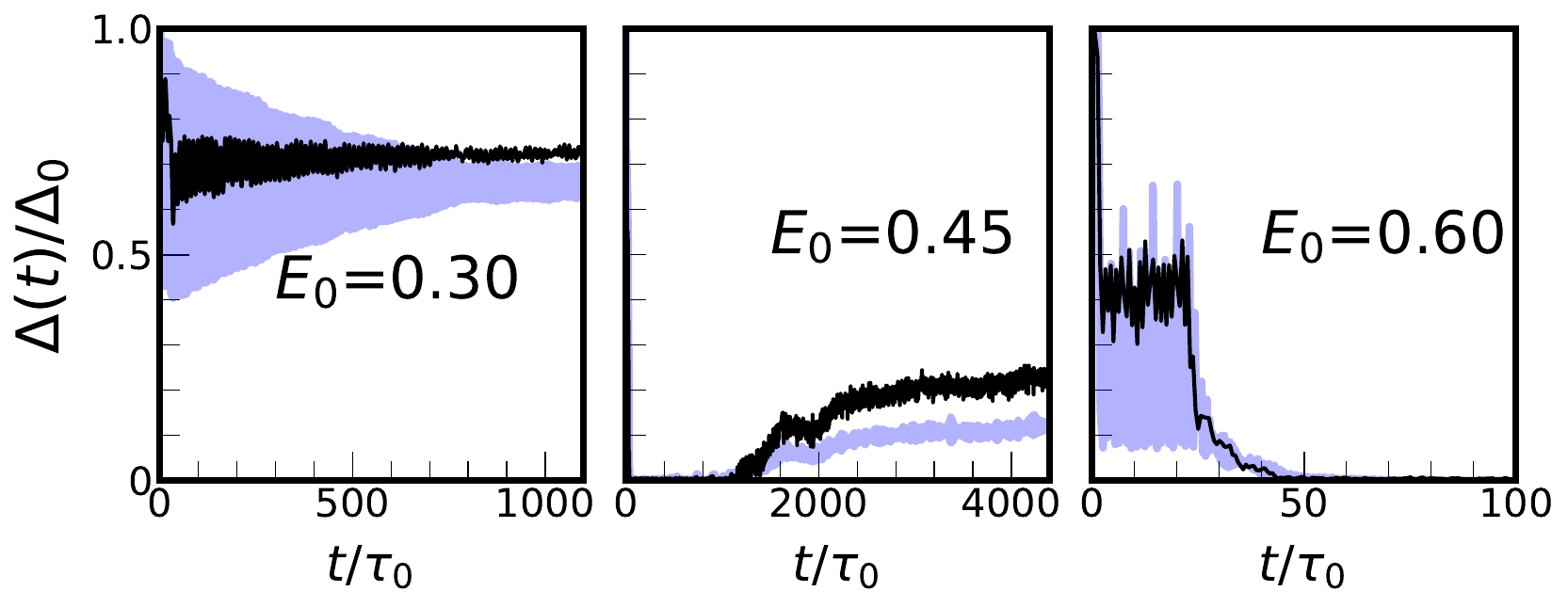}
}
\caption{Behaviour of the gap in the electronic DOS (black)
computed from the instantaneous spectrum at different $E_0$. In weak 
pulse regime, $E_0 = 0.3$,  $\Delta(t)$ oscillates then relaxes to 
70\% of $\Delta_0$. In critical regime, $E_0\sim 0.45$, 
$\Delta(t)$ shows suppression and revival. In the strong pulse 
regime, $E_0 = 0.6$, $\Delta(t)$ goes to zero by $\sim 50\tau_0$.
$\Delta(t)$ is closely correlated with $|x_{\bf{Q}}(t)|^2$ (blue).
Both $\Delta(t)$ and $|x_{\bf{Q}}(t)|^2$ were normalized by their 
initial equilibrium value.
}
\end{figure}
The ${\bf q}$ space picture reveals how energy is distributed 
over the BZ as a function of time. It does not explain whether
the destruction of CO, as in the
critical and  strong pulse regimes,
occur due to imperfect `phase correlations' between ordered
domains, or due to the destruction of charge modulations
itself. That requires examination of the spatial density
$n_i$ as a function of time.

\subsection{Spatial pattern and distribution functions}

Fig.12 top row shows maps of $n_i$ in the critical window. 
The main effect at intermediate time, $\sim (10^2 - 10^3)\tau_0$, 
is the homogenisation of the charge density. We can call it
`amplitude disordering', in contrast to `phase disordering' where 
modulations remain but interference between domains suppress order.
For $t \gtrsim 1500 \tau_0$ the density modulations reappear, 
though much weaker than before, and spatially organise.
At the longest time shown the 
$n_i$ field has a well organised alternating pattern, though
the modulations about $n=0.5$ are only $1/3$ of the $t=0$
value, and there is some amplitude inhomogeneity. 
The disordering and revival seems to be an amplitude
effect rather than a domain interference effect.
Phase slippage, i.e, boundaries between the two  
complementary charge order states, CO and CO',
is not playing an important role in the delayed recovery.
Rather, competition between the `metallic state' 
(homogeneous density) and suppressed CO is 
seemingly playing the crucial role.

When comparing the pump driven decay and revival to the
order parameter growth induced by thermal quench,
some differences become apparent. 
In the thermal situation the competition between 
complementary ordered phases CO and CO' plays a more 
significant role. 
In equilibrium, the dynamics of the local 
distortions $x_i(t)$ and density field $n_i(t)$ are 
closely correlated. At very low temperatures, 
the system assumes a CO state among the two symmetry-broken 
states (with ${\cal Z}_2$-like symmetry). However, at 
temperatures higher than the CO melting temperature 
$T_{CO}$, though local distortions are present, the presence 
of domains reduces long-range order, leading towards 
zero structure factor. Even if the system is 
thermally quenched, this domain growth alone is 
responsible for the delay in order parameter recovery.

In the photo-induction case, a laser pulse 
quickly melts this equilibrium distortion in 
$n_{\bf Q}(t)$, inducing an excess energy 
in $x_{\bf Q}(t)$, resulting in ringing in $x_{\bf Q}(t)$. 
However, the dynamics of $x_{\bf Q}(t)$ are not 
solely determined by the coupling to other 
phonon modes; electrons also play a crucial role, 
resulting in effective dissipation. 
Estimation of this dissipation is discussed in Sec. VI.

Due to this, the local distortions 
themselves reduce, and a combination of amplitude 
as well as domain growth plays an important 
role in the recovery of the order parameter. 
This is evident when we look at the destruction 
of the bimodality in the distribution 
function $P(n)$ discussed below.

The second row shows the distribution of densities, $P(n)$,
at different times. At $t=0$,  $P(n)$ shows a
bimodal structure with $\pm 0.3$ modulation around $n=0.5$.  
The post-pump state at $300 \tau_0$ shows a single peak centred 
at the mean density $n=0.5$. Around $t=1500\tau_0$ this starts 
to broaden, showing a hint of bimodality, and spatial configuration 
shows patches of charge order (CO). 
At $t=2000\tau_0$ we see a clear bimodal structure
which sharpens by $4000 \tau_0$. 

The third row shows the instantaneous DOS at different times obtained
by binning the eigenvalues $\epsilon_n(t)$ of the electronic
Hamiltonian in the background $x_i(t)$.
The pre-pump state has a gap $\sim 1.5t$
At $t \sim 300 \tau_0$ the gap has vanished
and the DOS shows a prominent peak at $\omega =0$. Beyond
this, with increasing $t$, as the lattice distortions 
increase and organise in the $(\pi, \pi)$ pattern
again, spectral weight is lost at low energy.
At $t = 2000 \tau_0$ a 
small gap shows up in the spectrum which
grows to about $\sim 1/3$ of the $t=0$ value at
the longest time, $4000 \tau_0$. 

In contrast to the equilibrium state at zero 
temperature, where electronic occupancy is
dictated by a sharp Fermi distribution, in the
post pump situation the occupancy of the instantaneous
eigenstates is time dependent and non zero for $\omega > 0$
states as well.
The bottom row 
illustrates the occupation of these levels derived from
$\rho_{nn}(t)$ by projecting $\rho_{ij}$ onto the 
instantaneous eigenvectors. At $t=0$, i.e before the
pulse, it is the usual Fermi function $\theta(-\omega)$.
The pump induces transitions to the upper
band. By $t \gtrsim 300\tau_0$  
the post-pump distribution  assumes a smooth form.
Subsequently, as the gap begins to open around 
$2000\tau_0$, a significant population of excited 
electrons $n_{exc}$ remains in the upper band. 
In this particular case, $n_{exc}$ amounts to 
approximately $25\%$.

Fig.13 quantifies the time 
dependence of the gap  $\Delta(t)$  inferred from the instantaneous
spectrum.
In the weak pulse regime the gap $\Delta(t)$ (normalized by $\Delta_0$)
shows oscillatory behaviour, followed be a relaxation to a slightly 
suppressed value. For example at
$E_0 \sim 0.3$ it stabilises to $\sim 70\%$ 
of it's original value. In the critical regime, the 
instantaneous gap shows an oscillatory behaviour but
averaged over a $10\tau_0$ time scale (black line) it shows 
suppression and revival dynamics. In strong pulse regime, the
gap goes to zero within $\sim 100\tau_0$. It can be noticed 
that this normalized gap is closely correlated 
with the normalized $|x_{\bf{Q}}(t)|^2$ as shown in the 
figure with faint blue line. 


\section{Results from a simple classical model}

Full mean field dynamics is expensive since 
it involves a ${\cal O}(N^2)$ process for 
every microscopic step $\delta t$.
The dynamics needs to be tracked to
a time $\tau_{max}$, that is needed to reach the
asymptotic state, and $\tau_{max}$ grows with system size.

The electron-phonon problem has a hierarchy of timescales,
$\delta t \ll {t_{hop}}^{-1} \ll \tau_0 \ll \tau_{max}$.
Typically $\delta t \sim 10^{-3} \tau_0$ and $\tau_{max}
\sim 10^4 \tau_0$. That indicates that the ${\cal O}(N^3)$
diagonalisation has to be done $\sim 10^7$ times. limiting
us to sizes $\sim 20 \times 20$. 
While the qualitative physics of suppression, revival, and
loss of order is accessible on this spatial and time scale
a simpler model, with much greater spatio-temporal reach, 
can be motivated from the lessons in the last section. 

These are:
(i)~The pump creates an `upper band' population 
that decays to non zero value over a short timescale.
This excited population $n_{exc}$ persists over a very long
time and can be approximated by a Fermi distribution with some
temperature $T_{el}$.
This affects the magnitude and ordering of the lattice
distortions.
(ii)~The effective temperature of the phonon system and the 
electron system are not the same on timescales that we
probe. $T_{el}$ is dictated by $n_{exc}$, we discuss $T_{ph}$
later.
(iii)~There is the usual growth and interference of CO domains
as with any ordering phenomena. This needs to be handled in a 
real space setting. 

This requires a model that captures large 
distortion physics non pertubatively, 
incorporates  intersite coupling that 
promotes a checkerboard pattern, and a 
mechanism to sense the population of excited electrons. 

\subsection{Constructing the model}

The simplest model for large distortion induced electron trapping 
is one electron on two sites (one electron on one site
is trivial since the hopping term is absent). For
this the electronic part of the model is:
$$
H_2^{el} = -t_{hop} (c^{\dagger}_1 c_2 + c^{\dagger}_2 c_1) - g(n_1 x_1 + n_2 x_2)
$$
Treating the $x$ as classical, the two eigenvalues of this are elementary:
$$
\lambda_{\pm} 
= - {{gx_+} \over 2} 
\pm {\sqrt{{g^2 x_{-}^2 \over 4} + t_{hop}^2}}
$$
where $x_{\pm} = x_1 \pm x_2$.
If the electrons sense an effective temperature $T_{el}$, that
creates the upper level population, then the free energy 
associated with the electronic part is
$$
F_{el}(x_1, x_2) 
= - T_{el} ~ln(e^{-\beta_{el}\lambda_{-}}  + e^{-\beta_{el}\lambda_{+}})
$$
Adding the stiffness terms, the effective `potential' dictating the 
dynamics of the $x_i$ is
$$
V(x_1, x_2) = {K \over 2}(x_1^2 + x_2^2) + F_{el}(x_1, x_2)
$$
A couple of comments before we generalise $V$ to the lattice, i.e, construct
an approximate $V(x_1, .. x_N)$.

(i)~The potential is symmetric under the interchange of $x_1$ and $x_2$.
(ii)~For $T_{el} \rightarrow 0$ we have $F_{el} \rightarrow \lambda_{-}$ and
minimising with respect to $x_+$ and $x_-$ leads to 
$$
{\bar x_+} = {g \over K},~~~
{\bar x_-} =  \pm {1 \over g} \sqrt{({{g^2} \over {K}} )^2 - t^2}
$$
(iii)~At large $g$ the lowest eigenvalue can arise from two $x$ configurations:
$(x_1 =g/K, x_2 =0)$ and its complement. 
The corresponding charge density would be $(1,0)$ or $(0,1)$. 
(iv)~We will see later that as $T_{el}$ increases the tendency to
have a modulation $x_-$ reduces.

\begin{figure}[b]
\centerline{
\includegraphics[width=8.25cm,height=4.6cm]{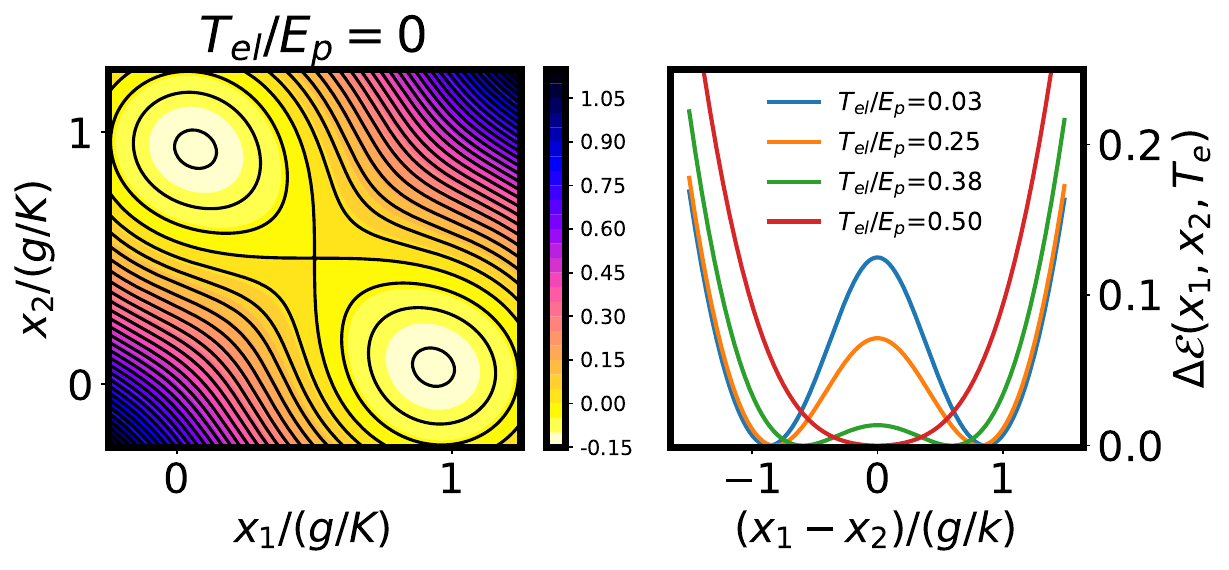}
}
\caption{(a)~Effective potential $V(x_1,x_2)$ at $T_{el} = 0$, and
(b)~the nature of $V(x_1-x_2)$ with increasing $T_{el}$.
}
\end{figure}

Putting these together we propose a phonon model of the form
\begin{eqnarray}
H^{eff}_{ph} ~~&=&~ \sum_i {{p_i^2} \over {2M}} + {K \over 2} \sum_i x_i^2
+ {1\over z} \sum_{\langle ij \rangle} V(x_i, x_j) \cr
\cr
V(x_i, x_j) &=&~
- T_{el} ln(e^{-\beta_{el} (\lambda_-(x_i,x_j)-\mu)}
+ e^{-\beta_{el} (\lambda_+(x_i,x_j)-\mu)})
\nonumber
\end{eqnarray}
Here $z$ is the number of nearest neighbours. In 2D, it is 4. 
$\mu$ is the chemical potential set to half filling 
condition in equilibrium. 
The parameters involved in $H^{eff}_{ph}$ are all defined
already in the Holstein model, except for $T_{el}$ - which
we extract by fitting the excited population in the full
MFD to a thermal distribution. That is, the MFD based
$n_{exc}(t,E_0)$ is mapped on to a $n_{exc}(t, T_{el})$.
Thankfully the time dependence of $T_{el}$ is simple
- it settles to its long time value quickly and the 
lattice dynamics plays out in the background of this
`final' electron temperature $T_{el}^f$. 

The effective potential for two sites can be
written as $V(x_{+},x_{-})=V_1(x_-)+V_2(x_+)$
and at half filling, this assumes the form 
\begin{eqnarray}
V_1(x_-) ~&=&~ \frac{1}{4}K x_{-}^2 
- T_{el}ln[e^{-\beta_{el}\bar{\lambda}}+e^{+\beta_{el}\bar{\lambda}}]\cr
V_2(x_+) ~&=&~ \frac{1}{4}Kx_+^2 - \frac{1}{2}gx_+\cr
\bar{\lambda}(x_-)~&=&~ \sqrt{\frac{1}{4}g^2x_-^2+t_{hop}^2} \nonumber
\end{eqnarray}
In Fig.14 we plot this 2 site effective potential 
as a function of $x_1$ and $x_2$ and we get two wells 
around ($0,~g/K$) and ($g/K,~0$) at $T_{el}=0$ (left).
When we plot the $V(x_-)$ with $T_{el}$ with
$x_+ = g/K$  we see a transition from double well to 
a single well around $T_{el}/E_p \sim 0.5$. 
Near this critical excitation $T_{el}^c$ 
the $x_-\rightarrow 0$, $x_+ \rightarrow 2\times x_{av}$,
where $x_{av}$ is the mean $x$ over the lattice ($g/2K$).
In this limit the function can be expanded in powers of 
$\xi_i = x_i - x_{av}$ and have the form
$$
V(\xi) = \sum_i (A \xi_i^2 + B \xi_i^4)  + C \sum_{ij} \xi_i \xi_j
$$
and higher order intersite terms. Here,
\begin{eqnarray}
    A(T_{el}) &=& \frac{1}{2}(K-\frac{g^2}{4t_{hop}}~\text{tanh}[\frac{t_{hop}}{T_{el}}])\cr
    B(T_{el})&=& \frac{g^4}{32t_{hop}^3}
    (~\text{tanh}[\frac{t_{hop}}{T_{el}}]-
    \frac{t_{hop}}{T_{el}}~\text{sech}^2[\frac{t_{hop}}{T_{el}}])\cr
    C(T_{el}) &=& \frac{g^2}{8t_{hop}}~\text{tanh}[\frac{t_{hop}}{T_{el}}]\nonumber
\end{eqnarray}

\begin{figure}[t]
\centerline{
\includegraphics[width=4.25cm,height=4.5cm]{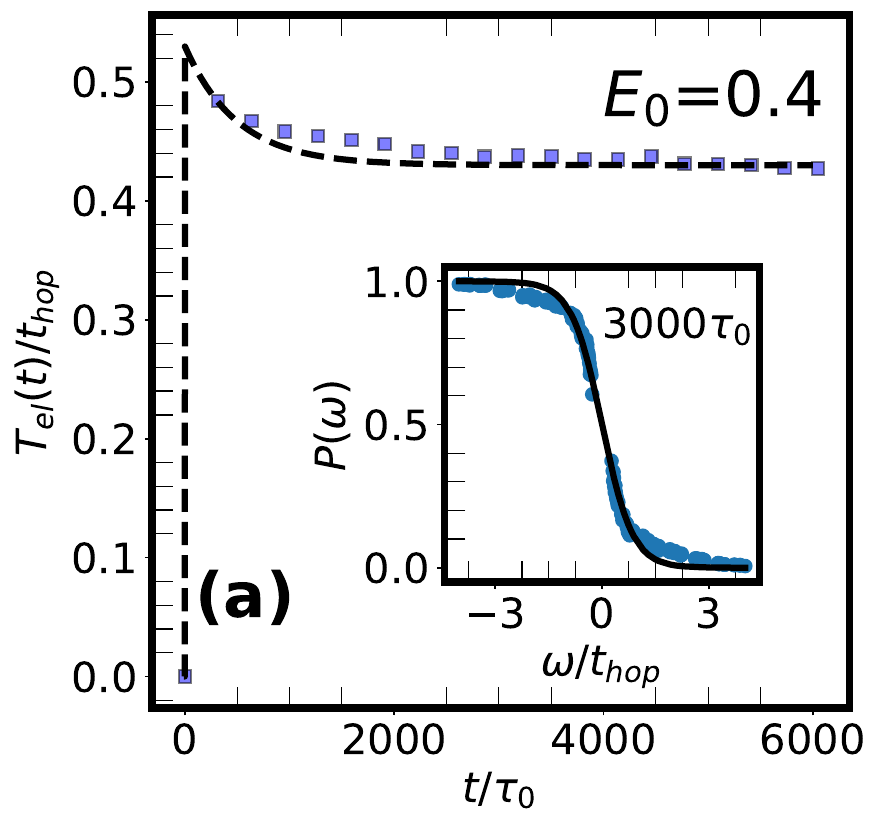}
\includegraphics[width=4.25cm,height=4.5cm]{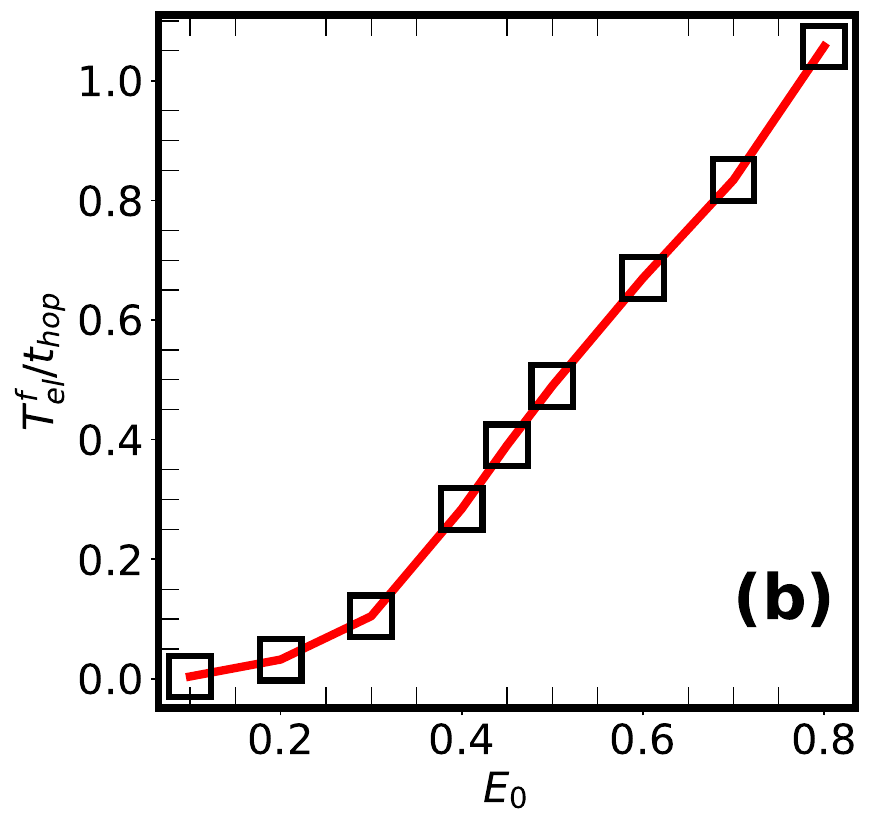}
}
\caption{Extracting an `electronic temperature': (a)~the
population function $P(\omega)$ obtained from MFD is fitted
to a Fermi function with temperature $T_{el}(t)$. $T_{el}$
settles down to a value $T_{el}^f$ over a timescale 
$\tau_{el}$ - the fit is shown (see comments in text). The inset
shows the Fermi function fit. (b)~The long time value of
$T_{el}$ for varying $E_0$. This is the crucial factor in
deciding the existence of long range order.
}
\end{figure}

\subsection{Scheme for dynamics}

With the excited electron population specified by $T_{el}$ 
the effective `force' driving $x_i$ is $-K x_i 
- ({\partial V}/{\partial x_i})$. 
If the $x_i$ sense a thermal 
environment with temperature $T_{ph}$ then the equation 
dictating $x_i$ dynamics would be:
$$
M\frac{d^2x_i}{dt^2}=  - {{\partial H^{eff}_{ph}} \over {\partial x_i}}  
-\gamma \frac{dx_i}{dt}  + \eta_i 
$$
where $\gamma$ is a dissipation parameter and $\eta$ is a random 
white noise satisfying:
$$
\langle \eta_i(t) \rangle = 0,~~~
\langle \eta_i (t)\eta_j (t')\rangle = 
2\gamma T_{ph} \delta_{ij} \delta(t-t')
$$
Both $T_{ph}$ and $\gamma$ enter
in our calculations as phenomenological constants. 
However, we can make an estimate of these phenomenological
constants from the MFD. The equation for oscillator in MFD is, 
$M\frac{d^2x_i}{dt^2}= - K x_i + g\rho_{ii}$.
This on-site density $\rho_{ii}$ can be written as 
the instantaneous term (calculated from the phonon background
with electronic population) and its fluctuation as 
$\rho_{ii}=n^{inst}_i +\delta n_i(t)$.
We assume this fluctuation to be of form 
$g\delta n_i(t)= -\gamma \frac{dx_i}{dt}+\eta_i(t)$.
At large frequency $|\delta n_i(\omega)|^2$ only contributes to 
$|\eta_i(\omega)|^2$ which, is a constant $\frac{2}{g^2}\gamma T_{ph}$.
From this we get an estimate for $\gamma T_{ph} \sim 10^{-5}t_{hop}$
at $E_0\sim0.4$.
In Fig.16, we get an estimate of $T_{ph}/t_{hop}$ which is 
between $10^{-3}$ to $10^{-4}$. So, our estimate for $\gamma$
is of order $10^{-2}$ to $10^{-1}$.

\begin{figure}[b]
\centerline{
\includegraphics[width=4.25cm,height=4.5cm]{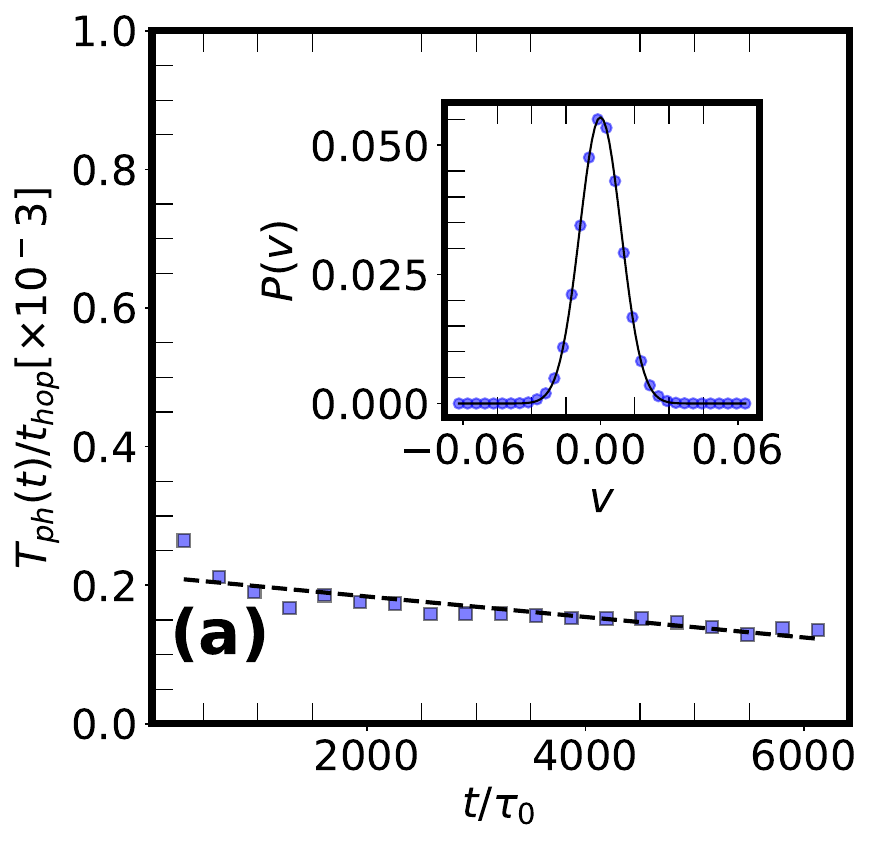}
\includegraphics[width=4.25cm,height=4.5cm]{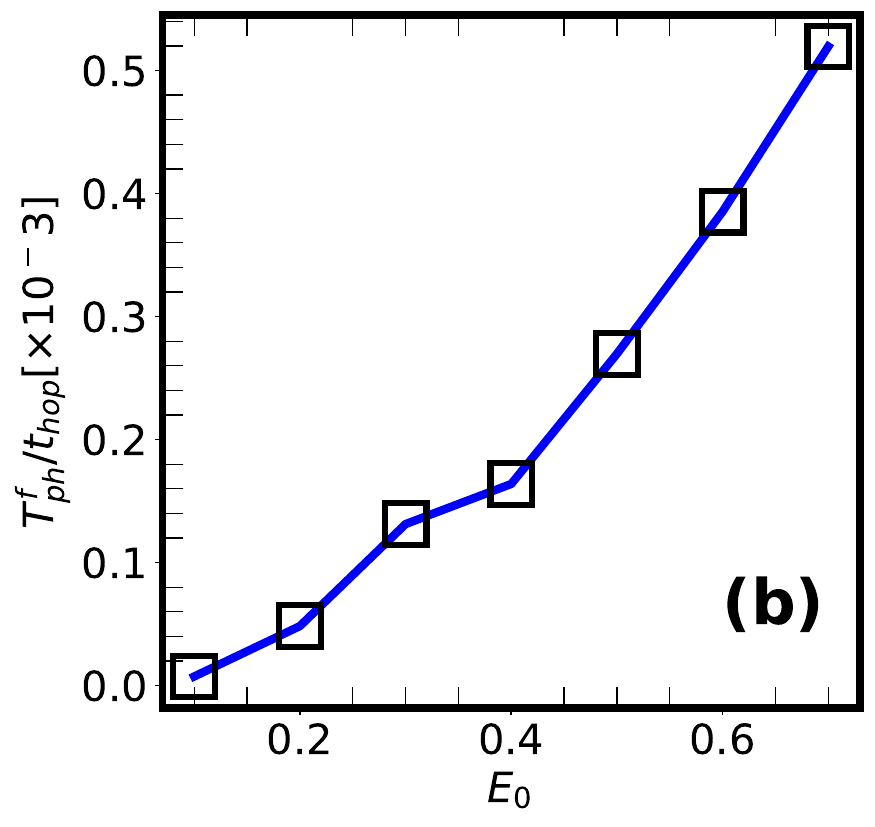}
}
\caption{Extracting an `phonon temperature': (a)~the
distribution function $P(v)$ constructed from velocity $\langle p_i\rangle/M$
which was obtained from MFD is fitted to a Maxwell-Boltzmann function
with temperature $T_{ph}(t)$. $T_{ph}$ is almost constant around a value
(For $E_0=0.4$, it is $2\times 10^{-4}$) which much lower than the $T_{el}^f$.
The inset shows the distribution function fit. (b)~The long time value of
$T_{ph}$ for varying $E_0$.
}
\end{figure}
\begin{figure}[t]
\centerline{
\includegraphics[width=8.5cm,height=4.5cm]{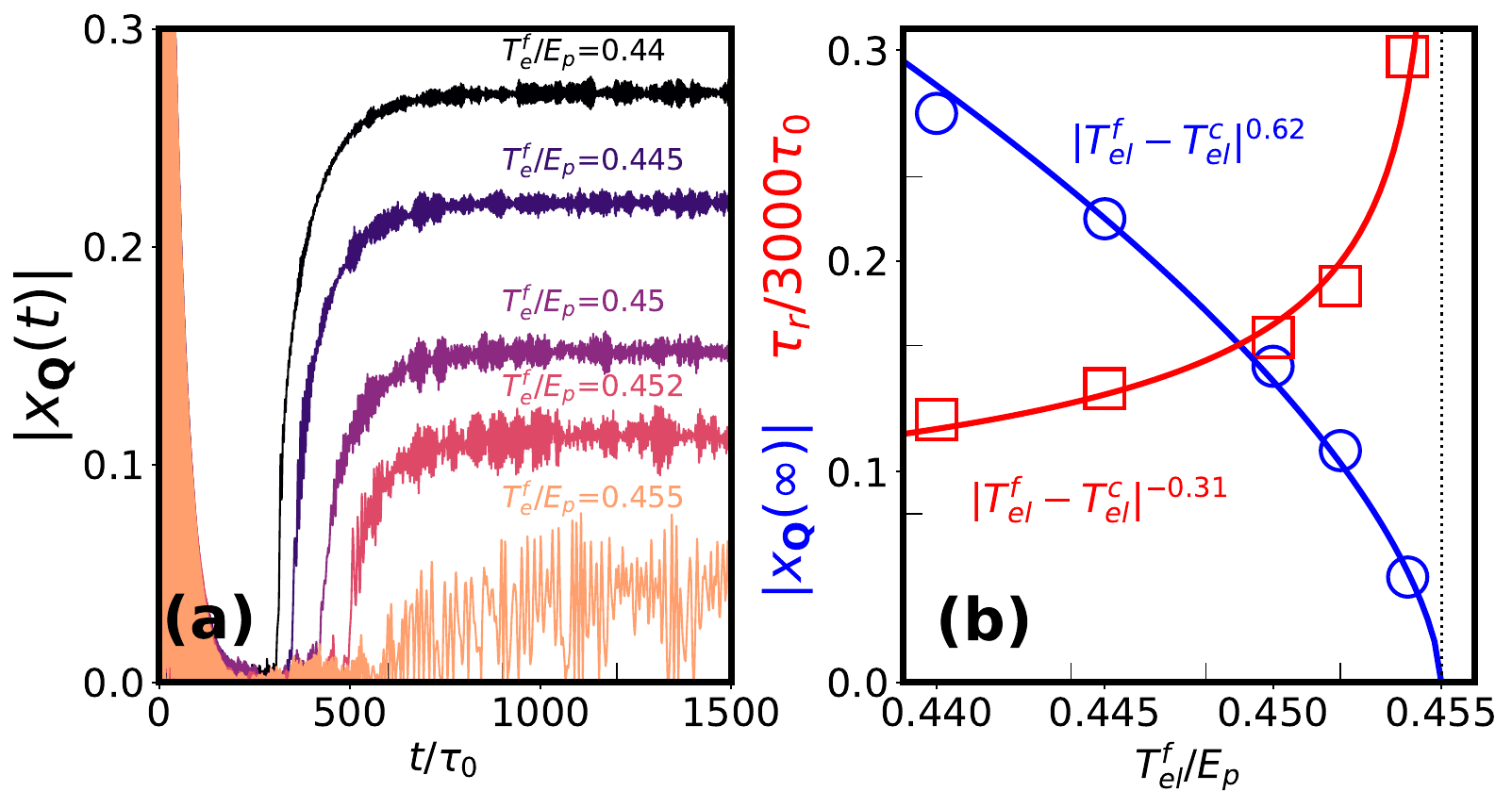}
}
\caption{In (a) the loss and revival dynamics of
$\vert x_{\bf Q}(t) \vert$ with varying $T_{el}^f$
is shown. (b)~As  $T_e^f/E_p \rightarrow$ $0.455$,
$\vert x_{\bf Q}(t) \vert
\rightarrow 0$ and recovery timescale
$\tau_{r}$ diverges.  These results
are on $40\times 40$ lattice size.
We show a fit $|x_{\bf{Q}}|\sim |T_e^f -T_e^c|^{\alpha}$
near the critical regime and we find $\alpha\sim {0.62}$.
The $\tau_{r}$ shows divergence as
$\tau_r \sim |T_e^f -T_e^c|^{-\nu}$ with
$\nu\sim 0.31$.
}
\end{figure}
\subsection{Electron population and Langevin parameters}

From the MFD we can track the excited population $n_{exc}(t, E_0)$
and map it to a $T_{el}(t, E_0)$. Fig.15(a) shows such a time
dependent profile for $T_{el}$ and the inset shows a long time
distribution function that is fitted by a Fermi function. 
The time dependence of $T_{el}$ can be fitted to:
$$
T_{el}(t, E_0)= T_{el}^i  e^{-t/\tau_{el}} + T_{el}^f(E_0)(1 - e^{-t/\tau_{el}})
$$
our MFD suggests $\tau_{el} \approx 10\tau_0$, while $T_{el}^f$
depends on $E_0$ in the manner shown in Fig.15(b).
We now set the rest of the parameters, 
$ t_{hop}=1$, $g=2$, $M=25$, $K=1$,
$\tau_0={2 \pi}/\sqrt{K/M}$, $E_p=g^2/{2K}$.
$\gamma =0.05$, $T_{ph}/t_{hop} =10^{-3}$.
$T_{el}^f$ is varied keeping $T_{el}^i/T_{el}^f=1.1$. 
We study $L \times L$ geometry with $L = 30, ~40,~60,~80,~100$.

\begin{figure*}[t]
\centerline{
\includegraphics[width=16cm,height=9.0cm]{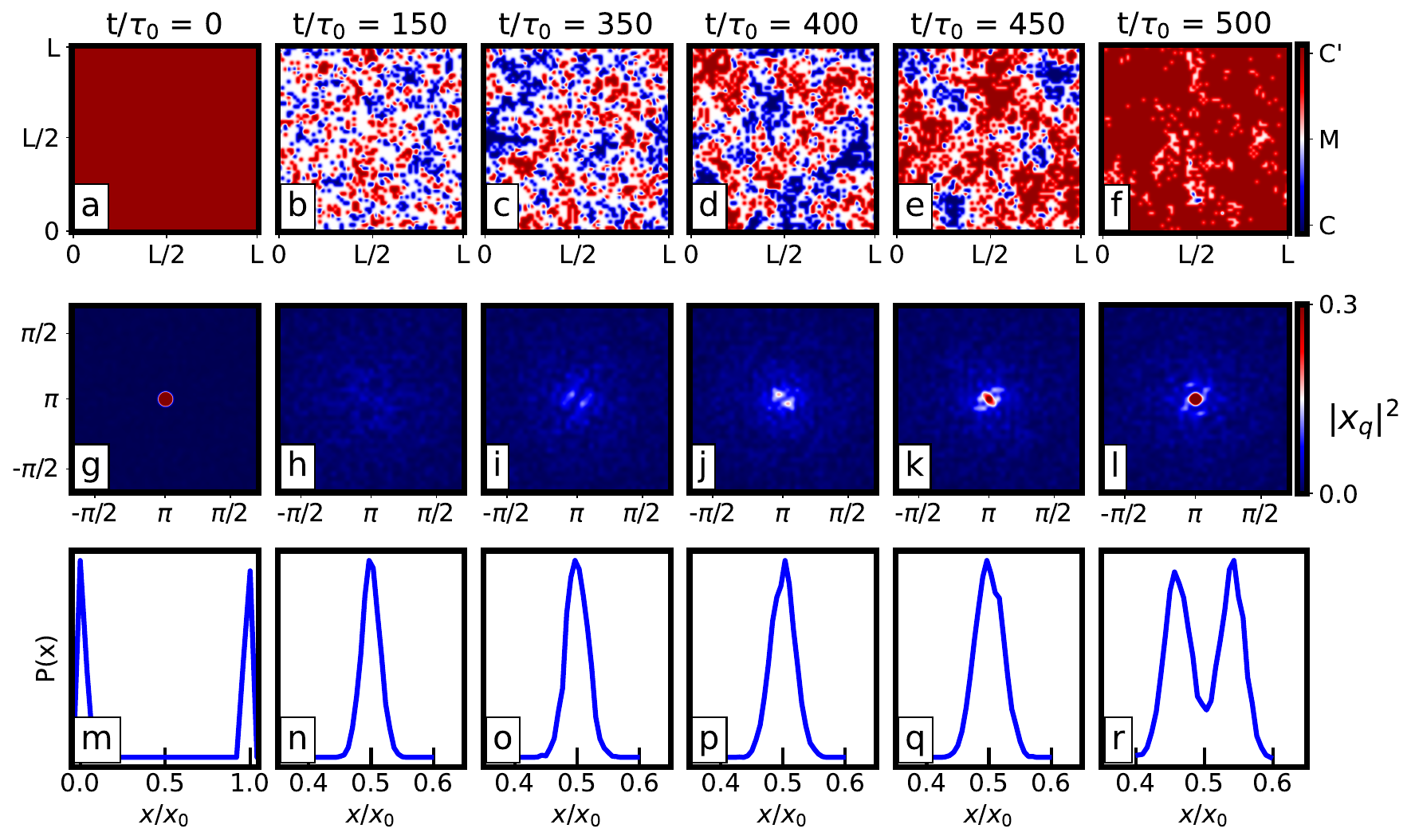}
}
\caption{Suppression and revival dynamics in the 2D classical model 
with $T_{el}^f = 0.45 E_p$ on a $60\times60$ lattice.
The top row shows charge order domains (see text). Red represents the 
$C$ type checkerboard pattern, blue denotes the phase shifted state $C'$, 
while white indicates no modulations - referred
to as the $M$ state. At $t=0$, a perfect $C$-state exists,
disrupted by the pump over time. At approximately $150\tau_0$,
equal amounts of red, white, and blue domains emerge.
Around $350\tau_0$, blue and red domains start to cluster,
reducing the $M$-state, leading to a CO state $C$ by
$500\tau_0$.  The middle panel shows $\vert x_{\bf q} \vert ^2$
with the suppression and recovery of the $(\pi, \pi)$ peak with
time.  The bottom row displays time dependence of the $P(x)$ 
distribution.  The sharp bimodal feature at $t=0$ homogenizes by 
$150\tau_0$.  This persists to $300\tau_0$, before a broadened 
distribution reemerges around $350\tau_0$. At $500\tau_0$, the 
`long time' distribution appears with a much smaller modulation 
in $x$ compared to $t=0$.
}
\end{figure*}

\subsection{Results on order parameter loss-recovery dynamics}

Our investigation of this classical model 
reveals three distinct regimes dependent on $T_{el}^f$, akin 
to the findings in MFD. Slight variations in 
$T_{el}^f$ (below $T_{el}^f \sim 0.3 E_p$) induce oscillations 
in $x_{\textbf{Q}(t)}$, similar to ringing. These oscillations 
dampen over time due to the phenomenological damping $\gamma$, 
with a time scale proportional to $\tau_w\sim 1/\gamma$. 
However, this time scale remains constant by design in 
this simplified model and cannot capture the variation of 
dissipation time with pump excitation.
Similarly, for $T_{el}^f>0.455~E_p$,
no recovery occurs, 
and the dissipation time scale is determined solely by 
$\gamma$.
We focus in Fig.17 on the critical dynamics window. 
In 17.(a), we observe the loss and recovery dynamics 
in $|x_{\bf{Q}}(t)|$. As $T_{el}^f\rightarrow 0.45E_p$, 
a transition becomes evident. 
We fit the recovery part of $|x_{\bf{Q}}(t)|$ with
$|x_{\bf{Q}}(t)|\sim X_0 ~ e^{-(\tau_r/t)^{\beta}}$
where $X_0=|x_{\bf{Q}}(\infty)|$ and $\beta$ ranges from $4-6$.
In 17 (b), we depict the dependence of $|x_{\bf{Q}}(\infty)|$
and $\tau_r$ on $T_{el}^f$. 
The recovered $|x_{\bf{Q}}|$ 
exhibits scaling $\sim |T_e^f-T_e^c|^{\alpha}$, 
with $\alpha={0.62}$.  The recovery time is fitted with 
$\tau_r\sim |T_e^f -T_e^c|^{-\nu}$ with $\nu=0.31$.

Fig.18 top panel illustrates the real space dynamics of domain
formation and growth in a system sized $60\times60$ with 
 $T_{el}^f=0.45~E_p$. In equilibrium, lattice distortions 
 exhibit a modulation of approximately $\sim g/2K$ 
 around the mean distortion, which is also $\sim g/2K$. 
 By subtracting the average distortion $g/{2K}$ 
at each site and multiplying by a periodic modulation 
$e^{i{\bf Q}\cdot \vec{r}_i}$,
  we homogenize the periodic structure to a positive ($C$-state) 
  or negative ($C'$-state) uniform field in the perfectly charge 
  ordered (CO) background. In the absence of a background 
  CO structure, the amplitude is zero
- we term it the $M$-state.

Following the `pump,' the initial $C$-state diminishes, and
 around $t\sim 150 \tau_0$, equal amounts of $C$, $C'$, 
 and $M$-type exist in a short range correlated manner.
 By $t\sim 350 \tau_0$, the $C$ and $C'$ states begin to
dominate, and a domain competition 
 ensues with $C$ eventually 
 prevailing around $t\sim 450 \tau_0$.
The middle panel depicts the structure factor $|x_{{\bf q}}|^2$, 
reflecting the loss and recovery of weight at ${\bf q}={\bf Q}$. 
The bottom row shows the time dependence of the $P(x)$
 distribution. The sharp bimodal feature at $t=0$ homogenizes 
 by $150\tau_0$ and persists until $300\tau_0$, before a 
 broadened distribution reemerges around $350\tau_0$.
  By $500\tau_0$, a `long time' distribution emerges with 
  a much smaller modulation in $x$ compared to $t=0$.

\subsection{Identification of three regimes from the
classical model}

As a result of increasing $T_{el}^f$ the $x_{\bf Q}(t)$ in 2D shows 
similar dynamical regimes- WOS, SSR, MS as shown in Fig. 19.
Similar analog of these regimes can be seen in $x_-(t)$ for 2 site.
We examine the energertics of the 2-site model first,
in response to an increase in $T_{el}$ and then move
to the lattice model. In the 2-site setup, the key variable is 
$x_-=x_1-x_2$, a crude mimic of $x_{\bf Q}$ on
the lattice. We denote $x_-$ by $y$ in what follows.
The effective potential 
$V_{eff}(y,T_{el})$, depicted in Fig.14, reveals a 
double-well structure with pre-pulse minima at 
$\pm y_{min}(T_{el} = 0)$. An abrupt change in $T_{el}$ 
alters the minima locations  to a  suppressed value 
$\pm y_{min}(T_{el})$, accompanied by a reduction in 
the `barrier height' $\Delta_b(T_{el})$
between the minima and the maximum.
Alteration of the potential makes the original minimum 
a `high energy' location now, with an  excess energy 
$$
\mathcal{E}(T_{el}^f)=V(y_{min}(0), T_{el}^f) -
V(y_{min}(T_{el}^f), 0)-\Delta \mathcal{E}(T_{el})
$$
Where, $\Delta \mathcal{E}(T_{el}) = V(0, T_{el}^f)-V(0, 0)$.
The two site problem cannot settle down into its new 
minimum unless this excess energy is dissipated.
We categorize the possibilities as follows:
\begin{figure}[!htbp]
\centerline{
\includegraphics[width=8cm,height=7cm]{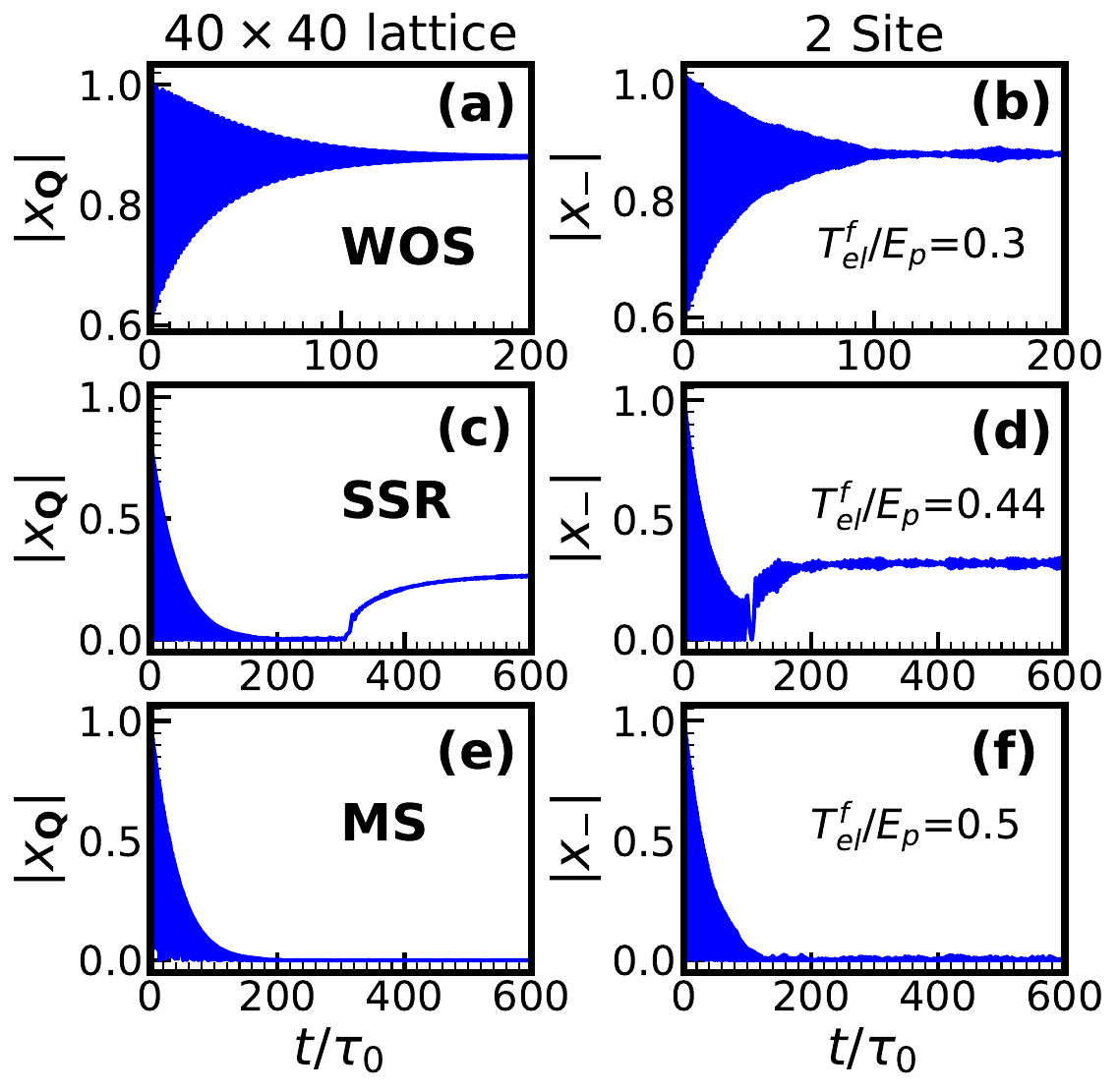}
}
\caption{In the first panel, we observe weak
oscillatory suppression (WOS) with $T_{el}^f/E_p=0.3$
in $x_{\bf{Q}}$ on a $40\times 40$  lattice (a).
In (b), we observe similar dynamics in $x_-$. Both
are normalized to their equilibrium value. Panels
(c) and (d) show strong suppression and recovery 
(SSR) dynamics at $T_{el}/E_p = 0.44$. While their
revival values are the same, their dynamics differ
in detail. At $T_{el}/E_p =0.5$, we observe monotonic
suppression (MS) in both the 2D and 2-site cases (e-f).
}
\end{figure}
(i)~Suppressed Oscillations ($0 \leq T_{el}^f \leq T_{el}^w$): 
As $T_{el}^f$ increases, the excess energy $\mathcal{E}(T_{el}^f)$ rises
while $\Delta_b(T_{el}^f)$ begins to decrease. At $T_{el}^f =
 T_{el}^w$, these values intersect ($\mathcal{E}(T_{el}^f)=
 \Delta_p(T_{el}^f)$). For $T_{el}^f < T_{el}^w$, oscillations in
  $|x_-(t)|$ are confined to a single well within the system. 
  Dissipation causes $\mathcal{E}(T_{el}^f)$ to decay as 
  $\sim \mathcal{E}(0) e^{-\gamma t}$, and $|x_{-}(t)|$ 
  stabilizes to the steady-state value $|x_{-}(T_{el}^f)|$ 
  over a time scale $\tau_w \sim {1\over \gamma}$.

(ii)~Monotonic Suppression ($T_{el}^f>T_{el}^c$): If 
$T_{el}^f$ surpasses the critical value such that 
$\Delta_b(T_{el}^f)=0$, no double-well structure exists 
in $V(x_1 - x_2, T_{el}^f)$. The $|x_1(t)-x_2(t)|$ decays 
to a homogeneous state with a similar time scale 
$\sim {1\over \gamma}$.

(iii)~Loss and Recovery ($T_{el}^w \leq T_{el}^f\leq T_{el}^c$): 
In this regime, despite the presence of a double well in 
$V(x_1 - x_2, T_{el}^f)$, the excess energy is significant 
enough to facilitate $x_1(t)-x_2(t)$ transitioning between 
wells. This $x_1(t)-x_2(t)$ can condense into one of the 
wells of $V(x_1 - x_2, T_{el}^f)$ when $\mathcal{E}(t
=\tau_r,T_{el}^f)=\Delta_b(T_{el}^f)$. Hence, $\tau_r = 
{1\over \gamma} \ln(\mathcal{E}(T_{el}^f)/\Delta_b(T_{el}^f))
\sim F(T_{el}^f) - {1\over \gamma} \ln(\Delta_b(T_{el}^f))$. 
Here, $F(T_{el}^f)$ is non-critical. This divergence of the
 timescale in this case is purely local. Nonetheless, this analysis 
  showcases how three regimes can arise due to changes 
  in $T_{el}$, even in a 2-site problem as shown in Fig 19.
  
In 2D, these regimes are also present and the corresponding 
$T_{el}$ values also match at regime boundaries. For `WOS' and 
`MS' we see similar dynamics in $x_{\bf Q}$ as shown in Fig.19.
In the 2D classical model, however, the relaxation
toward the steady state in `SSR' regime 
encompasses both amplitude growth and spatial 
organization, which are intricately intertwined, 
evident in the broad width of the distribution $P(x)$.
To analyze this, in Fig 20, we plot the recovery time $\tau_r$ 
against $\delta T_{el} =|T_{el}^f-T_{el}^c|$ for 
various system sizes $L=40,~60,~80,~100$. 
Notably, the disparity in recovery time escalates 
more rapidly with increasing system size, 
particularly pronounced for smaller values of $\delta T_{el}$
but the base value on top of which it grows is already very 
large indication a weak system size dependence.

\begin{figure}[t]
\centerline{
\includegraphics[width=5.2cm,height=5.0cm]{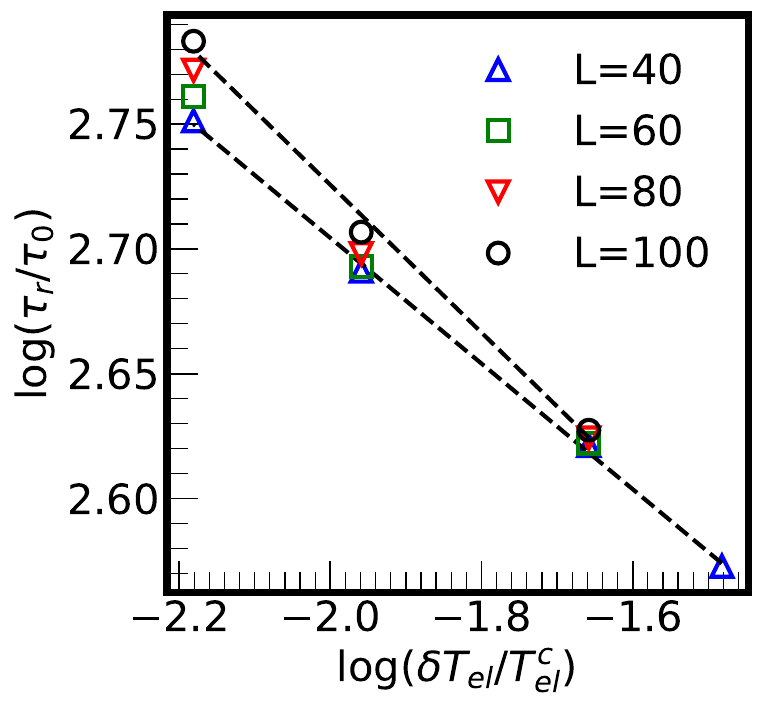}
}
\caption{Log-log plot of the recovery time $\tau_r$
vs $\delta T_{el} =|T_{el}^f-T_{el}^c|$ for different
system sizes. The difference in recovery time increases
with system size faster for small values of $\delta T_{el}$.
We plot this for L=40,60,80,100. Dotted lines correspond to 
linear fittings to L=40 and L=100 system sizes. 
}
\end{figure}

\begin{figure}[!bp]
\centerline{
\includegraphics[width=8.5cm,height=4.6cm]{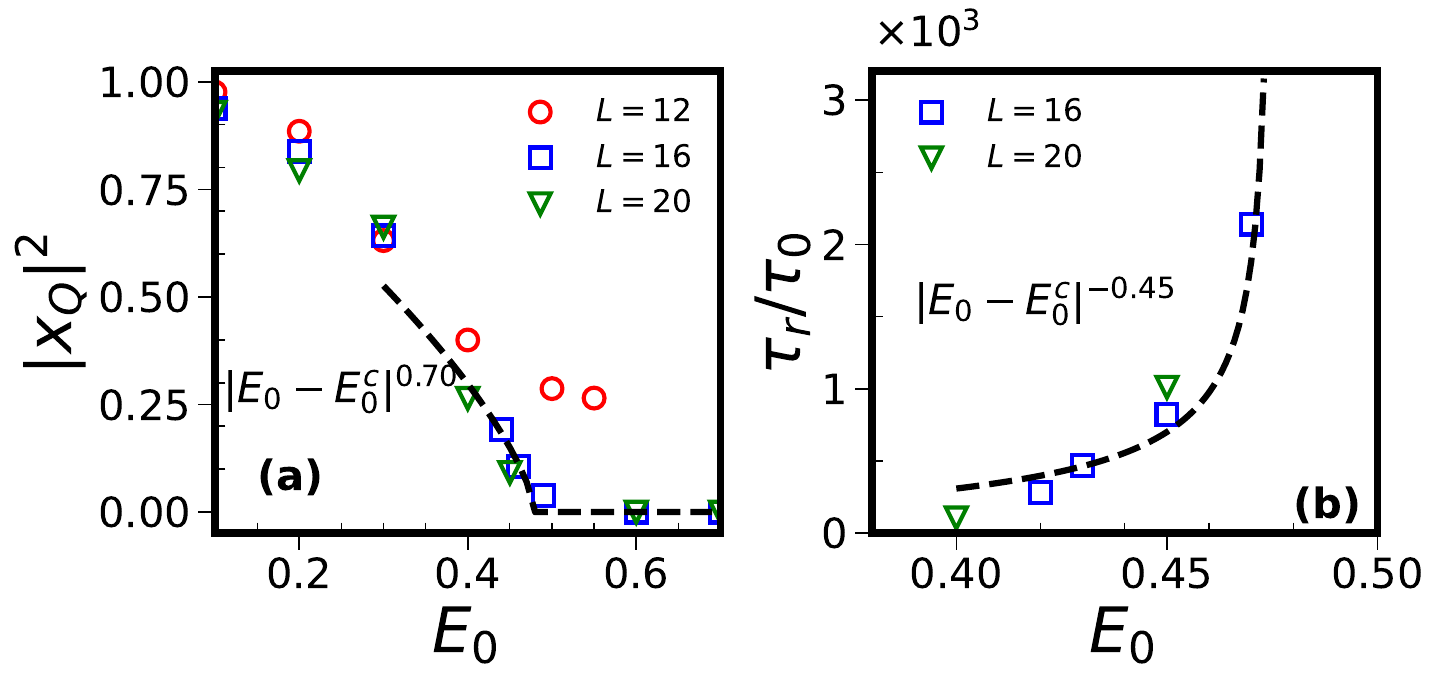}
}
\caption{We show (a)~$|x_{\bf{Q}}|^2$
and (b)~recovery time $\tau_{r}$ (b) obtained from
MFD for various system sizes. Fitting
$|x_{\bf{Q}}|^2$ to  $|E_0-E_0^c|^{\alpha}$
suggests a critical value $E_0^c \sim 0.48$ with
$\alpha=0.7$. For a given $\tau_{max}$, capturing
$\tau_{r}$ for larger systems ($L=20$) proves
challenging, while for $L=12$, the system exhibits
unclear loss-recovery dynamics. Notably, for $L=16$,
$\tau_r$ demonstrates a good fit with
$|E_0 - E_0^c|^{-\nu}$,
where $\nu\sim 0.45$.
}
\end{figure}

\section{Discussion}

\subsection{Size effects in MFD: locating the critical point}

The MFD computation time scales as $N^2 N_{\tau}$, where $N$ is the number of lattice sites and $N_{\tau}$ is the number of integration timesteps in the interval $[0-\tau_{max}]$. Within resource limits, we could access $N \sim 400$ and $N_{\tau} \sim 10^6$ (involving about $10^4-10^5$ phonon oscillations). While this was adequate in the weak and strong pulse regimes, accessing the critical regime, where the recovery time $\tau_{cr}$ diverges as $E_0 \rightarrow E_0^c$, was difficult. We had to extract the $t \rightarrow \infty$ value of $x_{\bf Q}$ by fitting. Fig. 21 shows the result of $x_{\bf Q}(t \rightarrow \infty)$ as a function of $E_0$ for different $L$.

We fit $|x_{\bf{Q}}|^2$ to $|E_0-E_0^c|^{\alpha}$, which suggests $E_0^c \sim 0.48$ with $\alpha = 0.7$. Similarly, $\tau_r$ demonstrates a good fit with $|E_0 - E_0^c|^{-\nu}$, where $\nu \sim 0.45$. However, due to the size limitations of our simulations, we must acknowledge that these exponents are not definitive. We have done our best to fit the data to our current knowledge, but larger system sizes would be required to precisely determine the critical exponents.

\begin{figure}[t]
\centerline{
\includegraphics[width=5.65cm,height=5.0cm]{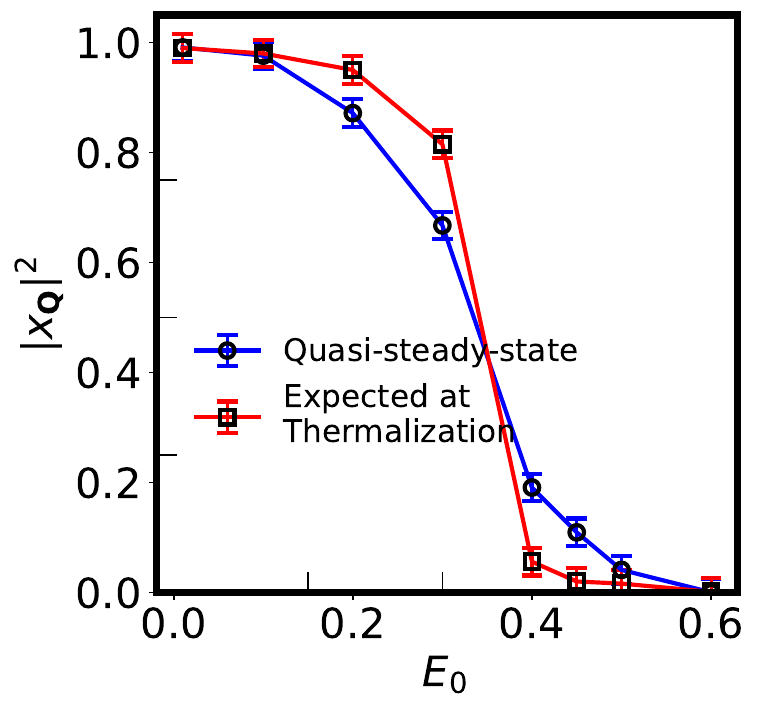}
}
\caption{Associated with the laser pulse amplitude $E_0$
we calculate the excess energy $\Delta\mathcal{E}$ and compare 
the structure factor $|x_{\bf{Q}}|^2$ at steady state (blue) with
the system with same $\Delta\mathcal{E}$ at thermal equilibrium (red).
This suggests if the system was able to redistribute it's excess
energy and reach an internal thermalisation, the structure factor 
would be similar though the nature of the suppression would change
from amplitude driven to fluctuation driven.}
\end{figure}

\subsection{Thermalisation}

Mean field treatment of a photo-pumped system often cannot capture
the eventual thermalisation, the effective electron and phonon
temperatures remain different. To that extent one could ask what
is the relevance of the long time state that emerges in our 
calculations.
There are two aspects we briefly comment on (i)~an estimate of
the thermalisation time based on work that has been done on
other gapped systems, and (ii)~the comparison of the state that
we obtain at a pump intensity $E_0$, and associated excess
energy $\Delta {\cal E}$, with the equilibrium state that would
have resulted if the system had excess {\it thermal energy}
$\Delta {\cal E}$.

Regarding thermalisation time specific calculations have been
done on Mott insulators. The conclusion there is that a pump
pulse excites electrons across the gap $\Delta$, creating
double occupancy, and these electrons deexcite by multimagnon
emission, each magnon having an energy $\sim J = 4t^2/U$,
where $U$ is the Hubbard repulsion. An early estimate of
the decay time was
provided by Strohmaier et al~\cite{Rajdeep}
who suggest a result of the form $\tau_D 
\sim {\hbar \over J} e^{(\alpha {U \over zJ})}$,
where $z$ is the coordination number of the lattice. 
The coefficient $\alpha \sim {\cal O}(1)$. The essence 
of this result is that the time for emitting multiple 
`bosons', each of energy $\sim zJ$, to dexcite an electron
at at energy $U \gg zJ$ is exponentially large.
The emission processes have to act in sequence. This
was established through an exact diagonalisation
calculation by Lenarcic and Prelovsek~\cite{Lenarcic}.
We are not aware of an equivalent calculation for the
Holstein model but the physical process invoked in the
Mott insulator suggests that electrons excited across the
charge ordering gap $\Delta_{CO}$ would deexcite by 
emitting phonons with energy $\sim \hbar \Omega_0$.
In model problem the gap is $\Delta_{CO} \sim 1.5t_{hop}$
and the phonon frequency is $\sim \hbar \Omega_0 
= 0.2 t_{hop}$. Even with intersite coupling the phonons
have a narrow band of energies around 
$\hbar \Omega_0$. 
As a result, following, Strohmaier et al,
we expect 
$\tau_D \sim \tau_0 exp({ \Delta_{CO} \over {\hbar \Omega_0}})$.
At our parameter point this is $\sim 1800 \tau_0$.
So, in a fuller calculation we expect  that 
the intermediate time dynamics that we see, on the 
scale of few hundred $\tau_0$,
would remain, but at longer times the upper band
population will relax leading to a common temperature
for phonons and electrons.
Our `divergent' timescale would be cut off.

We can compare the
long time state - characterised by an added energy $\Delta 
\mathcal{E}(E_0)$
- with what arises in a equilibrium thermal situation witn the
same energy $\Delta \mathcal{E}(T)$. The equilibrium `excess energy'
$\Delta \mathcal{E}(T) = \mathcal{E}(T) - \mathcal{E}(0)$ can be
worked out from Monte Carlo or Langevin calculation on the
electron-phonon system. Once the map from $E_0$ to $T$ is
obtained we can compare the order parameter in the long time
pumped state with a thermal state at the same excess energy.
The comparison is in Fig.22. While the critical $E_0$ 
estimate is reasonable, 
the actual thermal transition and the
pump drive transition have the following difference.
The suppression in $|x_{\bf{Q}}|^2$ in the photo-pump case
is due to the amplitude suppression but the suppression 
in $|x_{\bf{Q}}|^2$ in thermal case is more Ising like,
where the $x_i$ amplitude distribution remains roughly
similar at zero and finite temperature but interference 
between domains leads to loss of long range order.


\section{Conclusions}

We have studied the effect of a short laser pulse on a charge
ordered system, realised in the two dimensional half filled
spinless Holstein model. We work at intermediate electron-phonon
coupling, roughly  $\sim 40\%$ of the coupling needed for a single 
polaron formation, and study the coupled dynamics of the lattice
variables $(x_i)$ and the electronic correlator $\rho_{ij}(t)
= \langle c^{\dagger}_i(t) c_j(t) \rangle$ within a mean-field
dynamics scheme.
The method is nonperturbative in electron-phonon
coupling and handles spatial correlations exactly. The
dynamics can be categorized into three regimes with
increasing pulse strength. 
At weak pulse strength we find a small oscillatory 
suppression of the order parameter to a finite long time
value. At intermediate pulse strength (which we label
as the critical regime) the dynamics show strong
suppression and revival of the order parameter. This
involves a rapid drop of the order parameter to almost zero,
where the system stays for a time $\sim \tau_{cr}$, and
then a power law rises to a finite long-time value. 
We find that $\tau_{cr} \rightarrow \infty$ and the
long time value $\rightarrow 0$, as the pulse strength
$E_0$ tends to a critical value $E_0^c$. This defines
a nonequilibrium phase transition. We have established the
transition by studying the system on accessible lattice
size (up to $20 \times 20$) and time windows. For a strong
pulse, the CO order parameter decays monotonically to
zero. Associated with the CO order parameter behavior 
is an gap closing and re-opening transition in the
critical regime and a gap closing transition
in the strong pulse regime. 
Since full mean field dynamics is
expensive on large lattices we constructed an effective
classical model for the phonons where the phonon dynamics
is driven by a nonequilibrium electron population derived
from MFD. This model captures all the three regimes that
one sees in MFD and provides better access to critical
properties than MFD does. Both MFD and the classical model
suffer from lack of equilibriation between electrons
and phonons and our results would be modified when these
`beyond mean field' effects are incorporated. We provide an
estimate of this timescale.

\vspace{.3cm}

{\it Acknowledgment:} We acknowledge the use of the HPC 
clusters at HRI.
\bibliography{copp}

\end{document}